\documentclass[twocolumn,amsmath,aps,nofootinbib,superscriptaddress]{revtex4}
\usepackage{graphicx}
\usepackage{subfigure}
\usepackage{bm}
\usepackage{amsmath}
\usepackage{mathrsfs}
\usepackage{amsthm}
\usepackage{fixltx2e}
\usepackage{xcolor}
\usepackage{amssymb}
\usepackage{url}

\usepackage[normalem]{ulem}   
\usepackage[title]{appendix}
\usepackage{color}

\makeatletter
\newcommand{\rmnum}[1]{\romannumeral #1}
\newcommand{\Rmnum}[1]{\expandafter\@slowromancap\romannumeral #1@}
\makeatother

\newtheorem{proposition}{Proposition}[section]

\begin{document}

\title{Propagating Instability for Wave Dark Matter}

\author{Ui-Han Zhang}
\email{uihanzhang@ntu.edu.tw}
\email{zhanguihan@gmail.com}
\affiliation{Institute of Astrophysics, National Taiwan University. 10617, Taipei, Taiwan}

\author{Tak-Pong Woo}
\email{bonwood@phys.ntu.edu.tw}
\affiliation{Institute of Astrophysics, National Taiwan University. 10617, Taipei, Taiwan}
\affiliation{Department of Physics, National Taiwan University, 10617, Taipei, Taiwan}

\author{Tzihong Chiueh}
\email{chiuehth@phys.ntu.edu.tw}
\affiliation{Institute of Astrophysics, National Taiwan University. 10617, Taipei, Taiwan}
\affiliation{Department of Physics, National Taiwan University, 10617, Taipei, Taiwan}
\affiliation{Center for Theoretical Physics, National Taiwan University, 10617, Taipei, Taiwan}


\begin{abstract}
\label{abstract}
In the early Universe, large-scale flows were omnipresent, and the flow collisions produced sheets and filaments. This phenomenon occurs for particle dark matter, and so does it for wave dark matter. But for the latter, these sheets and filaments are the modulations of even finer-scale, large-amplitude interference fringes. This work aims to investigate the instability of the interference fringes arising from colliding waves. Two colliding streams in classical collisionless fluid systems can often produce small-scale unstable oscillations with a finite complex frequency, a situation identified as propagating instabilities. In fact, propagating unstable oscillations have never been observed in the conventional quantum system due to its being Sturm-Liouville property. For example, quantum fluid equations with Madelung variables only exhibit either Jeans instability, a purely growing unstable mode, or stable oscillations, for which the squared frequency, $\omega^2$, is real. Despite that, this work discovers that quantum interference fringes can indeed generate propagating unstable oscillations with a complex $\omega^2$ when the gravitational feedback perturbation is included. The presence of local density nulls in the background density is shown to be the necessary condition for such an instability. We establish a phase diagram separating the propagating instability, Jeans instability, and stable oscillation regions, and is verified by computer simulations. Generally speaking, Jeans instabilities tend to occur for long-wave density perturbations as expected; propagating instabilities on the other hand tend to occur for short density waves with wavelengths comparable to the fringe size, i.e., near the center of the Bloch zone; lastly, both instabilities diminish for very low density fringes. The propagating unstable fluctuation may possibly collapse into halos of small sizes, potentially seeding the formation of proto-globular clusters.
\end{abstract}

\maketitle
\section{Introduction}
\label{sec:introduction}

Wave dark matter with an extremely small particle mass of $10^{-22}$ eV described by the Schr\"{o}dinger-Poisson equation has emerged as a promising candidate for the dark matter \cite{Hu2000, WC2009, Schive2014, MP2015, CS2016, GMPU2017}. A novel feature is the universal presence of a smooth massive core, the soliton core, at the central kpc of a galaxy\cite{Schive2014,LM2014}. This characteristic mitigates the discrepancy between observed data \cite{Moore1994, Goerdt2006, Gilmore2007, WP2011, AAE2013} and the predictions of the conventional cold dark matter (CDM) model, which forecasts a density singularity at the halo center \cite{NFW1997}.

On the other hand, large-scale structure has revealed the presence of cosmic filaments and sheets \cite{CAPHM2014, BMGBMSWCNLetal2021}. These filaments and sheets are also evident in the cosmological simulation for wave dark matter \cite{WC2009, Schive2014}, which additionally exhibit fine periodic density variations on the top of sheets and filaments due to quantum interference \cite{Schive2014}. Furthermore, these sheets and filaments have been known to be unstable and are sites for collapsed halos in both particle and wave dark matter models. We are so motivated to understand how these unstable fine fringes may seed the formation of small bound objects around and within a much larger galaxy halo. The early formation of globular clusters has so far been evaded from a sound explanation. The potential association of early globular clusters \cite{ABVCWDMOSAetal2024} and quantum interference fringes offers an interesting possibility.

The periodic density variation inside sheets and filaments can be perceived as the superposition of two colliding plane waves in quantum mechanics. Two colliding streams also occur in classical collisionless fluid systems, including plasma \cite{Bellan2006} and gravitational systems \cite{FP_first1984, FP_second1984, CMS1998, ACP2004}. We observed that counter-streaming classical collisionless systems often exhibit propagating instability, characterized by a complex frequency with a non-zero imaginary part that yields unstable oscillations in the linear regime \cite{Bellan2006, FP_second1984, CMS1998, ACP2004}. 

In contrast, conventional quantum systems are Hermitian systems, and it is impossible for it to be unstable even with a perturbed potential. Upon considering the self-consistency of the perturbed potential, the system can also be made Hermitian in the Sturm-Liouville sense \cite{AlGwaiz2008} and obeys the variational principle, as illustrated by \cite{CH2023}. Though the self-consistent perturbation can indeed be unstable, the instability can never be a propagating instability. We are now facing a dilemma, in that a standing wave in Schr\"{o}dinger equation is a result of counter propagating waves and one would expect it to possess propagating instabilities as in the classical system, but the very quantum nature of the system suggests that this is impossible. In this work, we will resolve the dilemma. We will show that propagating instabilities can only occur in very specific quantum systems, where the variational principle breaks down, and we will provide a general criterion for this to occur.

In detail, the paper by \cite{CH2023} theoretically examines the soliton core existing at the Galactic center. Due to the absence of local density null, a soliton can be represented by well-defined fluid variables via Madelung transformation. The stability of the soliton can be analyzed using classical fluid equations, which conform to Sturm-Liouville type equations. These equations exhibit either stable oscillations, characterized by a real frequency, or Jeans instability, characterized by a purely imaginary frequency \cite{FP_second1984}.   
 
The periodic density pattern in the interference fringe is nothing but a standing wave, and it has local density nulls, which turn out often to be the sites of singular velocity perturbations if a velocity exists at those sites. Consequently, unlike the soliton case, stability analysis using classical fluid equations cannot be directly applied to these periodic fringes, rendering it beyond the scope of Sturm-Liouville theory. In this work, a novel approach has been developed for analyzing the stability of configurations with local nulls. This approach not only provides evidence for the existence of propagating instability in periodic sheets, but also establishes general necessary conditions for unstable oscillations in quantum systems. 

This paper is organized as follows: Sec. (\ref{sec: General Properties of the linearized Schrodinger-Poisson equation}) presents the general mathematical properties of the linearized Schrödinger-Poisson equation, providing the necessary conditions for the presence of propagating instability. In Sec. (\ref{sec: Nonlinear periodic equilibrium construction}), a periodic equilibrium state resembling local interference fringes within cosmic sheets is constructed. Sec. (\ref{sec: General Version of Bloch's theorem for the Stability analysis to the Periodic equilibrium Background}) provides a new mathematical approach related to Bloch's theorem for the linear stability analysis of the periodic background configuration. Results from both theoretical calculations and simulations are presented in Sec. (\ref{sec: Comparison Between Theoretical Prediction and Simulating result}), and the conclusion is given in Sec. (\ref{sec: Conclusion}). 

Key features of the periodic background configuration and the corresponding mathematical proofs are given in Appendix \ref{appendix: Properties for the background configuration}, including the Fourier series of the background configuration, the non-zero radius of convergence, and the local null locations for the background wave function. Appendix \ref{appendix: Generalized Bloch's Theorem for the Operator ThetaH} offers a generalized version of Bloch's theorem, while Appendix \ref{appendix: The General Property of the Matrix Representation} outlines the mathematical derivations for the matrix representation of the relevant operators. Rigorous proofs establishing an upper bound on the perturbed self-gravity energy tensor are provided in Appendix \ref{appendix: Uniform Boundedness of the Perturbed Self-Gravity Energy Tensor in the Operator Theta}. Finally, the stability of two-component wave dark matter is studied in Appendix \ref{appendix: General Property for the dispersion relation of two-component wave dark matter}, which is a much simpler straightforward model to be compared with the results presented in Sec. (\ref{sec: Comparison Between Theoretical Prediction and Simulating result}).  

The following units are applied throughout this paper: $\hbar=4\pi G=1$ where $\hbar$ is the reduced Planck constant and $G$ is the gravitational constant. Additionally, the particle mass $m$ is appropriately chosen as the unit throughout this paper, except in Appendix \ref{appendix: General Property for the dispersion relation of two-component wave dark matter}.

\section{General Properties of Linear perturbation framework}
\label{sec: General Properties of the linearized Schrodinger-Poisson equation}

In this section, we will explore some general properties of the linearized Schr\"{o}dinger-Poisson equation. The conclusions and results presented here are applicable to perturbation around any equilibrium state of the Schr\"{o}dinger-Poisson equation.

The equilibrium background wave function, denoted $\Psi_{eq}(\vec{x})$, satisfies the following time-independent Schr\"{o}dinger-Poisson equation:

\begin{equation}
\label{equ: Schrodiger-Poisson equation for single component}
\left\{\begin{aligned}
& E_{eq}\Psi_{eq}(\vec{x}) = \Big (-{\nabla^2\over{2}} + V_{eq}(\vec{x}) \Big )\Psi_{eq}(\vec{x}), \\
& \nabla^2 V_{eq}(\vec{x}) = |\Psi_{eq}(\vec{x})|^2 - \Lambda, \\
\end{aligned}
\right.
\end{equation}
Here, $V_{eq}(\vec{x})$ and $E_{eq}$ are the background gravitational potential and eigenenergy, respectively. The quantity $\Lambda$ represents the cosmological constant that exerts a repulsive force opposing the gravity \cite{CMS1998}. The $\Lambda$ term, whose magnitude equals the mean mass density $\Big \langle |\Psi_{eq}(\vec{x})|^2 \Big \rangle$, is needed here to establish a force-balance equilibrium state. In the context of cosmology, it is the ram pressure of the Hubble flow instead that counters the gravity. Without a loss of generality, the background wave function $\Psi_{eq}(\vec{x})$ can be assumed to be a real function, as indicated by Eq. (\ref{equ: Schrodiger-Poisson equation for single component}). 

Next, the perturbed wave function should be expressed as $\delta \Psi(\vec{x},t) \times \exp{(-\textbf{\textit{i}}E_{eq}t)}$ to isolate the wave oscillation from the background eigenenergy. In this formulation, the perturbed wave function $\delta \Psi(\vec{x},t)$ satisfies the following linearized Schr\"{o}dinger-Poisson equation:

\begin{equation}
\label{equ: perturbed Schrodiger-Poisson equation for single component}
\left\{\begin{aligned}
& \begin{aligned}
\textbf{\textit{i}} {{\partial \delta\Psi(\vec{x},t)}\over{\partial t}} = & \Bigg (-{\nabla^2\over{2}} + V_{eq}(\vec{x}) - E_{eq} \Bigg )\delta\Psi(\vec{x},t) +  \\
&\delta V(\vec{x},t)\Psi_{eq}(\vec{x}),
\end{aligned} 
\\
& \nabla^2 \delta V(\vec{x},t) = \Psi_{eq}(\vec{x})\delta\Psi^{*}(\vec{x},t) + \Psi^{*}_{eq}(\vec{x})\delta\Psi(\vec{x},t). \\
\end{aligned}
\right.
\end{equation}
Here, $*$ means the complex conjugate operator, and $\delta V(\vec{x},t)$ is the perturbed gravitational potential, a real function, and its source must also be real. It should be noted that the perturbed Poisson equation, the second equation in Eq. (\ref{equ: perturbed Schrodiger-Poisson equation for single component}), implies that such perturbations do not introduce additional mass into the system and adhere to the mass conservation constraint.

Let the perturbed wave function $\delta \Psi(\vec{x},t) = R(\vec{x},t)+\textbf{\textit{i}}I(\vec{x},t)$, where $R(\vec{x},t)$ and $I(\vec{x},t)$ are real functions representing the real and imaginary parts of the perturbed wave function, respectively.  Since Equation (\ref{equ: perturbed Schrodiger-Poisson equation for single component}) is time-translation invariant, we apply a Fourier transformation with respect to time, resulting in:

\begin{equation}
\label{equ: perturbed Schrodiger-Poisson equation for single component by real and imagine part}
\left\{\begin{aligned}
& \omega^2\eta(\vec{x}) =  \Bigg (-{\nabla^2\over{2}} + V_{eq}(\vec{x}) - E_{eq} \Bigg )R(\vec{x}) + \delta V(\vec{x})\Psi_{eq}(\vec{x}); \\
& R(\vec{x})=\Bigg (-{\nabla^2\over{2}} + V_{eq}(\vec{x}) - E_{eq} \Bigg )\eta(\vec{x}); \\
& \nabla^2\delta V(\vec{x}) = 2\Psi_{eq}(\vec{x})R(\vec{x}),
\end{aligned}
\right.
\end{equation}
where $\omega$ is the angular frequency and $\eta \equiv -\textbf{\textit{i}}I/\omega$. For convenience, we use the same notation for physical quantities in both the time and frequency spaces. 

The square of the angular frequency, $\omega^2$, determines the stability of this quantum system, i.e., linearly stable if and only if $\omega^2$ is positive. Additionally, there are two types of instability when $\omega^2$ is not longer a positive value. When $\omega^2$ is a complex number with a non-zero imaginary part, it corresponds to a frequency with a non-zero imaginary part, leading to unstable oscillations, or the propagating instability. Conversely, a negative value of $\omega^2$ indicates a pure imaginary frequency and corresponds to a purely growing unstable mode. Therefore, throughout the paper we define the three types of stability behavior as follows:
\\\\
\textbf{1. $\omega^2 \in \mathbb{C} \backslash \mathbb{R}$(a complex number with a non-zero imaginary part) is referred to as propagating instability.}
\\
\textbf{2. $\omega^2 \in \mathbb{R}$ and $\omega^2<0$(a negative real number) is referred to as Jeans instability.}
\\
\textbf{3. $\omega^2 \in \mathbb{R}$ and $\omega^2>0$(a positive real number) is referred to as stable.}
\\\\
Here, $\mathbb{R}$ represents the set of all real numbers, while $\mathbb{C}$ denotes the set of all complex numbers. The set $\mathbb{C} \backslash \mathbb{R}$ consists of all complex numbers with non-zero imaginary parts.

Finding the dispersion relation for Eq. (\ref{equ: perturbed Schrodiger-Poisson equation for single component by real and imagine part}) is challenging due to the inhomogeneity of the background configuration. Therefore, a new methodology for studying the stability must be developed. To analyze properties for Eq. (\ref{equ: perturbed Schrodiger-Poisson equation for single component by real and imagine part}), the following operators are defined:

\begin{equation}
\label{equ: definition of H and Theta}
\left\{\begin{aligned}
& \hat{\mathbf{H}} \equiv  -{{1}\over{2}} \nabla^2 + V_{eq}(\vec{x}) - E_{eq}; \\
& \hat{\mathbf{\Theta}} \equiv  \hat{\mathbf{H}} +  \hat{\mathbf{G}}; \\
& \hat{\mathbf{G}} \equiv 2\Psi_{eq}(\vec{x}) \nabla^{-2} \Psi_{eq}(\vec{x}).
\end{aligned}
\right.
\end{equation}
Here, $\nabla^{-2}$ denotes the inverse Laplacian operator. It is evident that operators $\hat{\mathbf{\Theta}}$ and $\hat{\mathbf{H}}$ correspond to the operators on the right-hand side of the first and second equations in Eq. (\ref{equ: perturbed Schrodiger-Poisson equation for single component by real and imagine part}), respectively.

Using these symbols, Eq. (\ref{equ: perturbed Schrodiger-Poisson equation for single component by real and imagine part}) can be rewritten as follows:

\begin{equation}
\label{equ: linearized Schrodinger Poisson equation with abstract symbol}
\left\{\begin{aligned}
& \omega^2\eta = \hat{\mathbf{\Theta}}R; \\
& R = \hat{\mathbf{H}}\eta,
\end{aligned}
\right.
\end{equation}
from which the following relations can be established:

\begin{equation}
\label{equ: symbolic notation for linearized Schrodinger equation}
\left\{\begin{aligned}
& \omega^2\eta = \hat{\mathbf{\Theta}}\hat{\mathbf{H}}\eta; \\
& \omega^2 R = \hat{\mathbf{H}}\hat{\mathbf{\Theta}}R.
\end{aligned}
\right.
\end{equation}
Hence, Eq. (\ref{equ: perturbed Schrodiger-Poisson equation for single component by real and imagine part}) transforms into an eigenvalue problem with $\omega^2$ being the eigenvalue. The functions $\eta$ and $R$ serve as the corresponding eigenfunctions of the different operators $\hat{\mathbf{\Theta}}\hat{\mathbf{H}}$ and $\hat{\mathbf{H}}\hat{\mathbf{\Theta}}$, respectively.  According to Eq. (\ref{equ: linearized Schrodinger Poisson equation with abstract symbol}), the operators $\hat{\mathbf{H}}\hat{\mathbf{\Theta}}$ and $\hat{\mathbf{\Theta}}\hat{\mathbf{H}}$ share identical eigenvalues. Consequently, the stability analysis can be conducted using either operator $\hat{\mathbf{\Theta}}\hat{\mathbf{H}}$ or $\hat{\mathbf{H}}\hat{\mathbf{\Theta}}$ whenever it is convenient.

We adopt the conventional inner product in quantum mechanics.  The inner product is given as $\langle g|f \rangle = \int g^{*}(\vec{x})f(\vec{x})d^3\vec{x}$ for $L^2$-integrable functions $f$ and $g$, meaning $\int |f(\vec{x})|^2dx$ and $\int |g(\vec{x})|^2dx$ are finite quantities. For any operator $\hat{\mathbf{O}}$ and $L^2$-integrable functions $f$ and $g$, we define $\langle g | \hat{\mathbf{O}} | f \rangle \equiv \int g^{*}(\vec{x})\hat{\mathbf{O}}f(\vec{x})d^3\vec{x}$. If the operator $\hat{\mathbf{O}}$ is Hermitian, then $\langle g | \hat{\mathbf{O}} | f \rangle = \langle f | \hat{\mathbf{O}} | g \rangle^{*}$. Moreover, $\langle f | \hat{\mathbf{O}} | f \rangle$ is referred to as the expected value of the Hermitian operator $\hat{\mathbf{O}}$ for the $L^2$-integrable function $f$.

Since the Laplacian operator is Hermitian, the operators $\hat{\mathbf{H}}$ and $\hat{\mathbf{G}}$ are separately also Hermitian, as is the operator $\hat{\mathbf{\Theta}}$, as shown below. For any normalized trial function $f$, i.e., $\langle f|f \rangle =1$, $\langle f| \hat{\mathbf{H}} |f \rangle = \Big ( \langle \nabla f |  \nabla f \rangle/2 + \langle f | V_{eq} | f \rangle \Big ) - E_{eq}$ and $\langle f| \hat{\mathbf{\Theta}} |f \rangle = \langle f |\hat{\mathbf{H}} | f \rangle + \langle f |\hat{\mathbf{G}} | f \rangle$ with $\langle f |\hat{\mathbf{G}} | f \rangle = - \langle \nabla \delta V_f | \nabla \delta V_f \rangle/2$, where $\delta V_f$ is the perturbed gravitational potential  satisfying $\nabla^2 \delta V_f = 2\Psi_{eq}f$. Note that we have included $E_{eq}$ in the background Hamiltonian $\hat{\mathbf{H}}$ to shift its eigenvalue.

Here, $\langle \nabla f |\nabla f \rangle/2$, $\langle f | V_{eq} | f \rangle$, and $\langle f |\hat{\mathbf{G}} | f \rangle = - \langle  \nabla \delta V_f | \nabla  \delta V_f \rangle/2$ represent the expected kinetic energy, the external potential energy (the background gravitational potential treated as an external potential), and the perturbed self-gravity energy for the normalized trial function $f$, respectively. It should be noted that the operator $\hat{\mathbf{G}}$, a self-gravity Hermitian operator arising from the perturbed gravitational potential, consistently yields a non-positive the expected perturbed self-gravity energy.

For a given normalized trial function $f$, the expected value of $\hat{\mathbf{H}}$ measures the energy difference between the expected mechanical energy (the sum of kinetic and potential energy) without self-gravity, $\langle \nabla  f | \nabla f \rangle/2 + \langle f | V_{eq} | f \rangle$, and the eigenenergy of the background wave function, $E_{eq}$. 

Similarly, the operator $\hat{\mathbf{\Theta}}$ is analogous to the Hamiltonian operator. For a given normalized trial function $f$, the expected value of $\hat{\mathbf{\Theta}}$ represents the energy difference between the expected mechanical energy with self-gravity, $\langle \nabla f | \nabla f \rangle/2 + \langle f | V_{eq} | f \rangle - \langle  \nabla \delta V_f | \nabla \delta V_f \rangle/2$, and the eigenenergy of the background wave function.

Moreover, the sign of the expected values of $\hat{\mathbf{H}}$ and $\hat{\mathbf{\Theta}}$ indicates whether the expected mechanical energy is greater or lesser than the eigenenergy of the background wave function, respectively.

Both operators $\hat{\mathbf{H}}$ and $\hat{\mathbf{\Theta}}$ have real eigenvalues due to their Hermitian nature. However, the lowest eigenvalue of the operator $\hat{\mathbf{\Theta}}$ cannot be greater than that of the Hamiltonian operator $\hat{\mathbf{H}}$ since the perturbed self-gravity energy is always non-positive. Additionally, the background wave function, satisfying $\hat{\mathbf{H}}\Psi_{eq}=0$, is an eigenfunction of the Hamiltonian operator with a zero eigenvalue.  The lowest eigenvalue of the operator $\hat{\mathbf{\Theta}}$ must be negative.

The operator $\hat{\mathbf{H}}$ generally does not commute with $\hat{\mathbf{\Theta}}$, as $\hat{\mathbf{G}}\hat{\mathbf{H}} \neq \hat{\mathbf{H}}\hat{\mathbf{G}}$. Such a non-commuting character renders the system to possess complex eigenvalues. To determine the condition for the presence of propagating instability, we begin with Eq. (\ref{equ: symbolic notation for linearized Schrodinger equation}) and derive

\begin{equation}
\label{equ: the almost real number of eigenvalue}
\left\{\begin{aligned}
 & \omega^2 \langle \eta | \hat{\mathbf{H}} | \eta \rangle =  \langle \eta | \hat{\mathbf{H}} | \omega^2 \eta \rangle = \langle \eta | \hat{\mathbf{H}}\hat{\mathbf{\Theta}}\hat{\mathbf{H}} | \eta \rangle; \\
 & \omega^2 \langle R | \hat{\mathbf{\Theta}} | R \rangle = \langle R | \hat{\mathbf{\Theta}} | \omega^2 R \rangle = \langle R | \hat{\mathbf{\Theta}}\hat{\mathbf{H}}\hat{\mathbf{\Theta}} | R \rangle.
\end{aligned}
\right.
\end{equation}
On the other hand, consider the equations $\omega^2_j\eta_j = \hat{\mathbf{\Theta}}\hat{\mathbf{H}}\eta_j$ and $\omega^2_jR_j = \hat{\mathbf{H}}\hat{\mathbf{\Theta}}R_j$ for $j=1,2$. A straightforward calculation yields:

\begin{equation}
\label{equ: the almost orthogonality of the eigenfunction} 
\left\{\begin{aligned}
& (\omega^2_1 - \omega^{2^{*}}_2)\langle \eta_2| \hat{\mathbf{H}}|\eta_1 \rangle = 0; \\
& (\omega^2_1 - \omega^{2^{*}}_2)\langle R_2| \hat{\mathbf{\Theta}}|R_1 \rangle = 0.
\end{aligned}
\right.
\end{equation}

Equation (\ref{equ: the almost real number of eigenvalue}) indicates that the eigenvalue $\omega^2$ is real when $\langle \eta | \hat{\mathbf{H}} | \eta \rangle \neq 0$ or $\langle R | \hat{\mathbf{\Theta}} | R \rangle \neq 0$. This leads to the following theorem:
\\\\
\noindent \textbf{(\rmnum{1}): If $\bm{\omega^2 \in \mathbb{C} \backslash \mathbb{R}}$ so that $\bm{\omega^2\neq\omega^{2^{*}}}$, then the corresponding eigenfunctions $\bm{\eta}$ and $\bm{R}$ must satisfy $\bm{\langle \eta | \hat{\mathbf{H}} | \eta \rangle=0}$ and $\bm{\langle R | \hat{\mathbf{\Theta}} | R \rangle=0}$.}

\textit{This statement is arrived by setting $\omega^2_1 = \omega^2_2=\omega^2$, $\eta_1 = \eta_2 = \eta$, and $R_1 = R_2 = R$ in Eq. (\ref{equ: the almost orthogonality of the eigenfunction}).} 
\\\\

The quantity $\eta$ is proportional to the imaginary part of the perturbed wave function. From Eq. (\ref{equ: linearized Schrodinger Poisson equation with abstract symbol}), the condition $\langle \eta | \hat{\mathbf{H}} | \eta \rangle = \langle R | \hat{\mathbf{\Theta}} | R \rangle=0$ implies that the real and imaginary parts of the eigen perturbed wave function with eigenvalue $\omega^2 \in \mathbb{C} \backslash \mathbb{R}$ are mutually orthogonal.

The demand for zero expected values also provides the necessary conditions for the existence of the propagating instability, and it leads to the following theorem:
\\\\
\noindent \textbf{(\rmnum{2}): The necessary condition for $\bm{\omega^2 \in \mathbb{C} \backslash \mathbb{R}}$ is that both operators $\hat{\mathbf{H}}$ and $\hat{\mathbf{\Theta}}$ must possess both positive and negative eigenvalues.} 

\textit{Let $\eta$ be the eigenfunction of the operator $\hat{\mathbf{\Theta}}\hat{\mathbf{H}}$, corresponding to an eigenvalue $\omega^2 \in \mathbb{C} \backslash \mathbb{R}$, and it can be expanded as $\sum^{\infty}_{j=0} a_j h_j$, where $a_j$ are constants, and $h_j$ is the eigenfunction of $\hat{\mathbf{H}}$ with associated eigenvalues $e_j$. Using the orthogonality of the eigenfunctions for $\hat{\mathbf{H}}$, we derive $\langle \eta | \hat{\mathbf{H}} | \eta \rangle = \sum^{\infty}_{j=0} e_j |a_j|^2 \langle h_j | h_j \rangle$. According to Theorem (\rmnum{1}), the fact that $\langle \eta | \hat{\mathbf{H}} | \eta \rangle = 0$ implies that some $e_j$ must be negative and others positive. Otherwise, if all $e_j\geq 0$, it would be impossible to achieve $\langle \eta | \hat{\mathbf{H}} | \eta \rangle = 0$, except when $\eta = h_0$ with $e_0 = 0$. However, this would imply that $\omega^2\eta=\hat{\mathbf{\Theta}}\hat{\mathbf{H}} \eta =0$, which would force $\omega^2=0$, contradicting the assumption that $\omega^2 \in \mathbb{C} \backslash \mathbb{R}$.  Conversely, if (ii) fails to hold, i.e., all $e_j \geq 0$, we would have $\langle \eta | \hat{\mathbf{H}} | \eta \rangle \geq 0$ and the fact that $e_j \geq 0$ implies $\Psi_{eq}$ is the lowest energy state of  $\hat{\mathbf{H}}$. The same arguments can also be applied to the eigenfunction $R$ of $\hat{\mathbf{\Theta}}$.} 

When the Theorem (\rmnum{2}) does not hold, we have the corresponding eigenvalue $\omega^2$ must be real. We have shown $\langle \eta |\hat{\mathbf{H}}| \eta \rangle \geq 0$, the system is described by the Sturm-Liouville theory and can be addressed using the variational method, as $\langle \eta_2 |\hat{\mathbf{H}}| \eta_1 \rangle$ can be regarded as an inner product. We shall illustrate it with the operator $\hat{\mathbf{H}}$. (A similar approach also be applied to the operator $\hat{\mathbf{\Theta}}$). 
 
Following the standard approach of the variational principle, any trial function $\eta$ can be expressed as $\sum_{j} a_j \eta_j$, where $a_j$ are constants. Each $\eta_j$ (different from $h_j$ defined above) satisfies the equation $\omega^2_j \eta_j = \hat{\mathbf{\Theta}}\hat{\mathbf{H}} \eta_j$, with $\omega^2_j$ being a real eigenvalue.  It follows:

\begin{equation}
\label{equ: variational method for the background wave function is the ground state}
\begin{aligned}
\langle \eta | \hat{\mathbf{H}}\hat{\mathbf{\Theta}}\hat{\mathbf{H}} | \eta \rangle & = \sum_j \omega^2_j |a_j|^2\langle \eta_j | \hat{\mathbf{H}} | \eta_j \rangle \\
& \geq \omega^2_{lowest}\sum_j |a_j|^2\langle \eta_j | \hat{\mathbf{H}} | \eta_j \rangle \\
& = \omega^2_{lowest} \langle \eta | \hat{\mathbf{H}} | \eta \rangle,
\end{aligned}
\end{equation}
where $\omega^2_{lowest}$ is the lowest eigenvalue. Here, the equalities are derived from Eq. (\ref{equ: the almost orthogonality of the eigenfunction}), which implies $\langle \eta_j | \hat{\mathbf{H}} | \eta_i \rangle = 0$ for $i \neq j$.

The lowest eigenvalue $\omega^2_{lowest}$ can therefore be estimated using the variational principle as summarized by the following theorem:
\\\\
\noindent \textbf{(\rmnum{3}): If $\bm{\Psi_{eq}}$ is the lowest energy ground state of $\hat{\mathbf{H}}$ , then for any trial function $\bm{\eta}$ with $\bm{\langle \eta | \hat{\mathbf{H}} | \eta \rangle>0}$, the inequality $\bm{\omega^2_{lowest} \leq \langle \eta | \hat{\mathbf{H}}\hat{\mathbf{\Theta}}\hat{\mathbf{H}} | \eta \rangle/\langle \eta | \hat{\mathbf{H}} | \eta \rangle}$ holds.}

\textit{The inequality in Eq. (\ref{equ: variational method for the background wave function is the ground state}) holds because all eigenvalues $\omega_j^2$ are real, $\bm{\langle \eta | \hat{\mathbf{H}} | \eta \rangle>0}$ , and $\Psi_{eq}$ is the ground state of the operator $\hat{\mathbf{H}}$. }
\\\\

A concrete example is given in \cite{CH2023}, where the stability of the soliton solution was analyzed through the construction of $\omega^2 \langle \eta | \hat{\mathbf{H}} | \eta \rangle$. The key mathematical step of the construction is to show  $\langle \eta | \hat{\mathbf{H}} | \eta \rangle >0$, and the following manipulation procedures reveal an insight to the above theorems:

\begin{equation}
\label{equ: the rigorous proof that the background wave function without nodes is the ground state}
\begin{aligned}
& \langle \eta | \hat{\mathbf{H}} | \eta \rangle \\
& =  \int \eta^{*}(\vec{x})\Big ( -{{1}\over{2}}\nabla^2 + V_{eq}(\vec{x}) - E_{eq} \Big )\eta(\vec{x})d^3\vec{x} \\
& = \int {{\eta^{*}(\vec{x})}\over{\Psi_{eq}(\vec{x})}}\Psi_{eq}(\vec{x})\Big ( -{{1}\over{2}}\nabla^2 + V_{eq}(\vec{x}) - E_{eq} \Big )\eta(\vec{x})d^3\vec{x} \\
& = -{{1}\over{2}} \int {{\eta^{*}(\vec{x})}\over{\Psi_{eq}(\vec{x})}} \nabla \cdot \Bigg ( \Psi^2_{eq}(\vec{x}) \nabla \Big (  {{\eta(\vec{x})}\over{\Psi_{eq}(\vec{x})}} \Big ) \Bigg ) d^3\vec{x} \\
& \begin{aligned} = &  {{1}\over{2}} \int \Psi^2_{eq}(\vec{x}) \Bigg |\nabla \Bigg ( {{\eta(\vec{x})}\over{\Psi_{eq}(\vec{x})}} \Bigg ) \Bigg |^2 d^3\vec{x} - \\
& {{1}\over{2}} \oint \eta^{*}(\vec{x})\Psi_{eq}(\vec{x}) \nabla \Big (  {{\eta(\vec{x})}\over{\Psi_{eq}(\vec{x})}} \Big ) d\vec{S}
\end{aligned} \\
& = {{1}\over{2}} \int \Psi^2_{eq}(\vec{x}) \Bigg |\nabla \Bigg ( {{\eta(\vec{x})}\over{\Psi_{eq}(\vec{x})}} \Bigg ) \Bigg |^2 d^3\vec{x} \geq 0.
\end{aligned}
\end{equation}
The expectation value of $\hat{\mathbf{H}}$ is not only positive definite but also finite.  However, the last equality holds provided that $\Psi_{eq}$ has no local null; if it does and $\eta$ is non-zero there, the last equality cannot hold since the pole in the surface integral at the $\Psi_{eq}$'s local null would give a major contribution and cannot be dropped.  On the contrary, if $\eta$ assumes a zero value there, the last equality is valid. Unfortunately, we have empirically found that most eigenfunctions, $\eta$'s, do not assume zero values there, meaning that the variational principle can no longer hold when $\Psi_{eq}$ has local nulls.

Some details about the above derivation of Eq. (\ref{equ: the rigorous proof that the background wave function without nodes is the ground state}) is given below. The third equality directly follows from $\hat{\mathbf{H}}\Psi_{eq}=0$; the fourth equality is derived using integration by parts; finally, the surface integral in the fourth equality can be neglected for any $L^2$-integrable function.  

We shall emphasize that having one, or more, null in the background wave function $\Psi_{eq}$ strongly suggests that $\Psi_{eq}$ is an excited state of $\hat{\mathbf{H}}$, and there exists a lower-energy eigenfunction orthogonal to $\Psi_{eq}$ so that Theorem (\rmnum{2}) can be satisfied. On the other hand, if $\Psi_{eq}$ has no local null, the last equality of Eq. (\ref{equ: the rigorous proof that the background wave function without nodes is the ground state}) holds and it means $\Psi_{eq}$ must be the ground state of $\hat{\mathbf{H}}$. 

\section{Nonlinear periodic equilibrium construction}
\label{sec: Nonlinear periodic equilibrium construction}

This section is divided into two parts. First, in Sec. (\ref{subsec: Taylor expansion construction}), we utilize the Taylor expansion to develop a periodic equilibrium background configuration that resembles interference fringes within cosmic sheets. This series approach of course has a limited range of validity. We further estimate the range. Second in Sec. (\ref{subsec: Runge-Kutta Scheme to Validate the Accuracy of the Taylor Expansion}), the accuracy of this series expansion is validated with numerical solutions using the fourth-order Runge-Kutta scheme. In addition, we empirically find that the solution can be extended from that with the series expansion to the highly nonlinear solution outside the valid range of series expansion.

\subsection{Taylor Expansion Construction}
\label{subsec: Taylor expansion construction}

To mimic the periodic interference fringes, the background wave function should consist of two colliding plane waves, producing a periodic standing wave.  We choose the rest frame, and this standing wave oscillates along the $x$-axis. We assume, without loss of generality, that the periodic spatial domain is $[0,2\pi]$ and that the standing wave function has the even symmetric form: $4E^{(0)}_{eq}\sqrt{\Tilde{\rho}^{(0)}_{eq}}\cos{\Big ( n^{(0)}_{eq}x \Big )}$, where $n^{(0)}_{eq} \in \mathbb{N}$, $E^{(0)}_{eq} \equiv \Big ( n^{(0)}_{eq} \Big )^2/2$ with $\Tilde{\rho}^{(0)}_{eq}>0$. 

For the two colliding plane waves, $n^{(0)}_{eq}$ and $E^{(0)}_{eq}$ represent momentum and kinetic energy of the plane wave, respectively, and the corresponding quantum wavelength is $2\pi/n^{(0)}_{eq}$. Additionally, the amplitude of the plane wave is $2E^{(0)}_{eq}\sqrt{\Tilde{\rho}^{(0)}_{eq}}$, and the corresponding Jeans wavelength is $2\pi n^{(0)}_{eq}/(2E^{(0)}_{eq}\sqrt{\Tilde{\rho}^{(0)}_{eq}}) = 2\pi/(n^{(0)}_{eq}\sqrt{\Tilde{\rho}^{(0)}_{eq}})$. Therefore, the parameter $\Tilde{\rho}^{(0)}_{eq}$ represents the square ratio of quantum to Jeans wavelengths.    

The solitary standing wave is in a force-equilibrium state, where a self-consistent gravitational potential is generated to balance the kinetic energy.  Note that there is only one dimensionless parameter, $\Tilde{\rho}^{(0)}_{eq}$, in the force-balanced standing wave. We adopt Taylor expansions of background eigenenergy, wave function, gravitational potential and cosmological constant with respect to $\Tilde{\rho}^{(0)}_{eq}$ to construct the force equilibrium. These four quantities are expanded as follows:

\begin{equation}
\label{equ: Tayler series of the eigenenergy, wave function and gravitional potential}
\left\{\begin{aligned}
& \Tilde{E}_{eq} \equiv {{E_{eq}}\over{E^{(0)}_{eq}}} = \sum^{\infty}_{n=0}(\Tilde{\rho}^{(0)}_{eq})^n \Tilde{E}_n, \\
& \Tilde{\Psi}_{eq}(x) \equiv {{\Psi_{eq}(x)}\over{4E^{(0)}_{eq}\sqrt{\Tilde{\rho}^{(0)}_{eq}}}} = \sum^{\infty}_{n=0}(\Tilde{\rho}^{(0)}_{eq})^n \Tilde{\Psi}_n(x). \\
& \Tilde{V}_{eq}(x) \equiv {{V_{eq}(x)}\over{E^{(0)}_{eq}}}= \sum^{\infty}_{n=1}(\Tilde{\rho}^{(0)}_{eq})^n \Tilde{V}_n(x), \\
& \Tilde{\Lambda} \equiv {{\Lambda}\over{\Big ( 4E^{(0)}_{eq}\Big )^2\Tilde{\rho}^{(0)}_{eq}}} = \sum^{\infty}_{n=0}(\Tilde{\rho}^{(0)}_{eq})^{n} \Tilde{\Lambda}_n.
\end{aligned}
\right.
\end{equation}
Here, $\Tilde{\Psi}_0(x) = \cos{\Big ( n^{(0)}_{eq}x \Big )}$ and $\Tilde{E}_0=1$. The normalized dimensionless background eigenenergy $\Tilde{E}_{eq}$, gravitational potential $\Tilde{V}_{eq}(x)$, wave function $\Tilde{\Psi}_{eq}(x)$, and cosmological constant $\Tilde{\Lambda}$ are introduced. The Taylor series expansions for the gravitational potential and the cosmological constant begins with $\Tilde{\rho}^{(0)}_{eq}$ because they are associated with the Jeans wavelength. By substituting the aforementioned infinite series into Eq. (\ref{equ: Schrodiger-Poisson equation for single component}) and collecting terms of order $\Big ( \Tilde{\rho}^{(0)}_{eq} \Big )^n$, it follows that:

\begin{equation}
\label{equ: background configuration for each order}
\left\{\begin{aligned}
& \Bigg ( {{1}\over{2E^{(0)}_{eq}}}{{d^2}\over{dx^2}} + 1 \Bigg ) \Tilde{\Psi}_n(x)=\sum^{n}_{j=1}\Tilde{\Psi}_{n-j}(x)\Big [ \Tilde{V}_j(x) - \Tilde{E}_j  \Big ], \\
& {{1}\over{2E^{(0)}_{eq}}}{{d^2 \Tilde{V}_n(x)}\over{dx^2}} =   8\Bigg [  \sum^{n}_{j=1}\Tilde{\Psi}_{n-j}(x)\Tilde{\Psi}_{j-1}(x) - \Tilde{\Lambda}_{n-1} \Bigg ], \\
& \Tilde{\Lambda}_{n-1} = {{1}\over{2\pi}}\int^{2\pi}_{0} \sum^{n}_{j=1}\Tilde{\Psi}_{n-j}(y)\Tilde{\Psi}_{j-1}(y)dy.
\end{aligned}
\right.
\end{equation}
Here, $\Tilde{\Lambda}_{n-1}$ represents the normalized mean mass density for $n$-th degree of $\Tilde{\rho}^{(0)}_{eq}$. To fix the gauge of the potential, each order of the background gravitational potential has a zero mean, i.e., $\int^{2\pi}_{0} \Tilde{V}_n(x)dx=0$. Additionally, each order of the wave function $\Tilde{\Psi}_n$ and the gravitational potential $\Tilde{V}_n$ are assumed to be periodic over the spatial domain $[0,2\pi]$, such that $\Tilde{\Psi}_n(x+2\pi) = \Tilde{\Psi}_n(x)$ and $\Tilde{V}_n(x+2\pi) = \Tilde{V}_n(x)$.

With these conditions, each order of the wave function and gravitational potential can be expanded as a Fourier series, and Eq. (\ref{equ: background configuration for each order}) becomes a set iterative equations for the coefficients of plane waves similar to the conventional high-order perturbation method of Quantum Mechanics, and the solutions of each order can be straightforwardly solved.

The iterative procedure is described as follows: Once $\Tilde{\Psi}_{j}(x)$, $\Tilde{V}_j(x)$, and $\Tilde{E}_j$ for all $j < n$ have been determined, $\Tilde{V}_n(x)$ can be obtained from the second equation in Eq. (\ref{equ: background configuration for each order}), as the right-hand side of this equation depends solely on $\Tilde{\Psi}_{j}(x)$ with $j<n$. After determining $\Tilde{V}_n(x)$, the value of $\Tilde{E}_n$ can be calculated by multiplying both sides of the first equation in Eq. (\ref{equ: background configuration for each order}) by $\Tilde{\Psi}_{0}(x)$ and integrating over the spatial domain $[0,2\pi]$. Finally, with the right-hand side of the first equation Eq. (\ref{equ: background configuration for each order}) now known, $\Tilde{\Psi}_{n}(x)$ can be determined.

Furthermore, by applying mathematical induction to the iterative procedure, each order of the eigenenergy, wave function, and gravitational potential takes the following form:

\begin{equation}
\label{equ: Fourier series of each order of background wave function and gravitational potential}
\left\{\begin{aligned}
& \Tilde{E}_n = {{\sum^{n}_{j=1} \int^{2\pi}_{0} \Tilde{\Psi}_{n-j}(x)\Tilde{V}_j(x)\Tilde{\Psi}_0(x) dx}\over{\int^{2\pi}_{0} \Tilde{\Psi}^2_0(x) dx }}; \\
& \Tilde{V}_n(x) =\sum^n_{l=1} \Tilde{v}_{n,l}\cos{\Big [(2l)n^{(0)}_{eq} x\Big ]}; \\
& \Tilde{\Psi}_n(x) = \sum^n_{l=1} \Tilde{\psi}_{n,l}\cos{\Big [(2l+1)n^{(0)}_{eq} x\Big ]},
\end{aligned}
\right.
\end{equation}
for some real constants $\Tilde{\psi}_{n,l}$ and $\Tilde{v}_{n,l}$.  The detailed derivation can be found in Appendix \ref{appendix: Properties for the background configuration}. Eq. (\ref{equ: Fourier series of each order of background wave function and gravitational potential}) demonstrates that each order of the wave function and gravitational potential consists of only a finite number of cosine Fourier modes. Specifically, the wave number is an odd integer multiple of $n^{(0)}_{eq}$ for the wave function and an even integer multiple for the gravitational potential.

The recursive relation for the Fourier coefficients $\Tilde{\psi}_{n,l}$ and $\Tilde{v}_{n,l}$, along with $\Tilde{E}_n$, is derived by substituting Eq. (\ref{equ: Fourier series of each order of background wave function and gravitational potential}) into Eq. (\ref{equ: background configuration for each order}). Consequently, the following relation is obtained:

\begin{equation}
\label{equ: recursive relation for E_n, v_nl, and pis_nl}
\left\{\begin{aligned}
& 2\Tilde{E}_n = \Tilde{v}_{n,1} + \sum^{n-1}_{m=1}\sum^{\infty}_{l^{'}=1} (\Tilde{\psi}_{n-m,l^{'}} + \Tilde{\psi}_{n-m,l^{'}-1})\Tilde{v}_{m,l^{'}}; \\
& -l^2\Tilde{v}_{n,l} = 
\left\{\begin{aligned}
& 1, \text{ if }n=l=1; \\
& \begin{aligned} 
 2 & (\Tilde{\psi}_{n-1,l} + \Tilde{\psi}_{n-1,l-1}) + \\
   & \sum^{n-1}_{m=2}\sum^{\infty}_{l^{'}=1}\Tilde{\psi}_{m-1,l^{'}} \Big (  \Tilde{\psi}_{n-m,l+l^{'}} +  \\
   & \Tilde{\psi}_{n-m,\min{\{|(l-1)-l^{'}|,|l-l^{'}|\}}} \Big), \text{ if } n \neq 1; 
\end{aligned}
\end{aligned}
\right.\\
& \begin{aligned}
-8l(l+1)\Tilde{\psi}_{n,l} = 
 ( & \Tilde{v}_{n,l} + \Tilde{v}_{n,l+1}) - 2\sum^{n-1}_{m=1}\Tilde{\psi}_{n-m,l}\Tilde{E}_{m} + \\
& \begin{aligned} & \sum^{n-1}_{m=1}\sum^{\infty}_{l^{'}=1}\Tilde{v}_{m,l^{'}} \Big (  \Tilde{\psi}_{n-m,l+l^{'}} + \\
& \Tilde{\psi}_{n-m,\min{\{|(l+1)-l^{'}|,|l-l^{'}|\}}} \Big); 
\end{aligned}
\end{aligned}
\end{aligned}
\right.
\end{equation}
Here, we impose the conditions $\Tilde{v}_{n,0} = \Tilde{\psi}_{n,0} = 0$ for all $n \in \mathbb{N}$ and $\Tilde{v}_{n,l} = \Tilde{\psi}_{n,l} = 0$ if $l>n$ to simplify the recursive relation. In Eq. (\ref{equ: recursive relation for E_n, v_nl, and pis_nl}), the operator $\bm{\min}$ selects the smallest value among its arguments. Some low-order values of $\Tilde{\psi}_{n,l}$, $\Tilde{v}_{n,l}$, and $\Tilde{E}_n$ up to $n=4$ are summarized in Table (\ref{Table: psi_n,k and v_n,k to n = 4}) after solving Eq. (\ref{equ: recursive relation for E_n, v_nl, and pis_nl}).

\begin{table*}
\centering
\caption{Values of $\Tilde{\psi}_{n,l}$, $\Tilde{v}_{n,l}$ and $\Tilde{E}_n$ up to $n=4$}{
\begin{tabular}{ccccccccccccccc} \hline \hline \\
& $\Tilde{\psi}_{n,1}$ & $\Tilde{\psi}_{n,2}$ & $\Tilde{\psi}_{n,3}$ & $\Tilde{\psi}_{n,4}$ & $\Tilde{v}_{n,1}$ & $\Tilde{v}_{n,2}$ & $\Tilde{v}_{n,3}$ & $\Tilde{v}_{n,4}$ & $\Tilde{E}_n$  \\ \hline \\
$n=1$ & ${{1}\over{2^4}}$ & $0$ & $0$ & $0$ & $-1$ & $0$ & $0$ & $0$ & $-{{1}\over{2}}$ \\ \hline \\
$n=2$ & ${{3}\over{2^9}}$ & ${{1}\over{2^9}}$ & $0$ & $0$ & $-{{1}\over{2^3}}$ & $-{{1}\over{2^5}}$ & $0$ & $0$ & $-{{3}\over{2^5}}$ \\ \hline \\
$n=3$ & $0$ & ${{19}\over{2^{11} \times 3^3}}$ & ${{11}\over{2^{13} \times 3^3}}$ & $0$ & $-{{3}\over{2^8}}$ & $-{{1}\over{2^8}}$ & $-{{1}\over{2^7 \times 3^2}}$ & $0$ & $-{{7}\over{2^9}}$ \\ \hline \\
$n=4$ & $-{{43}\over{2^{17} \times 3}}$ & ${{95}\over{2^{14} \times 3^5}}$ & ${{3067}\over{2^{20} \times 3^5}}$ & ${{11}\over{2^{20} \times 3^2}}$ & $-{{1}\over{2^{12}}}$ & $-{{19}\over{2^{12} \times 3^3}}$ & $-{{7}\over{2^9 \times 3^4}}$ & $-{{19}\over{2^{15} \times 3^3}}$ & $-{{9}\over{2^{13}}}$ \\ \hline \\
\end{tabular}
}
\label{Table: psi_n,k and v_n,k to n = 4}
\end{table*}

Equation (\ref{equ: Fourier series of each order of background wave function and gravitational potential}) indicates that each order of the wave function $\Tilde{\Psi}_n(x)$ is orthogonal to the leading order $\Tilde{\Psi}_0(x)$. Moreover, by substituting each order of the wave function $\Tilde{\Psi}_n(x)$ and the gravitational potential $\Tilde{V}_n(x)$ from Eq. (\ref{equ: Fourier series of each order of background wave function and gravitational potential}) into Eq. (\ref{equ: Tayler series of the eigenenergy, wave function and gravitional potential}), we obtain: 

\begin{equation}
\label{equ: Fourier series of the background wave and gravitational potential}
\left\{\begin{aligned}
& \Tilde{V}_{eq}(x) =\sum^{\infty}_{l=1} \Tilde{v}_{l}\cos{\Big [(2l)n^{(0)}_{eq} x\Big ]}, \text{ }\Tilde{v}_l=\sum^{\infty}_{n=l}\Tilde{v}_{n,l} \Big (\Tilde{\rho}^{(0)}_{eq} \Big )^n;\\
& \Tilde{\Psi}_{eq}(x) = \sum^{\infty}_{l=0} \Tilde{\psi}_{l}\cos{\Big [(2l+1)n^{(0)}_{eq} x\Big ]}, \text{ } 
\left\{\begin{aligned}
& \Tilde{\psi}_0 = 1; \\
& \Tilde{\psi}_l=\sum^{\infty}_{n=l}\Tilde{\psi}_{n,l} \Big ( \Tilde{\rho}^{(0)}_{eq} \Big )^n.
\end{aligned}
\right.
\end{aligned}
\right.
\end{equation}
Therefore, the background wave function comprises all standing waves with a cosine functional form, where the wave number is an odd integer multiple of $n^{(0)}_{eq}$,  and the gravitational potential is an even integer multiple, as shown in Eq. (\ref{equ: Fourier series of the background wave and gravitational potential}).

The above construction involves an infinite series with respect to $\Tilde{\rho}^{(0)}_{eq}$. Therefore, it is necessary to examine whether the radius of convergence for $\Tilde{\rho}^{(0)}_{eq}$ is non-zero to validate this series expansion. The radius of convergence is determined by the root test, which approximates the following three sequences $\{|\Tilde{E}_n|\}$, $\{\mathop{\max}_{x \in [0,2\pi]}|\Tilde{\Psi}_n(x)|\}$, and $\{\mathop{\max}_{x \in [0,2\pi]}|\Tilde{V}_n(x)|\}$ as geometric sequences. Here, $\mathop{\max}_{x \in [0,2\pi]}|\Tilde{\Psi}_n(x)|$ and $\mathop{\max}_{x \in [0,2\pi]}|\Tilde{V}_n(x)|$ are the maximum absolute values over the domain $[0,2\pi]$ for $\Tilde{\Psi}_n(x)$ and $\Tilde{V}_n(x)$, respectively. Therefore, the radius of convergence is the inverse of the largest value among $\lim_{n \rightarrow \infty}\sqrt[n]{|\Tilde{E}_n|}$,  $\lim_{n \rightarrow \infty}\sqrt[n]{\mathop{\max}_{x \in [0,2\pi]}|\Tilde{\Psi}_n(x)|}$, and $\lim_{n \rightarrow \infty}\sqrt[n]{\mathop{\max}_{x \in [0,2\pi]}|\Tilde{V}_n(x)|}$. Note that the Taylor series for the cosmological constant is not necessarily considered in the estimation of the radius of convergence. This is because $|\Tilde{\Lambda}_{n-1}| = |\int^{2\pi}_{0} \sum^{n}_{j=1}\Tilde{\Psi}_{n-j}(y)\Tilde{\Psi}_{j-1}(y)dy/(2\pi)| \leq \sum^{n}_{j=1}\mathop{\max}_{x \in [0,2\pi]}|\Tilde{\Psi}_{n-j}(x)|\mathop{\max}_{x \in [0,2\pi]}|\Tilde{\Psi}_{j-1}(x)|$ implies $\lim_{n \rightarrow \infty}\sqrt[n]{|\Tilde{\Lambda}_n|} \leq \lim_{n \rightarrow \infty}\sqrt[n]{\mathop{\max}_{x \in [0,2\pi]}|\Tilde{\Psi}_n(x)|}$. The detail of proving the non-zero radius of convergence is provided in Appendix \ref{appendix: Properties for the background configuration}. 

\begin{figure}
\includegraphics[scale=0.345, angle=270]{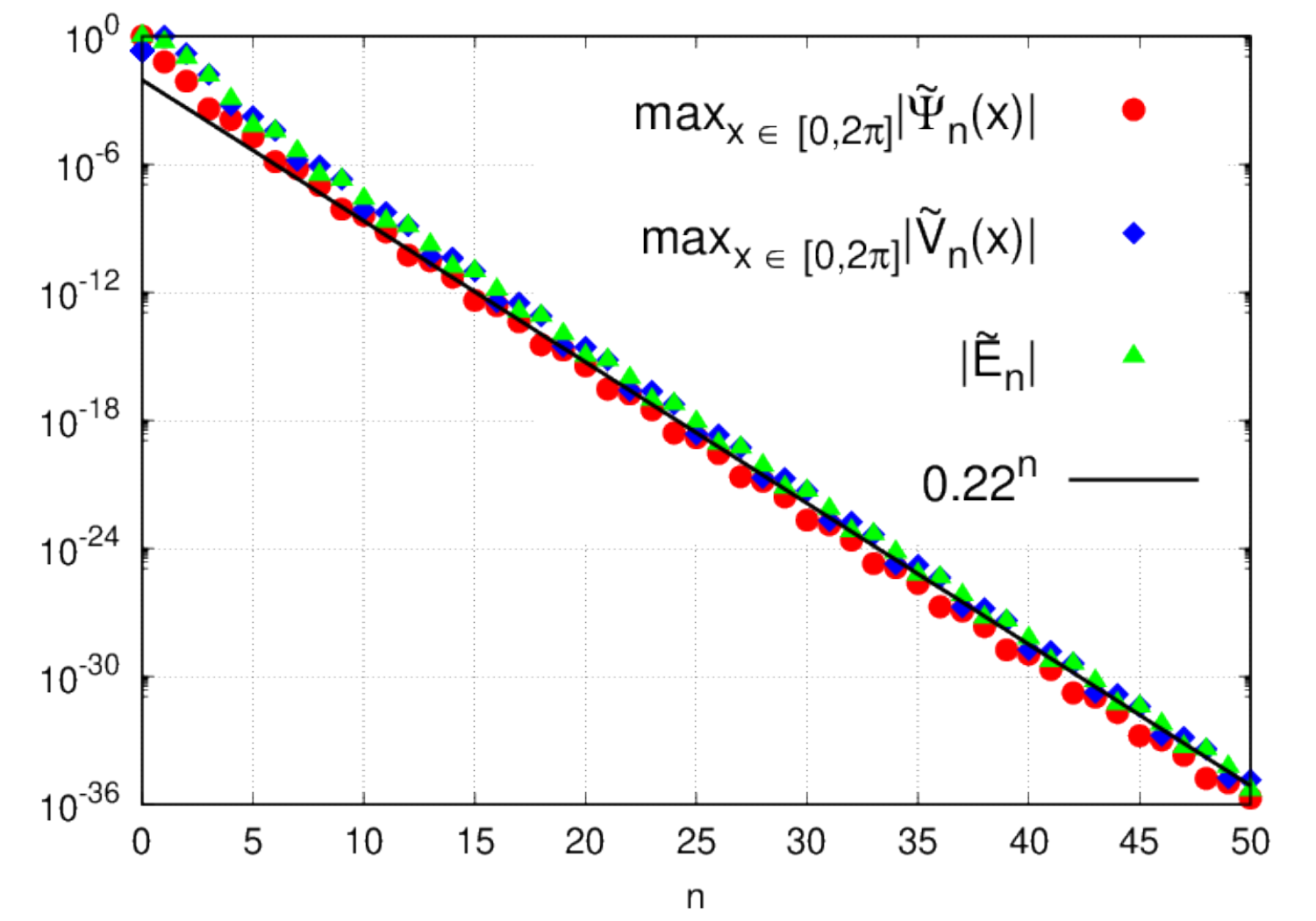}
\caption{\textit{The asymptotic behaviors of normalized $|\Tilde{E}_n|$ (green solid triangles), $\mathop{\max}_{x \in [0,2\pi]}|\Tilde{\Psi}_n(x)|$ (blue solid diamonds), and $\mathop{\max}_{x \in [0,2\pi]}|\Tilde{V}_n(x)|$ (red solid circles) as functions of large $n$, which exhibit convergence of  Taylor expansion solutions.  Here, $\mathop{\max}_{x \in [0,2\pi]}|\Tilde{\Psi}_n(x)|$ and $\mathop{\max}_{x \in [0,2\pi]}|\Tilde{V}_n(x)|$ are the maximum absolute values of $\Tilde{\Psi}_n(x)$ and $\Tilde{V}_n(x)$ over the interval $[0,2\pi]$. The horizontal axis is $n$, which corresponds to the degree of $\Tilde{\rho}^{(0)}_{eq}$ in the Taylor series for the background eigenenergy $E_{eq}$, wave function $\Psi_{eq}(x)$, and gravitational potential $V_{eq}(x)$. The figure shows that these three quantities follow geometric sequences with the same common ratio, indicated by the black solid line. The black solid line determines the radius of convergence, which is given by $\Tilde{\rho}^{(c)}_{eq} = 1/0.22 \approx 4.5$. Therefore, the background configuration constructed by the Taylor expansion is restricted to $\Tilde{\rho}^{(0)}_{eq} \lesssim \Tilde{\rho}^{(c)}_{eq}$.} }
\label{fig: convergence_of_radius}
\end{figure}

To determine the value of the radius of convergence, we plot Fig. (\ref{fig: convergence_of_radius}), which shows the behavior of $|\Tilde{E}_n|$, $\mathop{\max}_{x \in [0,2\pi]}|\Tilde{\Psi}_n(x)|$, and $\mathop{\max}_{x \in [0,2\pi]}|\Tilde{V}_n(x)|$ as functions of $n$ up to $n=50$. This figure clearly shows that these three quantities indeed behave as geometric series, all declining exponentially with $n$ as $0.22^n$. Therefore, the radius of convergence, denoted as $\Tilde{\rho}^{(c)}_{eq}$, is about $1/0.22 \approx 4.5$. Thus, this background configuration is valid only for $\Tilde{\rho}^{(0)}_{eq} \lesssim \Tilde{\rho}^{(c)}_{eq}$.    

Figure (\ref{fig: convergence_of_radius}) also indicates Fourier coefficients of the background gravitational potential $\Tilde{v}_l$ and wave function $\Tilde{\psi}_l$ can be approximated as follows: $|\Tilde{\psi}_l|$ and $|\Tilde{v}_l| \sim \Big ( \Tilde{\rho}^{(0)}_{eq}/\Tilde{\rho}^{(c)}_{eq} \Big )^l$ $\forall l \in \mathbb{N}$. We will apply this estimation in Appendix \ref{appendix: The General Property of the Matrix Representation}.

Equation (\ref{equ: Fourier series of the background wave and gravitational potential}) shows that this quantum system resembles a one-dimensional lattice.  From the in-phase ($0$ and $180$ degrees) sum of  Eq. (\ref{equ: Fourier series of the background wave and gravitational potential}), the unit lattice cell size is expected to be the period of the background gravitational potential,  $2\pi/2n^{(0)}_{eq}$. Moreover, Eq. (\ref{equ: Fourier series of the background wave and gravitational potential}) also reveals that $\Psi_{eq}(x+\pi/n^{(0)}_{eq})=-\Psi_{eq}(x)$ and $V_{eq}(x+\pi/n^{(0)}_{eq})=V_{eq}(x)$.   

In addition, the equilibrium wave function $\Psi_{eq}$ should possess local nulls since it has a zero spatial mean, and the null is expected to be located at $x=\pi/(2n^{(0)}_{eq})$ within the interval $[0,\pi/n^{(0)}_{eq}]$. Consequently, this background configuration will induce propagating instability.  The zeroth order background wave function, $\Tilde{\Psi}_0(x) = \cos{(n^{(0)}_{eq}x)}$, has exactly a null at the expected location at the center of the unit cell $[0,\pi/n^{(0)}_{eq}]$, and it is therefore crucial to demonstrate that the background wave function has no other nulls within the same cell. 

Assuming that there were other nulls of $\Psi_{eq}$ besides $x=\pi/(2n^{(0)}_{eq})$ in the interval $[0,\pi/n^{(0)}_{eq}]$ for some finite value of $\Tilde{\rho}^{(0)}_{eq}$,  the extra nulls must appear in pairs at the symmetric locations within the unit cell; moreover the adjacent two nulls must have opposite local slopes.  As one decreases $\rho^{(0)}_{eq}$ to a sufficiently low value, two adjacent nulls will be barely merging with zero local slope, i.e., $\Psi_{eq}\sim (x-x_0)^2$.   However, at this location the finite local effective potential $V_{eq}(x_0)-E_{eq}$ cannot balance the infinite kinetic energy $-(1/2\Psi_{eq})d^2\Psi_{eq}(x_0)/dx^2$. Hence such a configuration can never occur, and yields a contradiction to the assumption.

For a detailed proof, refer to Appendix \ref{appendix: Properties for the background configuration}. This result indicates that the background wave function has exactly one node in each unit cell, and this property can also be used to estimate the spectrum of  $\hat{\mathbf{H}}$ to be discussed in Sec. (\ref{sec: Comparison Between Theoretical Prediction and Simulating result}).

\subsection{Agreement between the Taylor Expansion solution with the Numerical Solution}
\label{subsec: Runge-Kutta Scheme to Validate the Accuracy of the Taylor Expansion}

The numerical solution of the nonlinear time-independent Schrödinger-Poisson equation, Eq. (\ref{equ: Schrodiger-Poisson equation for single component}), was computed using the fourth-order Runge-Kutta scheme \cite{Butcher2008} to evaluate the accuracy of the series expansion solutions. For the numerical solution, the initial conditions at $x=0$ must be specified. Without loss of generality, the background wave function at $x=0$ is set to $\Psi_{eq}(0) = 1$ and $\Psi^{'}_{eq}(0) = 0$ and the force $-V^{'}_{eq}(0)=0$ for all numerical solutions unless otherwise specified, leaving $E_{eq}$ and $\Lambda$ as the eigenvalues.  In addition, by adjusting $V_{eq}(0)$, we let $V_{eq}(x)$ to have a zero spatial mean.  The choice of initial conditions takes advantage of the scale invariance of Eq. (\ref{equ: Schrodiger-Poisson equation for single component}) under the transformations: $\Psi_{eq}(x) \rightarrow \Psi_{eq}(x)/\Psi_{eq}(0)$, $\Lambda \rightarrow \Lambda/\Psi^2_{eq}(0)$, $V_{eq}(x) \rightarrow V_{eq}(x)/|\Psi_{eq}(0)|$, $E_{eq} \rightarrow E_{eq}/|\Psi_{eq}(0)|$, and $x \rightarrow \sqrt{|\Psi_{eq}(0)|}x$.

For the normalization of the series-expansion solution consistent with that of the numerical solution, we convert the series expansion solution introduced in Sec. (\ref{subsec: Taylor expansion construction}) by the above transformation. With it, the normalized eigenenergy and the cosmological constant become:

\begin{equation}
\label{equ: E_eq and Lambda for Psi_eq(0) = 1}
\left\{\begin{aligned}
& E_{eq} = {{E^{(0)}_{eq} \Tilde{E}_{eq}}\over{|\Psi_{eq}(0)|}} = {{\sum^{\infty}_{n=0}(\Tilde{\rho}^{(0)}_{eq})^n \Tilde{E}_n}\over{4\sqrt{\Tilde{\rho}^{(0)}_{eq}}|\sum^{\infty}_{l=0}\Tilde{\psi}_l|}}; \\
& \Lambda =  { \Big ( {4E^{(0)}_{eq}\Big )^2\Tilde{\rho}^{(0)}_{eq}\Tilde{\Lambda}}\over{\Psi^2_{eq}(0)}} = {{\sum^{\infty}_{n=0}(\Tilde{\rho}^{(0)}_{eq})^{n} \Tilde{\Lambda}_n}\over{|\sum^{\infty}_{l=0}\Tilde{\psi}_l|^2}}; \\
& \Psi_{eq}(0) = 4E^{(0)}_{eq}\sqrt{\Tilde{\rho}^{(0)}_{eq}}\Tilde{\Psi}_{eq}(0) = 4E^{(0)}_{eq}\sqrt{\Tilde{\rho}^{(0)}_{eq}}\sum^{\infty}_{l=0}\Tilde{\psi}_l.
\end{aligned}
\right.
\end{equation}
Here, the values of $\Tilde{E}_n$ is determined by Eq. (\ref{equ: recursive relation for E_n, v_nl, and pis_nl}), while $\Tilde{\Lambda}_n$ is derived from Eqs. (\ref{equ: background configuration for each order}), (\ref{equ: Fourier series of each order of background wave function and gravitational potential}), and(\ref{equ: recursive relation for E_n, v_nl, and pis_nl}). The values of $\Tilde{\psi}_l$ are calculated using Eqs. (\ref{equ: recursive relation for E_n, v_nl, and pis_nl}) and (\ref{equ: Fourier series of the background wave and gravitational potential}). At the parameter $\Tilde{\rho}^{(0)}_{eq} = 1.6$,  the two normalized eigenvalues are found to be $E_{eq}\approx -1.7966272 \times 10^{-2}$ and  $\Lambda \approx 4.03135669 \times 10^{-1}$. These values are consistent with those obtained from the numerical solution. Additionally, The gravitational potential at $x=0$ is numerically determined to be $V_{eq}(0)=-3.64814660 \times 10^{-1}$ to ensure that the spatially averaged potential is zero; this value will affect the determination of $E_{eq}$.

\begin{figure}
\includegraphics[scale=0.345, angle=270]{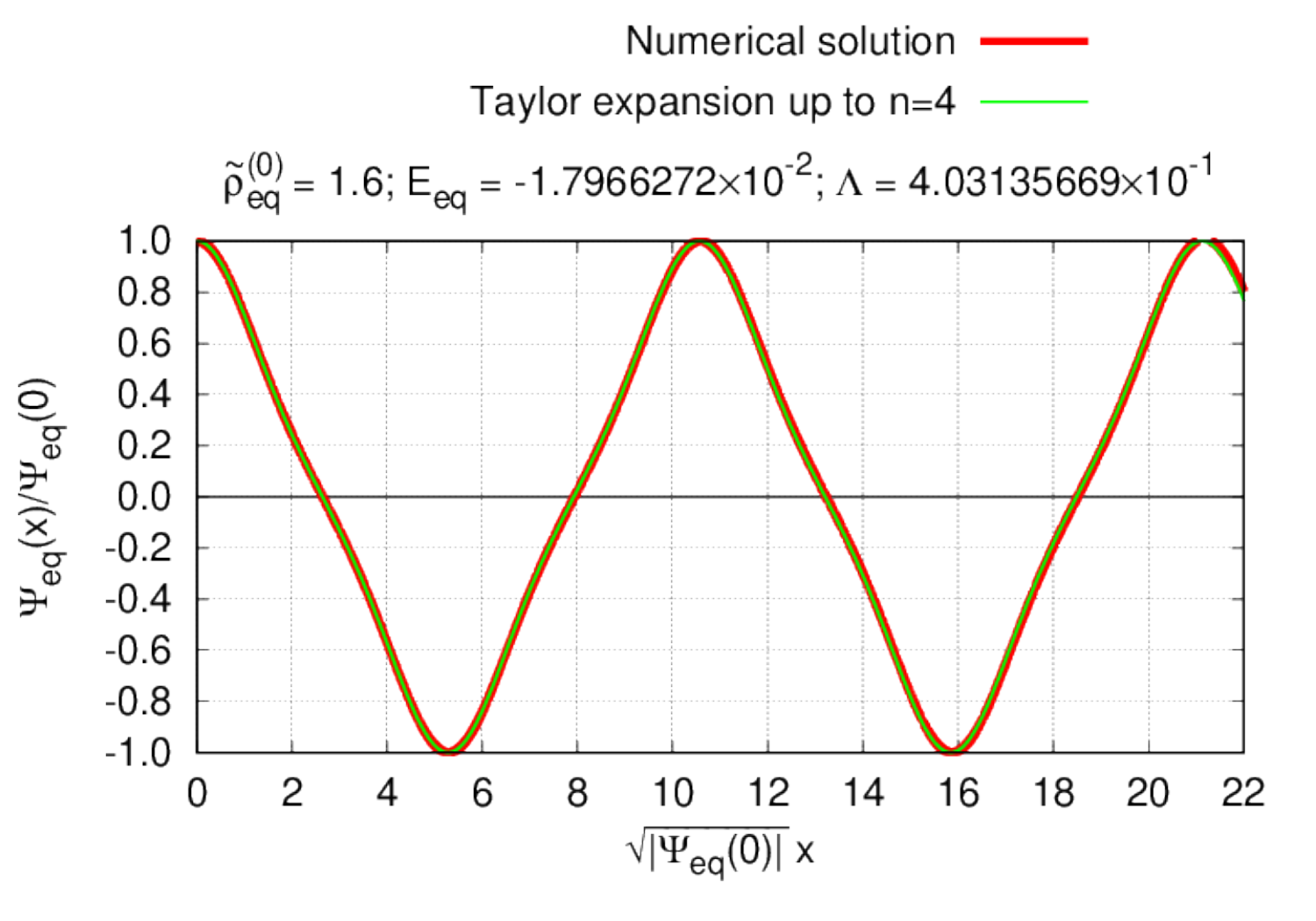}
\caption{\textit{An example of background wave functions with $\Tilde{\rho}^{(0)}_{eq} = 1.6$ that show the excellent agreement of the Taylor-expansion solution and the numerical solution. The red curve is the numerical solution solved from Eq. (\ref{equ: Schrodiger-Poisson equation for single component}). The green curve is derived from the Taylor expansion (Eqs. (\ref{equ: Tayler series of the eigenenergy, wave function and gravitional potential}), (\ref{equ: Fourier series of each order of background wave function and gravitational potential}), (\ref{equ: recursive relation for E_n, v_nl, and pis_nl}), and (\ref{equ: Fourier series of the background wave and gravitational potential})) up to $n = 4$.  The eigenenergy $E_{eq}$ and cosmological constant $\Lambda$ for $\Tilde{\rho}^{(0)}_{eq} = 1.6$ are shown at the top of the figure.}} 
\label{fig: Background_wave_function_compare_rho_1.6_Runge_Taylor}
\end{figure}

Figure (\ref{fig: Background_wave_function_compare_rho_1.6_Runge_Taylor}) presents the profile of $\Psi_{eq}$ for both numerical solution (red line) and series 
expansion solution (given by Eqs. (\ref{equ: Tayler series of the eigenenergy, wave function and gravitional potential}), (\ref{equ: Fourier series of each order of background wave function and gravitational potential}), and (\ref{equ: Fourier series of the background wave and gravitational potential})) up to $n=4$ (green line), for the case $\Tilde{\rho}^{(0)}_{eq} = 1.6$. The coefficients $\Tilde{\psi}_{n,l}$ used in the series expansion are listed in Table (\ref{Table: psi_n,k and v_n,k to n = 4}). Figure (\ref{fig: Background_wave_function_compare_rho_1.6_Runge_Taylor}) shows that the two lines closely track each other over two periods, demonstrating that the series expansion solution provides a sufficiently accurate approximation to the background configuration. 

\begin{figure}
\includegraphics[scale=0.345, angle=270]{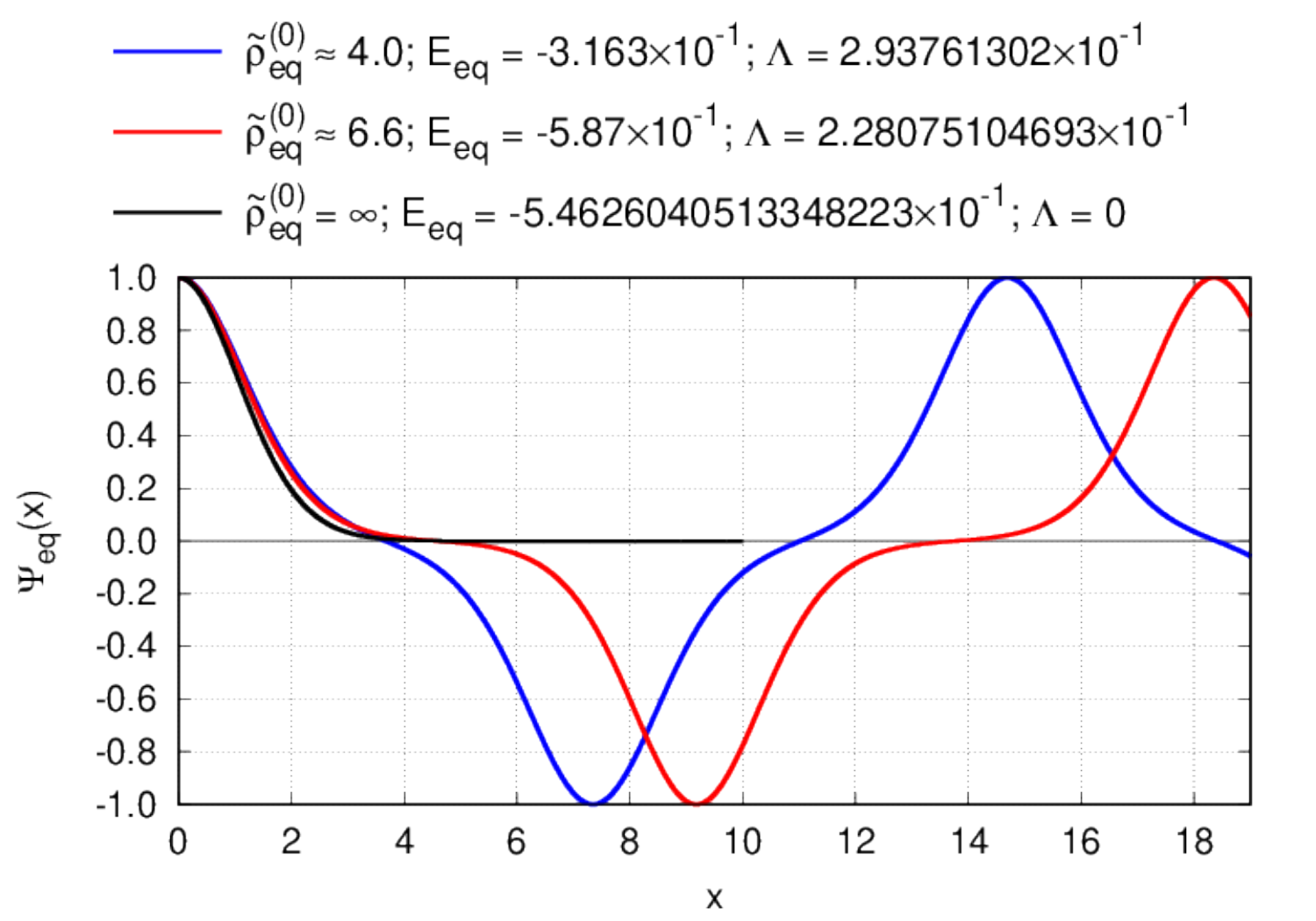}
\caption{\textit{Numerical solutions of background wave functions for $\Tilde{\rho}^{(0)}_{eq} = 4.0$(blue curve),  $\Tilde{\rho}^{(0)}_{eq} = 6.6$(red curve), and $\Tilde{\rho}^{(0)}_{eq} = \infty$(black curve), which demonstrate solutions within the radius of convergence $\Tilde{\rho}^{(c)}_{eq} =4.5$ and beyond. The similarity between the blue and red curves suggests that the background wave function can change smoothly as it crosses the radius of convergence $\Tilde{\rho}^{(c)}_{eq}$. Additionally, the black curve shows the one-dimensional soliton solution, which exhibits no local nodes.  For the blue and red curves, the gravitational potential has a zero spatial mean; whereas for the soliton solution(black curve), the gravitational potential is unbounded and increases with $x$.  Despite these differences, the values of $E_{eq} - V_{eq}(0)$ do not significantly vary among the cases, as the wave function profiles are similar within $x < 3$. Specifically, the values of $E_{eq} - V_{eq}(0)$ are about $3.64 \times 10^{-1}$, $3.85 \times 10^{-1}$, and $4.54 \times 10^{-1}$ for $\Tilde{\rho}^{(0)}_{eq}=4.0$, $6.6$, and $\infty$, respectively.}} 
\label{fig: Background_wave_function_large_rho_soliton_Runge}
\end{figure}

As the parameter $\Tilde{\rho}^{(0)}_{eq}$ increases, the period (unit cell) also becomes larger and the space average density decreases. At some point near the radius of convergence, the series expansion solution can no longer be valid. We nevertheless continue to explore numerical solutions near and beyond the radius of convergence to test if solutions continue to exist. Motivated by the existence of the 3D soliton solution \cite{Schive2014}, we have found the 1D soliton solution with relative ease because one of the two eigenvalues, $\Lambda$ (or mean density), takes a zero value. Soliton is the solution assuming an extreme value of $\Tilde{\rho}^{(0)}_{eq}$. Other two-eigenvalue solutions with moderate $\Tilde{\rho}^{(0)}_{eq}$ were subsequently identified.  

Solutions within and beyond the radius of convergence are found to be continuous, suggesting that such a radius is nothing but a mathematical artefact of the Taylor series expansion. The solution can in fact be analytically continued to a much larger domain \cite{Rudin1986}, joining the Taylor series solutions about $1/\Tilde{\rho}^{(0)}_{eq}\sim 0$, i.e., highly localized and extremely low average density solutions. This solution behavior is analogous to the smooth function $1/(1+x)$ for $x \geq 0$.  The Taylor series around $x=0$ has a domain of convergence of $x<1$.  Beyond the radius, the same function can be expressed as $(1/x)/((1/x)+1)$, and the Taylor expansion solution of  small $1/x$ can smoothly join the Taylor expansion solution of small $x$ at $x=1$. The class of series-expansion solutions about large $\Tilde{\rho}^{(0)}_{eq}$ is intriguing but will not be reported in this paper. Below, we present a sample of such numerical solutions to demonstrate the solution continuity. 

Aside from the soliton solution (black line), we present in Figure (3) two solutions below and above the radius of convergence, $\Tilde{\rho}^{(0)}_{eq}=4.0$ (blue line) and $6.6$ (red line). At the first glance, except for the peak separation, the two solutions look similar. When we increase $\Tilde{\rho}^{(0)}_{eq}$, the peak separation (or unit cell size) increases while the peak width remains intact. This observation is somewhat surprising, as it indicates that in the $1/\Tilde{\rho}^{(0)}_{eq}\ll 1$ regime, the asymtotic solution can be perceived as repeated soliton-anti-soliton pairs at a $\Tilde{\rho}^{(0)}_{eq}$-dependent separation $d$. Since the parameter $\Tilde{\rho}^{(0)}_{eq} \equiv(d/(2\pi))^4 \Big ( \int^{d}_{0} \Psi_{eq}(x)\cos{(2\pi x/d)}dx/d \Big )^2$, the separation scales as $d\propto (\Tilde{\rho}^{(0)}_{eq})^{1/4}$.

We fill in some details of the solutions here. The eigen-energy $E_{eq}$, cosmological constant, $\Lambda$, and $V_{eq}(0)$ for the three solutions are $E_{eq}=-3.163 \times 10^{-1}$, $-5.87 \times 10^{-1}$, and $-5.4626040513348223 \times 10^{-1}$; $\Lambda=2.93761302 \times 10^{-1}$, $2.28075104693 \times 10^{-1}$, and $0$; $V_{eq}(0)=-0.68$, $-0.972$, and $-1$ for $\Tilde{\rho}^{(0)}_{eq}=4.0$, $6.6$ and $\infty$, respectively. 

We note that the gravitational potential has a zero spatial mean for both cases with $\Tilde{\rho}^{(0)}_{eq}=4.0$ and $6.6$. However, for the soliton solution ($\Tilde{\rho}^{(0)}_{eq}=\infty$), the gravitational potential does not have a zero spatial mean, as the potential is unbounded and exhibits an asymptotic behavior proportional to $x$. Therefore, we adopt a different strategy to compare the eigenenergy of the soliton solution with those of other solutions. We define the new eigenvalue $E_{eq} - V_{eq}(0)$, where $V_{eq}(0)$ is the minimum of the gravitational potential. It turns out that the new eigenvalues are almost the same , i.e., $3.637 \times 10^{-1}$, $3.85 \times 10^{-1}$, and $4.5373959486651777 \times 10^{-1}$ for $\Tilde{\rho}^{(0)}_{eq}=4.0$, $6.6$, and $\infty$, respectively. Figure (\ref{fig: Background_wave_function_large_rho_soliton_Runge}) indicates that the background wave function profiles are similar within the main body of $\Psi_{eq}$ within the half periods ($x<3$) for $\Tilde{\rho}^{(0)}_{eq}=4.0$, $6.6$, and $\infty$. This explains why the new eigenvalues are similar.   
  
\section{Generalization of Bloch's theorem for the Stability analysis to the Periodic equilibrium Background}
\label{sec: General Version of Bloch's theorem for the Stability analysis to the Periodic equilibrium Background}

We will explore an approach through the Bloch's theorem for analyzing the stability of the periodic background configuration constructed in Sec. (\ref{sec: Nonlinear periodic equilibrium construction}).

Equation (\ref{equ: perturbed Schrodiger-Poisson equation for single component by real and imagine part}) is invariant under spatial translations perpendicular to the $x$-axis. Therefore, a Fourier transformation can be applied to Eq. (\ref{equ: perturbed Schrodiger-Poisson equation for single component by real and imagine part}) for the space perpendicular to the $x$-axis, the Laplacian operator is given by $\nabla^2 \equiv d^2/dx^2 - |\vec{k}_{\perp}|^2$ with $\vec{k}_{\perp}$ being the wave vector perpendicular to the $x$-axis, and the perturbed wave function depends only on the $x$ with a periodic spatial domain $[0,2\pi]$.  It is therefore natural to consider the perturbed wave function in the functional space $\mathbb{T}[0,2\pi]$ where  any function $f(x) \in \mathbb{T}[0,2\pi]$ satisfies $f(x+2\pi) = f(x)$. Moreover, the corresponding inner product in this functional space is defined as $\langle f | g \rangle \equiv \int^{2\pi}_{0} f^{*}(x)g(x)dx/(2\pi)$ for any functions $f(x)$ and $g(x) \in \mathbb{T}[0,2\pi]$. 

\subsection{Generalized Bloch's Theorem}
\label{subsubsec: Generalization to Bloch Theorem}

Since the background configuration can be treated as a lattice with the background gravitational potential period $\pi/n^{(0)}_{eq}$ as the lattices size, it relates to Bloch's theorem. Bloch's theorem states that the eigen wave function for a Hamiltonian operator is a Bloch function of the form $e^{\textbf{\textit{i}}qx}p(q,x)$. Of the Bloch functional form, $e^{\textbf{\textit{i}}qx}$ is a long wavelength plane wave modulation and $p(q,x)$ is the short wavelength periodic function satisfying $p(q,x+\pi/n^{(0)}_{eq}) = p(q,x)$, matching the period of the background gravitational potential. Here, $q$ is the quasi-momentum and a real number \cite{Kittel2004}.  Both operators $\hat{\mathbf{H}}$ and $\hat{\mathbf{\Theta}}$ are invariant under the spatial translation by a lattice cell. Therefore, the eigenfunction for the operator $\hat{\mathbf{\Theta}}\hat{\mathbf{H}}$ exhibits a similar property to Bloch's theorem, as follows:
\\\\
\noindent \textbf{(\rmnum{1}-1): The eigenfunction $\bm{\eta}$ for the operator $\hat{\mathbf{\Theta}}\hat{\mathbf{H}}$ can be expressed as $\bm{\eta(x)=e^{\textbf{\textit{i}}qx}p_{\eta}(q,x)}$, where $\bm{q}$ is a real number and $\bm{p_{\eta}}$ is a periodic function satisfying $\bm{p_{\eta}(q,x+\pi/n^{(0)}_{eq}) = p_{\eta}(q,x)}$.}

\textit{The proof is similar to the Floquet theory \cite{Chicone2006}, which is the general version of Bloch's theorem in a one-dimensional context. The key idea is that $\eta(x+\pi/n^{(0)}_{eq})$ is also the solution to the eigenvalue problem for the operator $\hat{\mathbf{\Theta}}\hat{\mathbf{H}}$ with eigenvalue $\omega^2$. A detailed proof can be found in Appendix \ref{appendix: Generalized Bloch's Theorem for the Operator ThetaH}.}
\\\\
The wavenumber $q$ represents the quasi-momentum of the plane wave. The sign of the quasi-momentum indicates the direction of propagation of this plane wave. Similar to Bloch's theorem, the range of the quasi-momentum can be confined to $[-n^{(0)}_{eq},n^{(0)}_{eq}]$ because the above statement remains unchanged under the transformation $q \rightarrow q \pm 2n^{(0)}_{eq}$. Consequently, any short-wavelength component smaller than the period of the background wave function in the eigenfunction is represented within the periodic function $p_{\eta}(q,x)$. In addition, the real part of wave function $R$ also has a similar invariance property as the imaginary part $\eta$ because $R=\hat{\mathbf{H}}\eta$.  Hence, the perturbed eigen wave function can be expressed in a form similar to the Bloch function. 

The primary distinction between the present case and conventional Bloch's theorem is that both perturbed density and gravitational potential can similarly be expressed as plane waves with quasi-momentum $q$, but they are modulated by anti-periodic functions. To demonstrate this, the perturbed density $\delta\rho(x)\propto\Psi_{eq}(x)R(x) =e^{\textbf{\textit{i}}qx}\Psi_{eq}(x)p_{R}(q,x)\equiv e^{\textbf{\textit{i}}qx}p_{\delta \rho}(q,x)$. Here, $p_{\delta \rho}(q,x+\pi/n^{(0)}_{eq}) = \Psi_{eq}(x+\pi/n^{(0)}_{eq})p_{R}(q,x+\pi/n^{(0)}_{eq})=-\Psi_{eq}(x)p_{R}(q,x)=-p_{\delta \rho}(q,x)$, since $ \Psi_{eq}(x)$ is anti-periodic. Similarly, the perturbed gravitational potential $\delta V(x)$ has the same property as the perturbed density via the Poisson's equation.     

In addition both left- and right-propagating plane waves coexist for a given eigenvalue $\omega^2$ , a statement is precisely put as follows: 
\\\\
\noindent \textbf{(\rmnum{1}-2): For a given eigenvalue $\bm{\omega^2}$, a pair of eigenfunctions with the Bloch functional form $\bm{\eta_{\pm}(x) = e^{\pm iqx} p_{\eta}(\pm q, x)}$ can be found for the operator $\hat{\mathbf{\Theta}}\hat{\mathbf{H}}$, provided that $\bm{\Psi_{eq}(x)}$ and $\bm{V_{eq}(x)}$ have even symmetry.}

\textit{Since $\Psi_{eq}(-x) = \Psi_{eq}(x)$ and $V_{eq}(-x) = V_{eq}(x)$,  it follows that $\eta(-x)$ is also a solution to the eigenvalue problem for the operator $\hat{\mathbf{\Theta}}\hat{\mathbf{H}}$. Based on this fact, the conclusion can immediately be reached by applying a similar argument as in Floquet theory. A detailed proof can be found in Appendix \ref{appendix: Generalized Bloch's Theorem for the Operator ThetaH}.}
\\\\

The Hamiltonian operator also has the same characteristic. Without loss of generality, the range of the quasi-momentum $q$ can be further confined to $[0,n^{(0)}_{eq}]$, similar to the band gap theory for conventional quantum mechanics. Here, $q=0$ and $q=n^{(0)}_{eq}$ form the boundaries of the $q$-space. Moreover, the quasi-momentum $q$ is discrete when we let the spatial domain $[0,2\pi]$ be periodic. Given that this domain contains $2n^{(0)}_{eq}$ lattice cells, the quasi-momentum $q$ must be quantized to discrete values $0,1,...,n^{(0)}_{eq}$ to ensure the periodicity of the plane wave $e^{\textbf{\textit{i}}qx}$ over the interval $[0,2\pi]$. As the quasi-momentum $q$ is restricted by $n^{(0)}_{eq}$, the ratio $q/n^{(0)}_{eq}$ covers all rational numbers within the interval $[0,1]$.

\subsection{Standing Wave Basis Expansion}
\label{subsubsec: Standing Wave Basis Expansion}

We introduce the normalized perturbed quantities as follows: $\Tilde{\omega} \equiv \omega/2E^{(0)}_{eq}$, $\Tilde{R} \equiv R/(E^{(0)}_{eq}\sqrt{\Tilde{\rho}^{(0)}_{eq}})$, $\Tilde{I} \equiv I/(E^{(0)}_{eq}\sqrt{\Tilde{\rho}^{(0)}_{eq}})$, $\widetilde{\delta V}  \equiv \delta V/E^{(0)}_{eq}$, and $\Tilde{\eta} \equiv \eta/\sqrt{\Tilde{\rho}^{(0)}_{eq}} = -\textbf{\textit{i}}\Tilde{I}/(2\Tilde{\omega})$. The normalized operators $\hat{\mathbf{H}}$ and $\hat{\mathbf{\Theta}}$ become

\begin{equation}
\label{equ: definition of H and Theta with normalization factor}
\left\{\begin{aligned}
& \hat{\mathbf{H}} \equiv {{1}\over{E^{(0)}_{eq}}} \Bigg [ -{{1}\over{2}} \Bigg ( {{d^2}\over{dx^2}} - |\vec{k}_{\perp}|^2 \Bigg ) + V_{eq}(x) - E_{eq} \Bigg ]; \\
& \hat{\mathbf{\Theta}} \equiv  {{1}\over{4}} \Big (  \hat{\mathbf{H}} + \hat{\mathbf{G}}  \Big ); \\
& \hat{\mathbf{G}} \equiv 16\Tilde{\rho}^{(0)}_{eq} \Tilde{\Psi}_{eq} (2E^{(0)}_{eq} \nabla^{-2}) \Tilde{\Psi}_{eq}.
\end{aligned}
\right.
\end{equation}
We also introduce $\Tilde{k}^2_{\perp} \equiv |\vec{k}_{\perp}|^2/2E^{(0)}_{eq} = \Big | \vec{k}_{\perp}/n^{(0)}_{eq} \Big |^2$. To simplify the notation in the subsequent sections, particularly in Secs. (\ref{subsubsec: Standing Wave Basis Expansion}) and (\ref{sec: Comparison Between Theoretical Prediction and Simulating result}), we use the same notations as in Eq. (\ref{equ: definition of H and Theta}) to represent the corresponding normalized operators, as shown in Eq. (\ref{equ: definition of H and Theta with normalization factor}).   

Note that any Bloch function with quasi-momentum $q$ can be represented by the Fourier series $\sum^{\infty}_{j=-\infty}c_m e^{\textbf{\textit{i}}(2jn^{(0)}_{eq}+q)x}$ for some constant $c_m$. Additionally, since the background configuration has no net momentum, a standing wave is more suitable than a plane wave for dealing with this quantum system. Based on these arguments, the functional space $\mathbb{T}[0,2\pi]$ should be considered as the direct sum of all functional subspaces $\mathbb{T}^{(cos)}_{q}[0,2\pi]$ and $\mathbb{T}^{(sin)}_{q}[0,2\pi]$ with $q=0,1,...,n^{(0)}_{eq}$. The definitions of these subspaces are 

\begin{equation}
\label{equ: the definition of sub-space}
\left\{\begin{aligned}
& \mathbb{T}^{(cos)}_{q}[0,2\pi] \equiv & span\Big( \Big \{  \cos{[(2j n^{(0)}_{eq} + q)x]} \Big \}^{\infty}_{j=-\infty}\Big ), \\    
& \mathbb{T}^{(sin)}_{q}[0,2\pi] \equiv & span\Big( \Big \{  \sin{[(2j n^{(0)}_{eq} + q)x]} \Big \}^{\infty}_{j=-\infty}\Big ). 
\end{aligned}
\right.
\end{equation}
Here, $span$ denotes the linear vector space formed by all linear combinations of vectors generated by the set in its argument. Note that these subspaces are mutually orthogonal. 

The size of the reciprocal lattice cell in reciprocal space is given by $2n^{(0)}_{eq}$, which is the inverse of the period of the background gravitational potential. Hence, the wave-number $2j n^{(0)}_{eq} + q$ is located in the $(|j|+1)$-the Brillouin zone in the reciprocal space. The positive (negative) sign of the index $j$ indicates that this Brillouin zone is on the right (left) side of the origin in reciprocal space.

Under this decomposition, each of the aforementioned subspaces remains decoupled from the others upon the application of the operators $\hat{\mathbf{H}}$ and $\hat{\mathbf{G}}$ operator, implying that the operator $\hat{\mathbf{\Theta}}$ possesses this property as well. This can be summarized as follows:
\\\\
\noindent \textbf{(\rmnum{2}-1): $\hat{\mathbf{H}}$f and $\hat{\mathbf{G}}$f $\bm{\in}$ $\bm{\mathbb{T}^{(cos)}_{q}[0,2\pi]}$($\bm{\mathbb{T}^{(sin)}_{q}[0,2\pi]}$) if f $ \in \bm{\mathbb{T}^{(cos)}_{q}[0,2\pi]}$($\bm{\mathbb{T}^{(sin)}_{q}[0,2\pi]}$).}

\textit{The Laplacian operator $\nabla^2$ in both operators do not generate new Fourier modes. In addition both $\Tilde{\Psi}_{eq}$ and $\Tilde{V}_{eq}$ are composed entirely of cosine functions, and from the trigonometric identities  $\cos{y}\cos{x} =(1/2)( \cos{(x+y)}+\cos{(x-y)})$ and $\cos{y}\sin{x} = (1/2)(\sin{(x+y)}+\sin{(x-y)})$, it follows that application of either operator $\hat{\mathbf{H}}$ and $\hat{\mathbf{G}}$ does not mix the two subspaces.}  
\\\\

The above statement further implies that both eigenfunctions, $\Tilde{\eta}$ of the operator $\hat{\mathbf{\Theta}}\hat{\mathbf{H}}$ and $\Tilde{R}$ of the operator $\hat{\mathbf{H}}\hat{\mathbf{\Theta}}$, reside either in the space $\mathbb{T}^{(cos)}_{q}[0,2\pi]$ or $\mathbb{T}^{(sin)}_{q}[0,2\pi]$ for some $q=0,1,...,n^{(0)}_{eq}$. Within each subspace, we are now ready to adopt the matrix approach to the eigenvalue problem with the standing wave of Eq. (\ref{equ: the definition of sub-space}) as the basis. To label these standing wave bases, we introduce the new notation $u_j(x)$, defined as follows:

\begin{equation}
\label{equ: the definition of the standing wave basis}
\begin{aligned}
& u_j(x) \equiv 
\left\{\begin{aligned}
& \sqrt{2} \cos{( k_jx ) }\text{, for } \mathbb{T}^{(cos)}_{q}[0,2\pi]; \\
& \sqrt{2} \sin{( k_jx ) }\text{, for } \mathbb{T}^{(sin)}_{q}[0,2\pi], \\
\end{aligned}
\right.\\
&k_j \equiv  2j n^{(0)}_{eq} + q.
\end{aligned}
\end{equation}
Here, the index $j \in \mathbb{N} \cup \{ 0\}$ for $q = 0, n^{(0)}_{eq}$, and $j \in \mathbb{Z}$ for $q =1,...,n^{(0)}_{eq}-1$, and $u_j(x)$ forms a complete orthogonal basis set within each subspace. The corresponding wave number $k_j$ is located in $(|j|+1)$-th Brillouin zone.

With the above, the operator $\hat{\mathbf{\Theta}}\hat{\mathbf{H}}$ can be represented as an infinite-dimensional matrix in the matrix representation, denoted as $\mathbf{F}$, with the matrix element $\mathbf{F}_{ij} \equiv \langle u_i | \hat{\mathbf{\Theta}}\hat{\mathbf{H}} | u_j \rangle$, representing the transition amplitude from the standing wave basis $u_j$ to the standing wave basis $u_i$. 
 
A simple calculation yields that the matrix element $\mathbf{F}_{ij}$ is:

\begin{equation}
\label{equ: the formula of F_ij}
\begin{aligned}
\mathbf{F}_{ij} & = \langle u_i | \hat{\mathbf{\Theta}}\hat{\mathbf{H}} | u_j \rangle =  {{1}\over{4}}\Big ( \langle u_i |\hat{\mathbf{H}}^2 | u_j \rangle + \langle u_i |\hat{\mathbf{G}}\hat{\mathbf{H}} | u_j \rangle \Big ) \\
                &  \begin{aligned}  
                   =  &  {{1}\over{4}}\langle u_i |\hat{\mathbf{H}}^2 | u_j \rangle + \\
                      &  4\Tilde{\rho}^{(0)}_{eq} (2E^{(0)}_{eq}) \sum_{m} {{\langle u_i | \Tilde{\Psi}_{eq} \nabla^{-2} \Tilde{\Psi}_{eq} | u_m \rangle \langle u_m | \hat{\mathbf{H}} | u_j \rangle}\over{\langle u_m | u_m \rangle }}.
                   \end{aligned} 
\end{aligned}
\end{equation}
The detailed calculations of the matrix elements $\langle u_m | \hat{\mathbf{H}} | u_j \rangle$, $\langle u_i |\hat{\mathbf{H}}^2 | u_j \rangle$, and $\langle u_i | \Tilde{\Psi}_{eq} \nabla^{-2} \Tilde{\Psi}_{eq} | u_m \rangle$ are provided in Appendix \ref{appendix: The General Property of the Matrix Representation}. 

The first term, $\langle u_i | \hat{\mathbf{H}}^2 | u_j \rangle/4$, in Eq. (\ref{equ: the formula of F_ij}) forms a symmetric matrix. However, the second term, $4\Tilde{\rho}^{(0)}_{eq}\sum_{m} \langle u_i | \Tilde{\Psi}_{eq} (2E^{(0)}_{eq}\nabla^{-2}) \Tilde{\Psi}_{eq} | u_m \rangle \langle u_m | \hat{\mathbf{H}} | u_j \rangle$ in Eq. (\ref{equ: the formula of F_ij}), is generally {\it not} symmetric, despite that $4\Tilde{\rho}^{(0)}_{eq}\langle u_i | \Tilde{\Psi}_{eq} (2E^{(0)}_{eq}\nabla^{-2}) \Tilde{\Psi}_{eq} | u_m \rangle$ and $\langle u_m | \hat{\mathbf{H}} | u_j \rangle$ are individually symmetric. The emergence of non-real eigenvalues is attributed to this second term. 

In conjunction with the detailed calculations in Appendix \ref{appendix: The General Property of the Matrix Representation}, Eqs. (\ref{equ: the formula of F_ij}) shows that the matrix element $\mathbf{F}_{ij} = \mathbf{F}_{ij} \Big ( q/n^{(0)}_{eq}, \Tilde{k}^2_{\perp}, \Tilde{\rho}^{(0)}_{eq}\Big )$. Therefore, the corresponding eigenvalue $\Tilde{\omega}^2 =\Tilde{\omega}^2 \Big ( q/n^{(0)}_{eq}, \Tilde{k}^2_{\perp}, \Tilde{\rho}^{(0)}_{eq}\Big )$.

To evaluate the eigenvalue $\Tilde{\omega}^2$, further approximation on the matrix representation $\mathbf{F}$ is necessary. The key idea for this approximation is to truncate the matrix to a finite dimension. What follows provide several properties of $\mathbf{F}_{ij}$, which offer insights on how to reduce the dimension.

Firstly, the diagonal element $\mathbf{F}_{jj}$ has the following property:
\\\\
\noindent \textbf{(\rmnum{2}-2): There exists a positive integer $\bm{N = N \Big (\Tilde{\rho}^{(0)}_{eq}, \Tilde{k}^2_{\perp}, q/n^{(0)}_{eq} \Big)}$ such that $\bm{\mathbf{F}_{jj} > 0}$ for $\bm{|j| \geq N}$. Moreover, the set $\bm{\{ \mathbf{F}_{jj}\}^{\infty}_{|j|\geq N}}$ has no the upper bound because the diagonal element $\bm{\mathbf{F}_{jj}}$ is roughly proportional to $\bm{\Tilde{k}^4_{j}}$ as the absolute value of the index $\bm{j}$ becomes sufficiently large. Here, $\bm{\Tilde{k}_{j} \equiv k_j/n^{(0)}_{eq} = 2j + q/n^{(0)}_{eq}}$.}

\textit{Initially, the term, $\langle u_j | \hat{\mathbf{H}}^2 | u_j \rangle/4$, in Eq. (\ref{equ: the formula of F_ij}) includes a factor of $\Tilde{k}^4_{j}$, which originates from the kinetic operator. In contrast, the perturbed self-gravity energy tensor in Eq. (\ref{equ: the formula of F_ij}) is always bounded by the total mass, $\langle \Tilde{\Psi}_{eq} | \Tilde{\Psi}_{eq} \rangle$, as demonstrated in Appendix \ref{appendix: Uniform Boundedness of the Perturbed Self-Gravity Energy Tensor in the Operator Theta}. Consequently, this property follows directly.} 
\\\\

On the other hand, the matrix representation $\mathbf{F}$ is not a diagonal matrix because the gravitational potential causes transitions between different standing wave bases. However, this coupling due to gravity suggests that the off-diagonal elements are at least of the order of $\Tilde{\rho}^{(0)}_{eq}$. Quantitatively, this order can be estimated as follows:
\\\\
\noindent \textbf{(\rmnum{2}-3): The off-diagonal element $\mathbf{F}_{\bm{ij}}$ with $\bm{i \neq j}$ is approximately bounded by $ \bm{\Bigg \{ |\Tilde{k}^2_i + \Tilde{k}^2_{\perp} - \Tilde{E}_{eq}|/4 + \Big [ 4\Tilde{\rho}^{(c)}_{eq} h(q/n^{(0)}_{eq}, \Tilde{k}^2_{\perp}) + 1/4 \Big ]\Big [ |\Tilde{k}^2_j + \Tilde{k}^2_{\perp} - \Tilde{E}_{eq}| + |i-j| + ( 1 + (\Tilde{\rho}^{(0)}_{eq}/\Tilde{\rho}^{(c)}_{eq})^2)/ ( 1 - (\Tilde{\rho}^{(0)}_{eq}/\Tilde{\rho}^{(c)}_{eq})^2) \Big ] \Bigg \}\Bigg ( {{\Tilde{\rho}^{(0)}_{eq}}\over{\Tilde{\rho}^{(c)}_{eq}}}\Bigg )^{|i-j|}}$ with the radius of convergence $\bm{\Tilde{\rho}^{(c)}_{eq}}$ introduced in Sec. (\ref{sec: Nonlinear periodic equilibrium construction}) and the function $\bm{h}$ defined in Eq. (\ref{equ: definition of function h}).}

\textit{To understand the factor $\Big ( \Tilde{\rho}^{(0)}_{eq}/\Tilde{\rho}^{(c)}_{eq}\Big )^{|i-j|}$, note that the matrix element $\mathbf{F}_{ij}$ represents the transition from the initial standing wave basis $u_j$ to another basis $u_i$. The orthogonality between $u_i$ and $u_j$ implies that only the Fourier mode with the wave number $|(2i n^{(0)}_{eq} -q) - (2jn^{(0)}_{eq}-q )|=2|i-j|n^{(0)}_{eq}$ for the background configuration can contribute a non-zero value to the matrix element $\mathbf{F}_{ij}$. Furthermore, the Fourier coefficient with such a wave number is approximately $\Big ( \Tilde{\rho}^{(0)}_{eq}/\Tilde{\rho}^{(c)}_{eq}\Big )^{|i-j|}$, as established in Sec. (\ref{sec: Nonlinear periodic equilibrium construction}).}
\\\\
The detail proof of above two properties involves extensive calculations and is provided in Appendix \ref{appendix: The General Property of the Matrix Representation}.  

Based on the aforementioned properties, the matrix elements satisfy the conditions $|\mathbf{F}_{ij}/\mathbf{F}_{jj}| \ll 1$ and $|\mathbf{F}_{ji}/\mathbf{F}_{jj}| \ll 1$ for $|j| \geq N$. This holds true for all $i \in \mathbb{N}$. To illustrate this, consider that the diagonal element $\mathbf{F}_{jj} \approx \Tilde{k}^4_j$. In contrast, the off-diagonal elements $\mathbf{F}_{ij}$ and $\mathbf{F}_{ji}$ can be approximated as $( \Tilde{k}^2_i + \Tilde{k}^2_j )(\Tilde{\rho}^{(0)}_{eq}/\Tilde{\rho}^{(c)}_{eq})^{|i-j|}$. Consequently, the exponential factor suppresses the off-diagonal elements towards zero as $|i-j| \rightarrow \infty$. When $|i-j| \approx O(1)$, the off-diagonal elements are approximately $O(\Tilde{k}^2_j)$. Thus, $|\mathbf{F}_{ij}/\mathbf{F}_{jj}|$ and $|\mathbf{F}_{ji}/\mathbf{F}_{jj}|$ are roughly inversely proportional to $\Tilde{k}^{2}_j$, and these ratios become sufficiently small when $\Tilde{k}^2_j$ is sufficiently large.  

The above conditions indicates that the matrix representation $\mathbf{F}$ is nearly diagonal when the absolute value of the element index is sufficiently large. Therefore, the diagonal element $\mathbf{F}_{jj}$ serves as a good approximation for the eigenvalue, and the corresponding eigenfunction can be approximated by the standing wave basis $u_j$ for $|j| \geq N$. Furthermore, these eigenvalues correspond to stable modes due to the positive definiteness of these diagonal elements.

Most stable modes have been identified, and this result provides a methodology to reduce the matrix dimensionality to $N$ whereby the dominant modes are kept and the most stable modes are discarded. There are a finite number of unstable dominant eigenvalues. To determine these eigenvalues, we recall that the eigenvalue problem $\Tilde{\omega^2 \Tilde{\eta}} = \hat{\mathbf{\Theta}}\hat{\mathbf{H}} \Tilde{\eta}$ can be transformed into the following equation:

\begin{equation}
\label{equ: the eigenvalue problem with matrix representation}
\left\{\begin{aligned}
&\Tilde{\omega}^2 a_i = \sum_{|j| < N}\mathbf{F}_{ij}a_j + \sum_{|l|\geq N}\mathbf{F}_{il}a_l, \text{ }|i| < N; \\
&0 = \sum_{|j| < N}\mathbf{F}_{ij}a_j + \sum_{|l| \geq N}( \mathbf{F}_{il} -\Tilde{\omega}^2 \delta_{il})a_l, \text{ }|i| \geq N.
\end{aligned}
\right.
\end{equation}
Here, $a_j = \langle u_j | \Tilde{\eta} \rangle$. To reduce the infinite number of equations to a finite set, we shall be confined to evaluation of the low-lying eigenvalues $\Tilde{\omega}^2$ and make use of the adiabatic condition where $|\Tilde{\omega}^2| \ll |\mathbf{F}_{ii}|$ for all $|i| \geq N$, i.e., the Wentzel–Kramers–Brillouin (WKB) approximation \cite{Shankar1994}. This condition is readily met as $\mathbf{F}_{ii}$ diverges to infinity with increasing $|i|$, ensuring the validity of the approximation. 

Combining the above propositions with the condition $|\mathbf{F}_{il}/\mathbf{F}_{ii}| \ll 1$ $|i| \geq N$ and for all $l \in \mathbb{N}$, we find the second equation in Eq. (\ref{equ: the eigenvalue problem with matrix representation}) can be approximated as $0 \approx \sum_{|j| < N}\mathbf{F}_{ij}a_j + \mathbf{F}_{ii}a_{ii}$. Consequently, the amplitude $a_i$ for $|i| \geq N$ can be  approximated by $a_i \approx - \sum_{|j| < N }\mathbf{F}_{ij}a_j/\mathbf{F}_{ii}$. Substituting this expression into the first equation of Eq. (\ref{equ: the eigenvalue problem with matrix representation}) gives:

\begin{equation}
\label{equ: the rest eigenvalue problem}
\left\{\begin{aligned}
& \Tilde{\omega}^2 a_i \approx \sum_{|j| < N} \Big ( \mathbf{F}_{ij} - \mathbf{\Delta F}_{ij} \Big )a_j, \text{ }|i| < N. \\
& \mathbf{\Delta F}_{ij} \equiv \sum_{|l| \geq N} {{\mathbf{F}_{il}\mathbf{F}_{lj}}\over{\mathbf{F}_{ll}}},  \text{ }|i|\text{, }|j| < N. 
\end{aligned}
\right.
\end{equation}
Here, the quantity $\mathbf{\Delta F}_{ij}$ is treated as feedback correction from the stable modes.

The feedback quantity $\mathbf{\Delta F}_{ij}$ can be estimated as follows: $\mathbf{F}_{ll} \sim \Tilde{k}^4_l$ since $|l| \geq N$. On the other hand, $|\mathbf{F}_{il}| \lesssim \Tilde{k}^2_l (\Tilde{\rho}^{(0)}_{eq}/\Tilde{\rho}^{(c)}_{eq})^{|i-l|}$ and $|\mathbf{F}_{lj}| \lesssim \Tilde{k}^2_l (\Tilde{\rho}^{(0)}_{eq}/\Tilde{\rho}^{(c)}_{eq})^{|l-j|}$. Hence, $|\mathbf{F}_{il}\mathbf{F}_{lj}/\mathbf{F}_{ll}| \lesssim (\Tilde{\rho}^{(0)}_{eq}/\Tilde{\rho}^{(c)}_{eq})^{|i-j|+2\min{\{|i - l|,|l - j|\}}}$. Consequently, the feedback correction $\mathbf{\Delta F}_{ij} \sim O \Big ( (\Tilde{\rho}^{(0)}_{eq}/\Tilde{\rho}^{(c)}_{eq})^{|i -j|+2\gamma} \Big )$ 
 with $\gamma \equiv \min_{|l|>N}{\{|i - l|,|l - j|\}}$. It is important to note that the exponential index $\gamma \neq 0$ since both $|i|$ and $|j| < N$.

The above estimation suggests that this correction $\mathbf{\Delta F}_{ij}$ can be further discarded for sufficiently large integre $N$. It can be understood as follows. The leading term for the diagonal element $\mathbf{F}_{jj}$ is $(\Tilde{k}^2_{j} + \Tilde{k}^2_{\perp} - \Tilde{E}_{eq})^2$; on the other hand, the off-diagonal element $\mathbf{F}_{ij}$ with $i \neq j$ is of the order of $(\Tilde{\rho}^{(0)}_{eq}/\Tilde{\rho}^{(c)}_{eq})^{|i - j|}$. It is straightforward to see that $|\mathbf{\Delta F}_{ij}/\mathbf{F}_{ij}| \approx O \Big ( (\Tilde{\rho}^{(0)}_{eq}/\Tilde{\rho}^{(c)}_{eq})^{2\gamma} \Big ) \leq O \Big ( (\Tilde{\rho}^{(0)}_{eq}/\Tilde{\rho}^{(c)}_{eq})^{2} \Big )$. Hence, this feedback correction can be discarded once the parameter $\Tilde{\rho}^{(0)}_{eq}$ is sufficiently far from the radius of convergence $\Tilde{\rho}^{(c)}_{eq}$. 

In conclusion, the dominant eigenvalues for the operator $\hat{\mathbf{\Theta}}\hat{\mathbf{H}}$ can be approximated by the standard eigenvalue problem 
\begin{equation}
\label{equ: the final eigenvalue problem} 
\Tilde{\omega}^2a_i \approx \sum_{ |j| < N } \mathbf{F}_{ij}a_j, \text{ }|i| < N,
\end{equation}
and Eq. (\ref{equ: the final eigenvalue problem}) can be solved using standard methods of linear algebra.

Recall that the wave number for the standing wave basis $u_j$ is located in $(|j|+1)$-th Brillouin zone. Thus, Eq. (\ref{equ: the final eigenvalue problem}) suggests that the stability of the periodic background, as constructed in Sec. (\ref{sec: Nonlinear periodic equilibrium construction}), is primarily determined by the interactions among the standing wave bases with wave numbers situated in the $j$-th Brillouin zone for all $j \leq N$.

In this section, we have thoroughly expounded the methodology employed to study the stability. The detailed formula of $\mathbf{F}_{ij}$ is given in Appendix C and the numerical results are presented in Sec. (\ref{subsec: Theoretical Predictions}). 

\section{Comparison Between Theoretical Predictions and Simulation Results}
\label{sec: Comparison Between Theoretical Prediction and Simulating result}

This section is divided into two parts. Simulation results are presented in Section (\ref{subsec: Simulating results}). Theoretical predictions, derived using the methodology proposed in Section (\ref{subsubsec: Standing Wave Basis Expansion}), are provided in Section (\ref{subsec: Theoretical Predictions}) for comparison with the simulation results.

In this section, we introduce the quasi-momentum of the perturbed density as the primary parameter rather than that of the perturbed wave function. This choice is motivated by the fact that density oscillations eliminate the background eigenenergy $E_{eq}$, allowing for the direct measurement of the eigenvalue $\Tilde{\omega}^2$. According to Sec. (\ref{subsubsec: Standing Wave Basis Expansion}), the perturbed density $\delta \rho \propto \Tilde{\Psi}_{eq}\Tilde{R} \in \mathbb{T}^{(cos)}_{p}[0,2\pi]$(or $\mathbb{T}^{(sin)}_{p}[0,2\pi]$) if $\Tilde{R} \in \mathbb{T}^{(cos)}_{q}[0,2\pi]$(or $\mathbb{T}^{(sin)}_{q}[0,2\pi]$), where $p = n^{(0)}_{eq} - q$. Consequently, the quasi-momentum of the perturbed density is given by $p$. The normalized quasi-momentum, denoted as $\Tilde{p}$, is defined as $p/n^{(0)}_{eq}$, where $\Tilde{p} \in [0,1]$.

\subsection{Simulation Results}
\label{subsec: Simulating results}

The simulation setup is as follows: the simulation space is a one-dimensional periodic interval with a length of $2\pi$. The initial wave function is provided by Eq. (\ref{equ: Tayler series of the eigenenergy, wave function and gravitional potential}) up to $n=4$ for a given constant $\Tilde{\rho}^{(0)}_{eq}$ and a positive integer $n^{(0)}_{eq} > 1$. In Eq. (\ref{equ: Tayler series of the eigenenergy, wave function and gravitional potential}), each order of the wave function $\Tilde{\Psi}_n(x)$ is determined by Eq. (\ref{equ: Fourier series of each order of background wave function and gravitational potential}), with the corresponding Fourier coefficients given in Table (\ref{Table: psi_n,k and v_n,k to n = 4}). The periodic interval is discretized with a uniform grid of length $0.01\pi/n^{(0)}_{eq}$ so that the shortest wavelength of the initial wave function is sampled by approximately $22$ grid points.

This initial setting is not an equilibrium state because all terms of $\Tilde{\rho}^{(0)}_{eq}$ higher than the fourth degree are ignored. However, these omitted terms can be treated as adding a perturbed wave function to the background wave function to compensate for the missing terms. Thus, this initial setting represents the background wave function with a seed perturbation of the order $(\Tilde{\rho}^{(0)}_{eq}/\Tilde{\rho}^{(c)}_{eq})^5$.  

Our simulation is one dimensional. Demonstration of the presence of instabilities in one dimension is sufficient for our purpose, as finite $\Tilde {k_\perp}$ modes are less unstable. In addition, one dimensional simulation offers great spatial resolution for accurate results, allowing for close examination of modes of various wave numbers $n^{(0)}_{eq}$ and amplitudes $\Tilde{\rho}^{(0)}_{eq}$.

\begin{figure}
\includegraphics[scale=0.44]{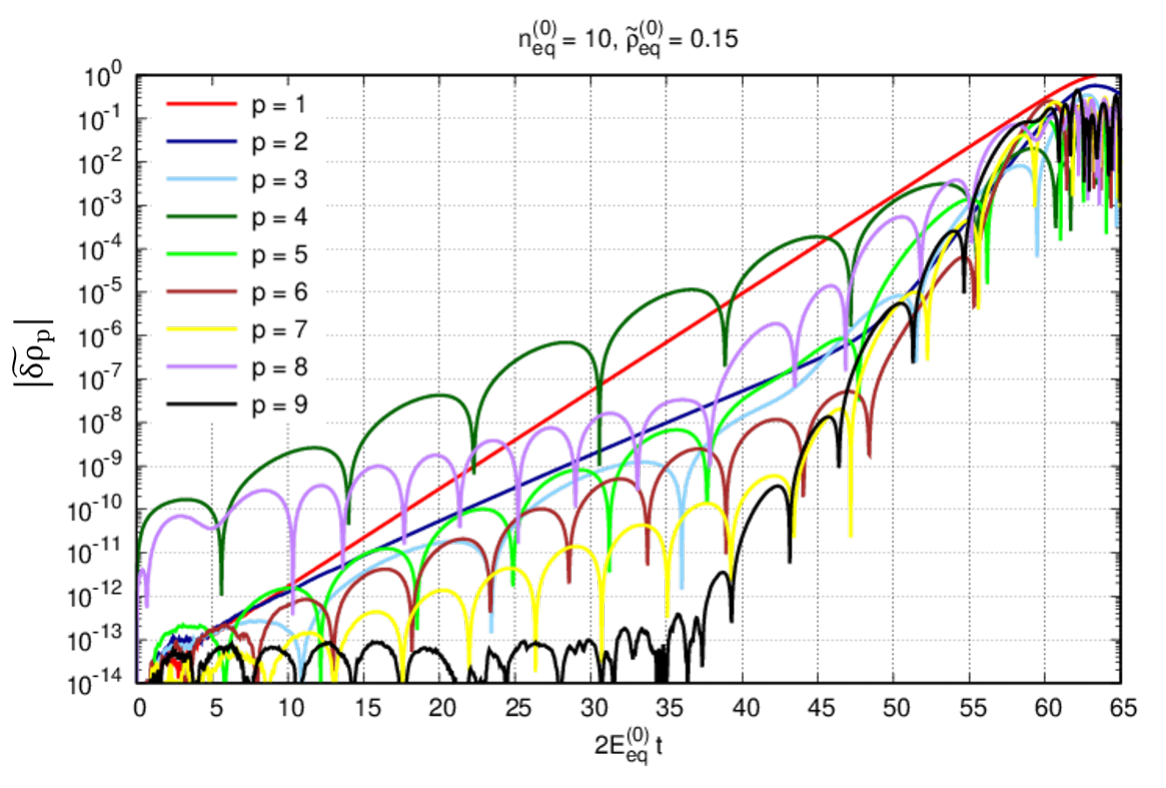}
\caption{\textit{An example of time evolution of perturbed density amplitudes for different cosine Fourier components, with $n^{(0)}_{eq}=10$, $\Tilde{\rho}^{(0)}_{eq} = 0.15$ and $p$ is the quasi-momentum of the perturbed density. The vertical axis shows the amplitude of $\widetilde{\delta \rho}_p$, defined as $\int^{2\pi}_{0} \Tilde{\Psi}_{eq}(x)\Tilde{R}(x)\cos{(px)}dx/\pi$. The system enters the nonlinear regime when $2E^{(0)}_{eq}t > 35$.  In the linear regime ($2E^{(0)}_{eq}t < 35$), three types of time-evolving behaviors are observed: (1) Jeans unstable modes ($\Tilde{\omega}^2<0$) for $p=1,2$, where the Fourier coefficients grow monotonically and exponentially; (2) Propagating unstable modes ($\Tilde{\omega}^2\in \mathbb{C} \backslash \mathbb{R}$) for $p=3,4,5,6,7,8$, where the Fourier coefficients oscillate while exhibiting exponential growth; and (3) Stable modes ($\Tilde{\omega}^2>0$) for $p=9$, where the Fourier coefficient oscillates without any exponential growth in amplitude.}} 
\label{fig: Density cos part evolution n 10 rho 0.15}
\end{figure}

Figure (\ref{fig: Density cos part evolution n 10 rho 0.15}) shows the time evolution of the normalized perturbed density with $n^{(0)}_{eq}=10$ and $\Tilde{\rho}^{(0)}_{eq} = 0.15$. Despite the eigenfunction is a mixed mode in the cosine basis, it nevertheless dominantly contains a particular cosine component. Hence, this time evolution plot can capture the most unstable eigen-mode.  Fig. (\ref{fig: Density cos part evolution n 10 rho 0.15}) also shows that the quantum system enters the non-linear regime when $2E^{(0)}_{eq}t > 35$. The following paragraph summarizes the time evolution for different quasi-momenta in the linear regime $2E^{(0)}_{eq}t < 35$.

There are three types of time-evolving behaviors in Fig. (\ref{fig: Density cos part evolution n 10 rho 0.15}), summarized as follows:
\\\\
\noindent(1) $p=1,2$: Perturbed density grows monotonically and exponentially. This behavior corresponds to a negative real eigenvalue $\Tilde{\omega}^2$ and we have identified it as the Jeans unstable mode.

\noindent(2) $p=3,4,5,6,7,8$:  Density oscillates at a certain frequency and has an exponentially growing envelope. This phenomenon indicates a complex eigenvalue $\Tilde{\omega}^2$ with a non-zero imaginary part, and it corresponds to the propagating unstable mode.

\noindent(3) $p=9$: Density oscillates without an increasing envelope, corresponding to a stable mode with a positive real eigenvalue $\Tilde{\omega}^2$.
\\\\

Figure (\ref{fig: Density cos part evolution n 10 rho 0.15}) provides evidence via simulations for the existence of propagating unstable modes, the main focus of this paper. The different time evolution of the perturbed density offer a useful tool to identify the stability type for different $\Tilde{\rho}^{(0)}_{eq}$ and $n^{(0)}_{eq}$. This identification allows for direct comparisons with theoretical predictions.

\subsection{Theoretical Predictions}
\label{subsec: Theoretical Predictions}

This section presents theoretical results, excluding the cases $q=0$ and $q=n^{(0)}_{eq}$. This exclusion is based on the smoothness of the eigenvalue $\Tilde{\omega}^2$ with respect to $q/n^{(0)}_{eq}$, except for the subspace $\mathbb{T}^{(cos)}_{q}[0,2\pi]$ at the boundary $q=n^{(0)}_{eq}$ under the condition $\Tilde{k}^2_{\perp}=0$, even though the corresponding matrix representation using the standing wave basis is not continuous at $q/n^{(0)}_{eq}=0$ and $1$. Verification of this continuity can be found in Appendix \ref{appendix: The General Property of the Matrix Representation}.

At the boundary $q=n^{(0)}_{eq}$, all eigenvalues in the subspace $\mathbb{T}^{(cos)}_{q}[0,2\pi]$ are real numbers. To see this, note that the background wave function $\Tilde{\Psi}_{eq}(x) \in \mathbb{T}^{(cos)}_{q=n^{(0)}_{eq}}[0,2\pi]$, and the nodes of $\Tilde{\Psi}_{eq}(x)$ coincide with those of any function within this subspace. Based on the Sturm comparison theorem \cite{AlGwaiz2008}, the Hamiltonian operator restricted to this subspace $\mathbb{T}^{(cos)}_{q=n^{(0)}_{eq}}[0,2\pi]$ has a positive spectrum. Therefore, the necessary condition for the existence of $\Tilde{\omega}^2 \in  \mathbb{C} \backslash \mathbb{R}$ is violated (See Sec. (\ref{sec: General Properties of the linearized Schrodinger-Poisson equation})), and it is not worthwhile to discuss this subspace.

Subspaces with $q=1,...,n^{(0)}_{eq}-1$ exhibit the following advantageous property:
\\\\
\noindent \textbf{(\rmnum{1}): For fixed $\bm{q=1,..,n^{(0)}_{eq}-1}$, both subspaces $\bm{\mathbb{T}^{(cos)}_{q}[0,2\pi]}$ and $\bm{\mathbb{T}^{(sin)}_{q}[0,2\pi]}$ share the same eigenvalue.}

\textit{The key idea is to demonstrate the identity of the matrix representation in both subspaces. A more detailed proof can be found in Appendix \ref{appendix: The General Property of the Matrix Representation}.}
\\\\
Hence, the following theoretical results are derived from the subspace $\mathbb{T}^{(cos)}_{n^{(0)}_{eq}}[0,2\pi]$ for $q=1,...,n^{(0)}_{eq}-1$, without loss of generality.

The background configuration, represented as a Taylor series with respect to $\Tilde{\rho}^{(0)}_{eq}$ (see Eq. (\ref{equ: Tayler series of the eigenenergy, wave function and gravitional potential})), must be truncated to a specific degree of $\Tilde{\rho}^{(0)}_{eq}$ when evaluating the matrix element $\mathbf{F}_{ij}$. Based on the computational analysis, we retain terms up to $(\Tilde{\rho}^{(0)}_{eq})^4$ for the background wave function $\Tilde{\Psi}_{eq}$ and up to $(\Tilde{\rho}^{(0)}_{eq})^5$ for the gravitational potential $\Tilde{V}_{eq}$ and eigenenergy $\Tilde{E}_{eq}$, ensuring that the eigenvalue $\Tilde{\omega}^2$ converges within approximately $1\%$ of the value obtained with higher degrees of truncation when $\Tilde{\rho}^{(0)}_{eq} \leq 0.36 \Tilde{\rho}^{(c)}_{eq}$. The difference in the truncation order between $\Tilde{\Psi}_{eq}$ and $\Tilde{V}_{eq}$ arises from the fact that $\Tilde{\Psi}_{eq}$ appears only in the perturbed self-gravity term, which already includes one degree of $\Tilde{\rho}^{(0)}_{eq}$ (see Eq. (\ref{equ: definition of H and Theta with normalization factor})).

Following Sec. (\ref{subsubsec: Standing Wave Basis Expansion}), we set the integer $N=4$, indicating that the stability is governed by the couplings among standing wave bases with wave numbers in the first, second, third, and fourth Brillouin zones. This value is chosen to ensure that $|\mathbf{F}_{ij}/\mathbf{F}_{jj}|$ and $|\mathbf{F}_{ji}/\mathbf{F}_{jj}| \lesssim 0.02$ for $|j|>N$ and for all $i$, a bound based on  $|\mathbf{F}_{ij}/\mathbf{F}_{jj}|$ and $|\mathbf{F}_{ji}/\mathbf{F}_{jj}| \lesssim 1/\Tilde{k}^2_{j} = (2j+q/n^{(0)}_{eq})^{-2} \geq (2|j|+1)^{-2}$. Notably, there are 7 standing wave bases whose wave numbers satisfy this condition. Consequently, the eigenvalue is determined by a $7 \times 7$ matrix.  

\begin{figure}
\includegraphics[scale=0.345, angle=270]{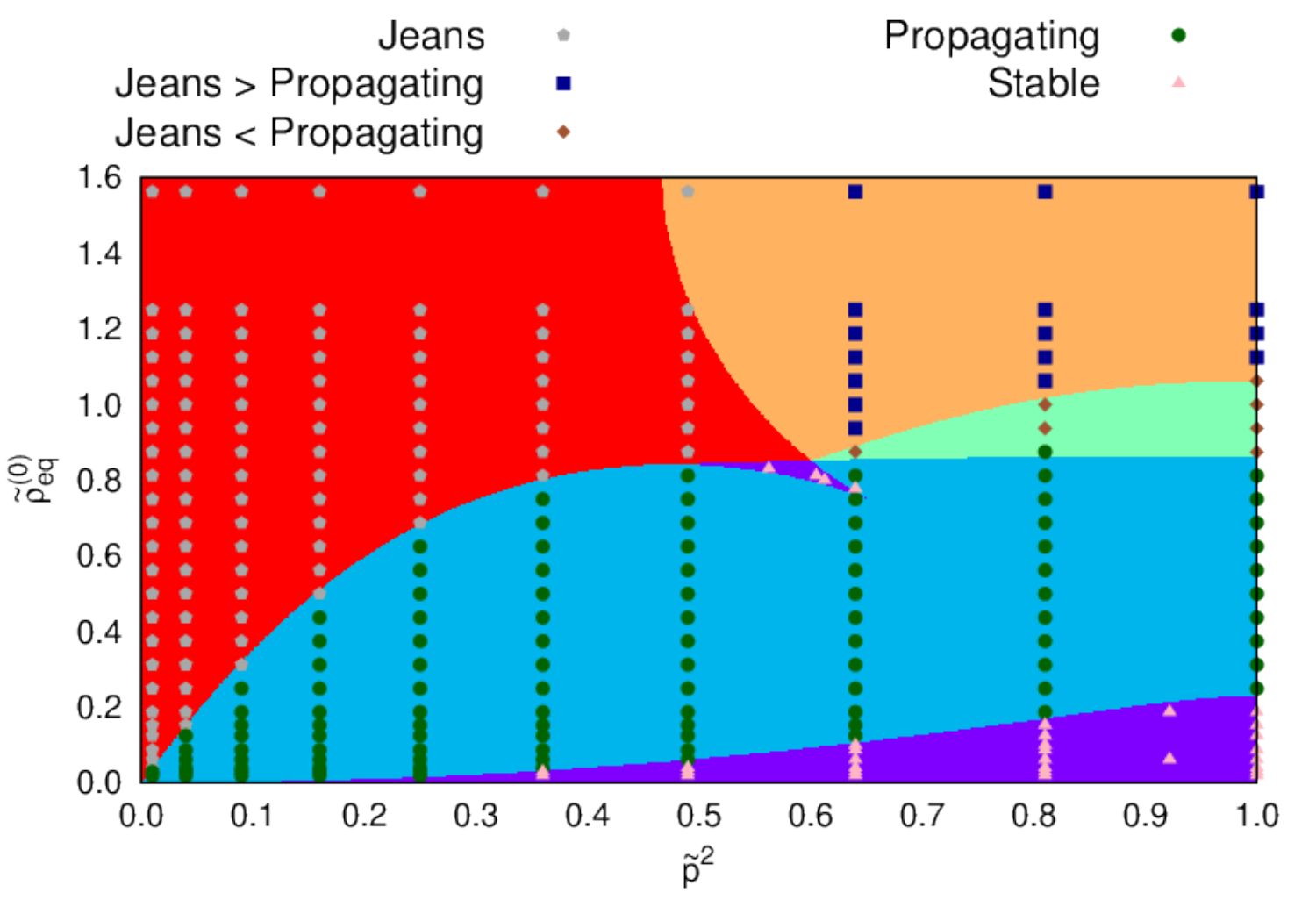}
\caption{\textit{Stability phase diagram for the case $\Tilde{k}_{\perp}=0$. The horizontal axis represents $\Tilde{p}^2$, where $\Tilde{p} = 1 - q/n^{(0)}_{eq}$ is the normalized quasi-momentum for the perturbed density, and $q$ is the quasi-momentum for the perturbed wave function. The vertical axis represents $\Tilde{\rho}^{(0)}_{eq}$. Different colored regions indicate various types of stability determined by theoretical calculations, and are summarized as follows: (1) Stable region (purple): All $\Tilde{\omega}^2$ are positive; (2) Propagating unstable region (blue): Instability is characterized by the complex $\Tilde{\omega}^2$ with non-zero imaginary parts; (3) Jeans unstable region (red): Instability arises from the negative real $\Tilde{\omega}^2$; (4) Coexistence of Jeans and propagating instabilities (green): Both complex $\Tilde{\omega}^2$ with non-zero imaginary parts and negative real $\Tilde{\omega}^2$ are present, with the growth rate of the propagating instability ($\Tilde{\omega}^2 \in \mathbb{C} \backslash \mathbb{R}$) being larger than that of the Jeans instability ($\Tilde{\omega}^2 < 0$); and (5) Similar coexistence region (yellow): Both complex and negative $\Tilde{\omega}^2$ are present, but the growth rate of the Jeans instability is larger than that of the propagating instability. All points in the diagram represent simulation results. Different types of points correspond to different stability behaviors, identified by the time evolution of the Fourier modes of the perturbed density, similar to those shown in Fig. (\ref{fig: Density cos part evolution n 10 rho 0.15}). The simulation results agree very well with theoretical predictions, even at the boundaries between different stability types.}} 
\label{fig: Phase diagram for zero k_perp}
\end{figure}

Figure (\ref{fig: Phase diagram for zero k_perp}) shows the stability comparison between theoretical calculations and simulation results with $\Tilde{k}^2_{\perp}=0$. The $(\Tilde{p}^2, \Tilde{\rho}^{(0)}_{eq})$ phase diagram is introduced to illustrate stability, where $\Tilde{p} = 1 - q/n^{(0)}_{eq}$ is the normalized quasi-momentum for the perturbed density. The use of $\Tilde{p}^2$ is based on the even symmetry of the background configuration.

The $(\Tilde{p}^2, \Tilde{\rho}^{(0)}_{eq})$-phase diagram is divided into five regions, labeled by different colors in Fig. (\ref{fig: Phase diagram for zero k_perp}). This division is based on theoretical eigenvalues obtained from the $7 \times 7$ matrix mentioned earlier. Additionally, it is found that $\mathbf{F}_{jj} = \langle u_j | \hat{\mathbf{\Theta}}\hat{\mathbf{H}} | u_j \rangle$ is always positive and can approximate the eigenvalue when the wave number of the standing wave basis $u_j$ is located in the third or fourth Brillouin zones where $\mathbf{F}_{jj}$ is significantly larger than $\mathbf{F}_{ij}$ and $\mathbf{F}_{ji}$ for $|j|=2$ and $3$. Consequently, there are already four positive eigenvalues for this $7 \times 7$ matrix, and the stability is determined by the remaining three eigenvalues.

The different color regions are summarized as follows:
\\\\
\noindent1. Purple region: All three remaining eigenvalues are positive, indicating that this region is stable. 
\\
\noindent2. Blue region: One of the three remaining eigenvalues is positive, and the other two are complex conjugates of each other with non-zero imaginary parts. This region represents propagating instability.
\\
\noindent3. Red region: All three remaining eigenvalues are real, with some being negative, indicating Jeans instability.
\\
\noindent4. Green region: One of the three remaining eigenvalues is negative, and the other two are complex conjugates with non-zero imaginary parts. This region indicates the coexistence of Jeans and propagating instabilities, with the growth rate for propagating instability being larger than that for Jeans instability.
\\
\noindent5. Yellow region: Similar to the green region, but the growth rate for propagating instability is smaller than that for Jeans instability.
\\\\

All points in Fig. (\ref{fig: Phase diagram for zero k_perp}) represent simulation results, with different symbol points indicating different stability types. These stability types are identified by the time-evolving behavior of the Fourier coefficients of the perturbed density, as demonstrated in Fig. (\ref{fig: Density cos part evolution n 10 rho 0.15}). The types of points are summarized as follows:
\\\\
\noindent1. triangle points: Stable mode.
\\
\noindent2. circle points: Propagating unstable mode.
\\
\noindent3. pentagon points: Jeans unstable mode.
\\
\noindent4. diamond points: Coexisting propagating and Jeans instability mode, with the growth rate for propagating instability being larger than that for Jeans instability.
\\
\noindent5. square points: Coexisting propagating and Jeans instability mode, with the growth rate for propagating instability being smaller than that for Jeans instability.
\\\\
It should be noted that points of coexisting Jeans and propagating instabilities can be further identified by initiating the appropriate perturbed wave function.

Figure (\ref{fig: Phase diagram for zero k_perp}) demonstrates that the simulation results are closely aligned with the theoretical analyses, which in particular predict the existence of propagating instability. Specifically, the propagating instability tends to occur near $p=n^{(0)}_{eq}$ (i.e., when $q=0$), whereas the Jeans instability is predominant near $p=0$ (i.e., when $q=n^{(0)}_{eq}$). This observation is expected because, as $p$ approaches zero, the density perturbation exhibits long-wavelength Fourier modes that can exceed the Jeans wavelength, leading to the Jeans instability.

Shown in Sec. (\ref{sec: General Properties of the linearized Schrodinger-Poisson equation}), when $\eta$ is the eigenfunction of the operator $\hat{\mathbf{\Theta}}\hat{\mathbf{H}}$ with $\omega^2 \in \mathbb{C} \backslash \mathbb{R}$, it must satisfy the condition $\langle \eta |\hat{\mathbf{H}}|\eta \rangle = 0$ . This requirement implies that $\eta$ must contain Fourier modes with wavelengths exceeding the period of the background wave function. Such a condition is more easily satisfied when the eigenfunction $\eta$ resides in a subspace where $q$ very close to zero. This explains why the propagating instability tends to occur near $q=0$.

Figure (\ref{fig: Phase diagram for zero k_perp}) indicates that Jeans instability dominates the region where $\Tilde{\rho}^{(0)}_{eq} > 1$. This result is expected because the parameter $\Tilde{\rho}^{(0)}_{eq}$ represents the square of the ratio between the quantum wavelength and the Jeans wavelength for the background wave function, as discussed in Sec. (\ref{sec: Nonlinear periodic equilibrium construction}). On the other hand, the domain $\Tilde{p}^2 < 1$ implies that the wavelength for the perturbed density is always greater than the quantum wavelength for the background wave function. Consequently, $\Tilde{\rho}^{(0)}_{eq} > 1$ means the wavelength for the perturbed density is also greater than the Jeans wavelength, leading to Jeans instability. 

The following two phenomena are observed in Fig. (\ref{fig: Phase diagram for zero k_perp}). First, Jeans and propagating instabilities coexist near the boundary at $\Tilde{p}=1$ when $\Tilde{\rho}^{(0)}_{eq}> 0.8$. Second, two separated stable regions are identified. The first stable region is located at $\Tilde{\rho}^{(0)}_{eq}<0.25$, which aligns with the expectation that the background gravitational potential diminishes as $\Tilde{\rho}^{(0)}_{eq}$ approaches zero, leading the wave function to behave similarly to that of a free particle. The second stable region emerges near $\Tilde{\rho}^{(0)}_{eq} \approx 0.8$ and $\Tilde{p}^2 \approx 0.55$. The presence of this stable region suggests that certain quasi-momenta remain stable under the pinning of a finite gravitational potential. 

\begin{figure*}
\centering \includegraphics[scale=0.96, angle=270]{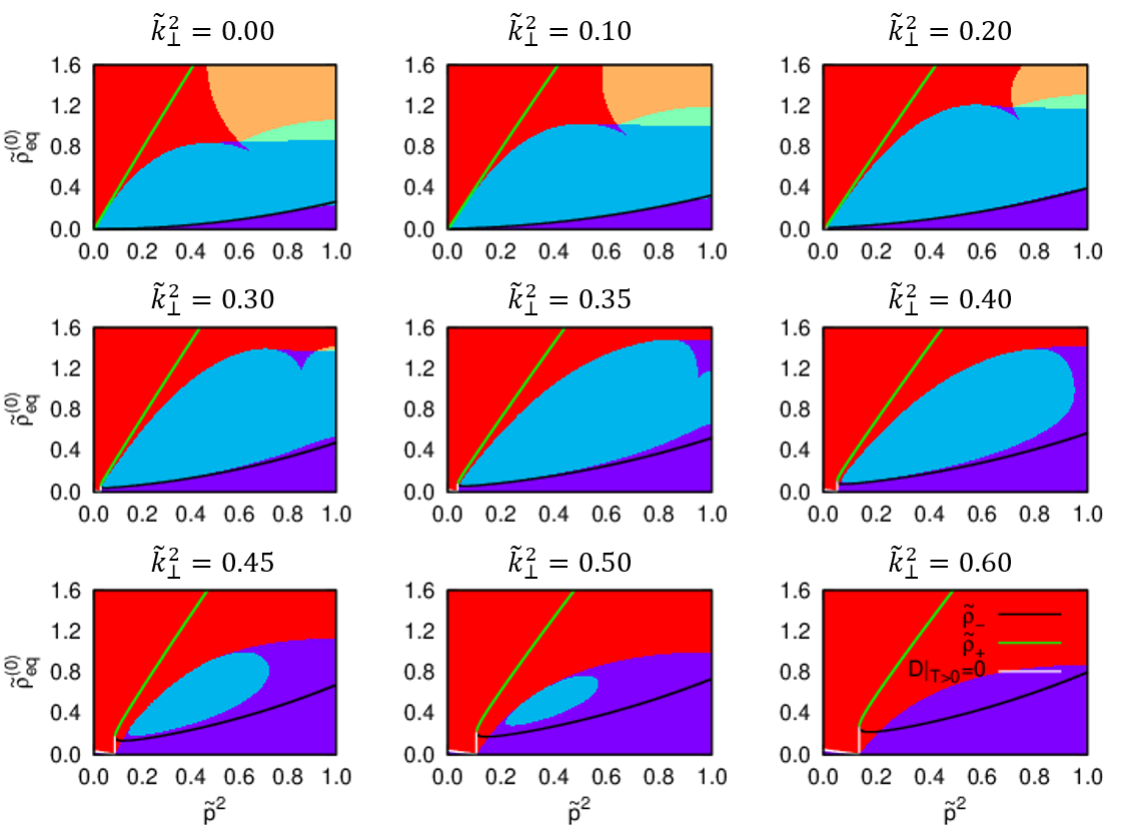}
\caption{\textit{Stability phase diagram for different values of the normalized perpendicular kinetic energy, $\Tilde{k}^2_{\perp}$ that shows the three-dimensional topology of the diagram. The figure has nine panels with $\Tilde{k}^2_{\perp}$ values of $0.00$, $0.10$, $0.20$, $0.30$, $0.35$, $0.40$, $0.45$, $0.50$, and $0.60$, arranged from left to right and top to bottom.  In each panel, the horizontal and vertical axes, as well as the colored stability regions, are identical to those shown in Fig. (\ref{fig: Phase diagram for zero k_perp}). As $\Tilde{k}^2_{\perp}$ increases up to $0.60$, three key changes are observed: (1) The propagating instability region becomes smaller and eventually disappears; (2) The region where Jeans and propagating instabilities coexist changes to a Jeans instability region; and (3) The stable regions, which are separate when $\Tilde{k}^2_{\perp} = 0$, connect along the boundary at $\Tilde{p} = 1$. For comparison with the two-component wave dark matter model, we also show the stability boundary for this case. The black and green curves in each panel delineate the boundaries between regions with real and non-real eigenvalues, respectively, for the two-component wave dark matter model, as given by $\Tilde{\rho}_{\pm}(\Tilde{p}^2, \Tilde{k}^2_{\perp})$ in Eq. (\ref{equ: phase diagram for two-component}). The white curve, labeled $D|_{T>0}=0$, marks the boundary between the Jeans unstable and stable regions for the two-component model. This curve represents the contour where $D(\Tilde{p}^2, \Tilde{k}^2_{\perp}, \Tilde{\rho}^{(0)}_{eq}) = 0$ with the condition $T(\Tilde{p}^2, \Tilde{k}^2_{\perp}, \Tilde{\rho}^{(0)}_{eq}) > 0$, with $D$ and $T$ are defined in Eq. (\ref{equ: the definition of trace and determinant for the two-component case}). The boundaries between different stability types can be approximated using the two-component wave dark matter model, but this approximation works best when $\Tilde{\rho}^{(0)}_{eq}$ is small. As $\Tilde{k}^2_{\perp}$ increases, the range of $\Tilde{\rho}^{(0)}_{eq}$ for which this approximation is valid decreases. For example, at $\Tilde{k}^2_{\perp} = 0.00$, the approximation is valid up to $\Tilde{\rho}^{(0)}_{eq} \approx 0.20$, but at $\Tilde{k}^2_{\perp} = 0.60$, it is only valid when $\Tilde{\rho}^{(0)}_{eq} < 0.05$.}}
\label{fig: Phase diagram for different k_perp}
\end{figure*}

These two separated stable regions in Fig. (\ref{fig: Phase diagram for zero k_perp}) connect through the other dimension $\Tilde{k}_{\perp}$, as revealed in Fig. (\ref{fig: Phase diagram for different k_perp}). Fig. (\ref{fig: Phase diagram for different k_perp}) consists of nine panels, each showing the stability phase diagram for a fixed value of the normalized perpendicular kinetic energy $\Tilde{k}^2_{\perp}$. The values of $\Tilde{k}^2_{\perp}$ for each panel are $0.00$, $0.10$, $0.20$, $0.30$, $0.35$, $0.40$, $0.45$, $0.50$, and $0.60$, arranged from left to right and top to bottom, respectively. The horizontal and vertical axes in each panel are the same as in Fig. (\ref{fig: Phase diagram for zero k_perp}). Different color regions indicate different stability types, consistent with those in Fig. (\ref{fig: Phase diagram for zero k_perp}). As $\Tilde{k}^2_{\perp}$ increases from $0.00$ to $0.60$, the upper stable region shifts to the boundary $\Tilde{p}=1$ and connects with the lower stable region.

As $\Tilde{k}^2_{\perp}$ increases to $0.60$, two phenomena are observed. First, the region exhibiting propagating instability shrinks and eventually vanishes. Second, the area where Jeans and propagating instabilities coexist is replaced by a region exhibiting only Jeans instability. The disappearance of propagating instability is due to the fact that a positive $\Tilde{k}^2_{\perp}$ makes the spectrum of the Hamiltonian operator more positive. Specifically, if $\Tilde{k}^2_{\perp} > \Tilde{E}_{eq} - \min_{x \in [0,2\pi]} \Tilde{V}_{eq}(x)$, where $\min_{x \in [0,2\pi]} \Tilde{V}_{eq}(x)$ is the minimum value of the background gravitational potential, the requirement is satisfied. Consequently, it is unsurprising that the region where Jeans and propagating instabilities previously coexisted is now replaced by a region exhibiting only Jeans instability.

Moreover, when $\Tilde{k}^2_{\perp}$ is sufficiently large, this quantum system will be stabilized, as stated in the following sentence:
\\\\
\noindent \textbf{(\rmnum{2}): $\bm{\Tilde{\omega}^2>0}$ when the perpendicular kinetic energy $\bm{\Tilde{k}^2_{\perp}}$ is sufficiently large.}

\textit{The key idea is to show that the expected value of $\hat{\mathbf{\Theta}}$ is always positive for any function once $\Tilde{k}^2_{\perp}$ is sufficiently large. Given any function $f(x) \in \mathbb{T}[0,2\pi]$, it can be expressed as $f(x) = \sum^{n^{(0)}_{eq}}_{q=0} \Big ( f^{(cos)}_q(x) + f^{(sin)}_q(x)\Big )$ with $f^{(cos)}_q(x) \in \mathbb{T}^{(cos)}_{q}[0,2\pi]$ and $f^{(sin)}_q(x) \in \mathbb{T}^{(sin)}_{q}[0,2\pi]$. The subspaces $\mathbb{T}^{(cos)}_{q}[0,2\pi]$ and $\mathbb{T}^{(sin)}_{q}[0,2\pi]$ are orthogonal to each other and are decoupled by the operator $\hat{\mathbf{\Theta}}$. Therefore, $\langle f|\hat{\mathbf{\Theta}} | f\rangle = \sum^{n^{(0)}_{eq}}_{q=0} \Big ( \langle f^{(cos)}_q |\hat{\mathbf{\Theta}} | f^{(cos)}_q \rangle + \langle f^{(sin)}_q |\hat{\mathbf{\Theta}} | f^{(sin)}_q \rangle \Big )$. The positive expected value of $\hat{\mathbf{\Theta}}$ for large perpendicular kinetic energy is achieved since the perturbed self-gravity energy has the upper bound in each subspace (as shown in Appendix \ref{appendix: Uniform Boundedness of the Perturbed Self-Gravity Energy Tensor in the Operator Theta}) and this upper bound is independent of the subspaces.}
\\\\ 

To delineate the boundaries between different stability regions, the analytical boundary formula derived from the two-component wave dark matter model, as presented in Appendix \ref{appendix: General Property for the dispersion relation of two-component wave dark matter}, is shown in Fig. (\ref{fig: Phase diagram for different k_perp}). In this model, two colliding plane waves arise from distinct components of wave dark matter and are coupled through the gravitational potential governed by the total mass density. As a result, there is no background gravitational potential, leading to spatial homogeneity in the two-component wave dark matter system. This spatial homogeneity enables the derivation of an analytical dispersion relation, with the corresponding stability behavior described by Eq. (\ref{equ: phase diagram for two-component}).

In Fig. (\ref{fig: Phase diagram for different k_perp}), the black and green curves in Fig. (\ref{fig: Phase diagram for different k_perp}) are $\Tilde{\rho}_{-}(\Tilde{p}^2,\Tilde{k}^2_{\perp})$ and $\Tilde{\rho}_{+}(\Tilde{p}^2,\Tilde{k}^2_{\perp})$ given by Eq. (\ref{equ: phase diagram for two-component}). These $\Tilde{\rho}_{\pm}$ curves are the boundaries between non-real and real eigenvalues for the two-component wave dark matter. Additionally, the white curve $D|_{T>0} = 0$ represents the contour for $D(\Tilde{p}^2,\Tilde{k}^2_{\perp}, \Tilde{\rho}^{(0)}_{eq}) = 0$ under the constraint $T(\Tilde{p}^2,\Tilde{k}^2_{\perp}, \Tilde{\rho}^{(0)}_{eq}) > 0$, where functions $D$ and $T$ are defined by Eq. (\ref{equ: the definition of trace and determinant for the two-component case}). The $D|_{T>0}=0$ curve serves as the boundary between the Jeans instability and stable regions for the two-component wave dark matter.

The boundaries between different stability regions can be further fitted by the two-component wave dark matter when $\Tilde{\rho}^{(0)}_{eq}$ is sufficiently small. This approximation is valid because, as $\Tilde{\rho}^{(0)}_{eq}$ approaches zero, the background configuration for this periodic equilibrium background approaches that of the two-component case, where the background gravitational potential vanishes. However, the range for which this approximation holds decreases as $\Tilde{k}^2_{\perp}$ increases. For instance, when $\Tilde{k}^2_{\perp} = 0.00$, the boundary can be fitted by the two-component case up to $\Tilde{\rho}^{(0)}_{eq} \approx 0.20$. However, when $\Tilde{k}^2_{\perp} = 0.60$, the boundary can only be captured by the two-component case when $\Tilde{\rho}^{(0)}_{eq} < 0.05$.     

There is a peculiar feature that Jeans instability occurs at a non-zero quasi-momentum for the perturbed density with $\Tilde{k}^2_{\perp} \neq 0$ when $\Tilde{\rho}^{(0)}_{eq}$ approaches zero. This non-zero quasi-momentum approximately follows the relation $(1-\Tilde{p})^2+\Tilde{k}^2_{\perp} \approx 1$. For example, when $\Tilde{k}^2_{\perp} = 0.60$, Jeans instability occurs near $\Tilde{p} \approx 0.37$ ($\Tilde{p}^2 \approx 0.14$) for $\Tilde{\rho}^{(0)}_{eq} < 0.05$, as shown in Fig. (\ref{fig: Phase diagram for different k_perp}). This feature can be explained by the two-component wave dark matter scenario. As detailed in Appendix \ref{appendix: General Property for the dispersion relation of two-component wave dark matter}, when $(1-\Tilde{p})^2 + \Tilde{k}^2_{\perp}=1$, the corresponding perturbed wave function can establish another equilibrium background, i.e., $\Tilde{\omega}^2=0$. The new background wave function consists of two wave-vectors for the leading order of $\Tilde{\rho}^{(0)}_{eq}$, one is $n^{(0)}_{eq}\hat{x}$ and the other is $q\hat{x}+\vec{k}_{\perp}$, where $q = n^{(0)}_{eq} - p$ with $\Tilde{p} = p/n^{(0)}_{eq}$ satisfying the above relation. Consequently, any additional perturbed wave function having the Fourier mode with the wave-vector near $q\hat{x}+\vec{k}_{\perp}$ produces perturbed density with large-scale fluctuations, which favor Jeans instability.   

\section{Conclusion}
\label{sec: Conclusion}

The presence of fine-scale density fringes resulting from colliding cosmic large-scale flows in the early Universe may result in fine-scale collapsed objects seeding the first generation proto-globular clusters when the fringes are unstable.  Motivated by this possibility, this paper investigates the linear stability of wave dark matter governed by the Schr\"{o}dinger-Poisson equation. We have characterized the stability of fine-scale fringes with a phase diagram of the squared wave number $\Tilde p^2$ versus the fringe density parameter $\Tilde{\rho}^{(0)}_{eq}$. The phase diagram marks the boundaries of three type of density perturbations, viz, stable, Jeans unstable and propagating unstable. 

Most importantly, we have theoretically found that the Sturm-Liouville properties of quantum systems can be disrupted in some specific situation, specifically when the background wave function has local nulls. A stationary periodic background density, which mimic interference fringes embedded in otherwise large-scale sheets in the cosmic web, always has local nulls. Occurrence of the propagating instability, characterized by a complex eigen-frequency with a non-zero imaginary part, in the quantum interference system is a consequence of the failure of the Sturm-Liouville theory that renders a real eigenvalue $\omega^2$.  

In addition to the propagating instability, we have also found the Jeans instability with a pure imaginary frequency and the stable oscillation with a real frequency, which are within the valid domain of Sturm-Liouville theory. When the background fringe density parameter $\Tilde{\rho}^{(0)}_{eq} \ll 1$, the system tends to be stable. When $\Tilde{\rho}^{(0)}_{eq}$ is moderate, the system becomes unstable. Among these unstable modes, long-wavelength density perturbations are Jeans unstable and short-wavelength density perturbations propagating unstable.  These small-scale unstable density modes are the focus of this work and expected to result in a population of small scale collapsed proto-globular clusters.

Mathematically, the system stability is governed by an eigenvalue problem involving a non-trivial product of two Hermitian operators, $\hat{\mathbf{H}}$, the Hamiltonian of the Schr\"{o}dinger equation and $\hat{\mathbf{G}}$, the self-gravity operator arising from the perturbed potential. The two operators do not commute, $\hat{\mathbf{G}}\hat{\mathbf{H}}\neq \hat{\mathbf{H}}\hat{\mathbf{G}}$, meaning that the product is \textit{not} Hermitian. Such a non-Hermitian nature subjects the system with complex eigenvalues. Despite that, we may strive to make the system almost Sturm-Liouville-like by defining a new inner product described in Sec. (\ref{sec: General Properties of the linearized Schrodinger-Poisson equation}). However, when the new inner product vanishes, the Sturm-Liouville theory will still fail and complex eigenvalues may result.

To connect local density nulls to propagating instabilities, it invokes some arguments. First, the necessary condition for the failure of the Sturm-Liouville theory is a vanishing new inner product, which turns out to be identical to the eigen-expectation value of the Hamiltonian $\langle \hat{\mathbf{H}} \rangle$. Since the eigen-function of this perturbed system is not the same as the eigen-function of $\hat{\mathbf{H}}$, when $\langle \hat{\mathbf{H}} \rangle=E_{eq}\equiv 0$ it implies that the eigen-energy $E_{eq}$ of the background wave function is not the minimum-eigenvalue of $\hat{\mathbf{H}}$. That is, the background wave function must be an \textit{excited state} of $\hat{\mathbf{H}}$, which very often has local nulls, and we suspect that no excited state can exist with no local nulls. Therefore, local nulls in the interference fringe serves as an indicator for the propagating instability.

In this paper, we conduct a set of wave dynamics simulations to verify the theoretical analyses. All three types of perturbations are identified in the simulations, and they agree well with theoretical predictions, as exemplified by the excellent agreement of the phase boundary in the phase diagram. 

We have also conducted a comparative study of a two-component case, where two independent wave functions of the same particle mass counter stream against each other. The detailed analysis is given in Appendix \ref{appendix: General Property for the dispersion relation of two-component wave dark matter}. We find that the two-component case also has the three types of stable and unstable modes. Moreover, the two cases agree well with each other when the density parameter $\Tilde{\rho}^{(0)}_{eq} \ll 1$.

Moderately nonlinear background density $\Tilde{\rho}^{(0)}_{eq}$, for which the large-scale attractive gravitational force is balanced by the repulsive force given by the cosmological constant, has been conducted for perturbation analysis. We have nevertheless explored the background fringe density fully even in highly nonlinear configurations. We note that in the extreme case, the interference fringe appears as a delta function density pulse train and the density null is located at the midpoint of two adjacent pulses. Looking into each individual pulse, the wave function is like an isolated soliton, which connects to two adjacent anti-solitons.  The soliton is the ground state and should be locally stable in its own right. Instability may nevertheless occur when the adjacent soliton-anti-soliton pair mutually interact. This situation resembles large-scale cosmic sheets sandwiching a vacuum and can be interesting for exploration. Since these solitons are separated at a great distance, we suspect the instability should begin to take off over a very long time scale. 

The theoretical framework developed in this paper can be applied to quantum systems with local interaction potentials, e.g., the ones described by Gross-Petaevskii Equation.  For such systems, the Jeans instability will be absent, but the propagating instability may still survive if the background density has local nulls.  

Returning to Astrophysics, when two halos collide, there will also be interference fringes. Unlike those fringes in the cosmic web that are superposed on a quiescent matter, the ambient background of such interference patterns is the granular, or hot, halo matter. The theoretical framework developed here cannot be extended to such a case, and one has to model the halo granules with a statistical approach. The Wigner function (quantum distribution function) can be adopted for this purpose \cite{B-OFT2021}. In a future work, we will explore such a colliding system.

\section*{Acknowledgements}

This work is supported in part by NSTC of Taiwan under Grant No. NSTC 112-2811-M-002-084 and NSTC 112-2112-M-002-039 

\onecolumngrid

\appendix
\renewcommand{\thesection}{\Alph{section}}
\setcounter{section}{0}

\renewcommand{\theequation}{A\arabic{equation}}
\setcounter{equation}{0}
\section{Key Features of the Background Configuration}
\label{appendix: Properties for the background configuration}

In this appendix, we highlight three significant features of the background configuration introduced in Sec. (\ref{sec: Nonlinear periodic equilibrium construction}) are displayed. These features include: (\textbf\textit{{1}}) Fourier modes of the background wave function and potential, (\textbf\textit{{2}}) non-zero radius of convergence with respect to the parameter $\Tilde{\rho}^{(0)}_{eq}$, and (\textbf\textit{{3}}) number of nodes for all nonlinear background wave functions. 
\\\\
\textbf{\textit{(1): Fourier Modes of the Background Wave Function and Potential}}
\\

The Fourier modes for the background wave function and potential are given by the following proposition:
\\\\

\begin{proposition}
\label{Pro: each order of Psi and V}
Let the function sequences $\Big\{\Tilde{\Psi}_n(x) \Big \}^{\infty}_{n=0}$, $\Big \{\Tilde{V}_n(x) \Big \}^{\infty}_{n=1}$, and the sequence $\Big \{ \Tilde{E}_n\Big \}^{\infty}_{n=1}$ satisfy Eq. (\ref{equ: background configuration for each order}) with $\Tilde{\Lambda}_{n-1} = \int^{2\pi}_{0} \sum^{n}_{j=1}\Tilde{\Psi}_{n-j}(y)\Tilde{\Psi}_{j-1}(y)dy/(2\pi)$, subject to the following conditions: $\Tilde{\Psi}_0(x) = \cos{\Big ( n^{(0)}_{eq}x \Big )}$; $\Tilde{\Psi}_n(x+2\pi) = \Tilde{\Psi}_n(x)$; $\Tilde{V}_n(x+2\pi) = \Tilde{V}_n(x)$; and $\int^{2\pi}_{0}\Tilde{V}_n(x)dx =0$ for all $n \in \mathbb{N}$. Then $\Big\{\Tilde{\Psi}_n(x) \Big \}^{\infty}_{n=0}$, $\Big \{\Tilde{V}_n(x) \Big \}^{\infty}_{n=1}$, and $\Big \{ \Tilde{E}_n\Big \}^{\infty}_{n=1}$ satisfy Eq. (\ref{equ: Fourier series of each order of background wave function and gravitational potential}).    
\end{proposition}

\renewcommand\qedsymbol{QED}
\begin{proof}
The validity of this statement is established through mathematical induction. First, it is straightforward to verify that this statement holds true for $n=1$. Next, assume that this statement is valid for all $n \leq m$ for some $m \in \mathbb{N}$. For $n = m+1$, we begin by noting that $\Tilde{\Psi}_n$ is orthogonal to $\Tilde{\Psi}_0$ for all $n=1,2,...,m$. By multiplying $\Tilde{\Psi}_0$ on both sides of the first equation in Eq. (\ref{equ: background configuration for each order}) for $n=m+1$ and integrating over the spatial domain $[0,2\pi]$, we obtain $\Tilde{E}_{m+1} = \sum^{m+1}_{j=1} \int^{2\pi}_{0}  \Tilde{\Psi}_{(m+1)-j}(x)\Tilde{V}_j(x) \Tilde{\Psi}_0(x) dx/\int^{2\pi}_{0}\Tilde{\Psi}^2_0(x)dx$.

Next, a straightforward calculation yields the following result: 

\begin{equation}
\label{equ:appendix: Fourier series for each order of the density}
\begin{aligned}
\sum^{m+1}_{j=1} \Tilde{\Psi}_{(m+1)-j}(x)\Tilde{\Psi}_{j-1}(x) = & \sum^{m}_{l=1} \Tilde{\psi}_{m,l}\Big \{ cos\Big [ (2l)n^{(0)}_{eq}x\Big ] + \cos{\Big [ 2(l+1)n^{(0)}_{eq}x\Big ]}  \Big \} + \\
& {{1}\over{2}}\sum^{m}_{j=2} \sum^{(m+1)-j}_{l_1=1}\sum^{j-1}_{l_2=1} \Tilde{\psi}_{(m+1)-j,l_1}\Tilde{\psi}_{j-1,l_2} \Big \{ \cos{\Big [ 2(l_1-l_2)n^{(0)}_{eq}x\Big ]} + \cos{\Big [ 2(l_1+l_2+1)n^{(0)}_{eq}x\Big ]} \Big \}.
\end{aligned}
\end{equation}
The right-hand side of Eq. (\ref{equ:appendix: Fourier series for each order of the density}) consists of cosine functions with wave numbers that are even integer multiples of $n^{(0)}_{eq}$. Furthermore, this even integer is at most $2(m+1)$, as determined by the constraints $l \leq m$, $l_1+l_2+1 \leq [(m+1)-j] +(j-1)+1 = m+1$, and $|l_1 - l_2| \leq  m-2$ due to $1 \leq l_1 \leq (m+1)-j$, $1 \leq l_2 \leq j-1$, and $2 \leq j \leq m$. Additionally, $\Tilde{\Lambda}_{m}$ serves as the Fourier coefficient for the zero wave number mode in Eq. (\ref{equ:appendix: Fourier series for each order of the density}). Consequently, the second equation in Eq. (\ref{equ: background configuration for each order}) implies that $\Tilde{V}_{m+1}(x)=\sum^{m+1}_{l=1}\Tilde{v}_{(m+1),l}\cos{[ (2l)n^{(0)}_{eq}x ]}$ with real constants $\Tilde{v}_{(m+1),l}$, as the Laplacian operator cannot generate new Fourier modes.

Finally, a straightforward calculation shows that the right-hand side of the first equation in Eq. (\ref{equ: background configuration for each order}) becomes:

\begin{equation}
\label{equ:appendix: Fourier series of the Schordinger equation}
\begin{aligned}
\sum^{m+1}_{j=1}\Tilde{\Psi}_{(m+1)-j}(x)\Big [ \Tilde{V}_j(x) - \Tilde{E}_j  \Big ] = & -\Tilde{E}_{m+1}\cos{[n^{(0)}_{eq}x]} + \\
& {{1}\over{2}}\sum^{m+1}_{l=1}\Tilde{v}_{(m+1),l}\Big \{ \cos{\Big [(2l-1)n^{(0)}_{eq}x\Big ]} + \cos{\Big [(2l+1)n^{(0)}_{eq}x \Big ]} \Big \} - \\
& \sum^{m}_{j=1}\Tilde{E}_j\sum^{(m+1)-j}_{l=1}\Tilde{\psi}_{(m+1)-j,l}\cos{\Big [ (2l+1)n^{(0)}_{eq} x\Big ]} + \\
& {{1}\over{2}}\sum^{m}_{j=1}\sum^{(m+1)-j}_{l_1=1}\sum^{j}_{l_2=1}\Tilde{\psi}_{(m+1)-j,l_1}\Tilde{v}_{j,l} \Big \{ \cos{ \Big [ (2(l_1+l_2) + 1)n^{(0)}_{eq}x\Big ]} + \cos{\Big [ (2(l_1-l_2) + 1)n^{(0)}_{eq}x\Big ]}\Big \}.
\end{aligned}
\end{equation}
Similar to the gravitational potential case, any wave number appearing in Eq. (\ref{equ:appendix: Fourier series of the Schordinger equation}) is an odd integer multiple of $n^{(0)}_{eq}$ and at most $(2(m+1)+1)n^{(0)}_{eq}$. Furthermore, Eq. (\ref{equ:appendix: Fourier series of the Schordinger equation}) cannot contain the Fourier mode with wave number $n^{(0)}_{eq}$, as it is orthogonal to $\Tilde{\Psi}_0=\cos{(n^{(0)}_{eq}x)}$. This orthogonality can be demonstrated by multiplying $\Tilde{\Psi}_0$ on both sides of the first equation in Eq. (\ref{equ: background configuration for each order}) and then integrating over the spatial domain $[0,2\pi]$. Similarly to the gravitational potential case, $\Tilde{\Psi}_{m+1}=\sum^{m+1}_{l=1} \Tilde{\psi}_{(m+1),l} \cos{\Big [ (2l+1)n^{(0)}_{eq}x\Big ]}$ for some real constants $\Tilde{\psi}_{(m+1),l}$. Thus, the statement is valid for $n = m+1$, and the conclusion is established by mathematical induction. 
\end{proof}

\noindent\textbf{\textit{(2): Non-zero Radius of Convergence with Respect to the Parameter $\bm{\Tilde{\rho}^{(0)}_{eq}}$}}
\\

Equation (\ref{equ: Tayler series of the eigenenergy, wave function and gravitional potential}) reveals that the background configuration is expressed as a Taylor series with respect to the parameter $\Tilde{\rho}^{(0)}_{eq}$. To make such an expression meaningful, it is necessary to show that these Taylor series have a non-zero radius of convergence. Before starting the proof, we introduce the $L^2$-norm of a periodic function $f$ on $[0,2\pi]$, denoted as $||f||_{L^2} \equiv \sqrt{ \int^{2\pi}_{0} |f(x)|^2dx /(2\pi)}$, to simplify the notation.

To demonstrate the non-zero radius of convergence, it is necessary to estimate the upper bound of each order of the background eigenenergy $\Tilde{E}_n$, wave function $\Tilde{\Psi}_n$, and gravitational potential $\Tilde{V}_n$. First, in accordance with Eq. (\ref{equ: Fourier series of each order of background wave function and gravitational potential}), $\Tilde{E}_n$ can be bounded by:

\begin{equation}
\label{equ: the bounded of the n-th order eigenergy}
\begin{aligned}
|\Tilde{E}_n| = {{|\sum^{n}_{j=1} \int^{2\pi}_{0}  \Tilde{\Psi}_{n-j}(x)\Tilde{V}_j(x) \Tilde{\Psi}_0(x) dx|}\over{\int^{2\pi}_{0} \Tilde{\Psi}^2_0(x) dx }} 
              & \leq {{\sum^{n}_{j=1} \int^{2\pi}_{0} |\Tilde{\Psi}_{n-j}(x)||\Tilde{V}_j(x)||\Tilde{\Psi}_0(x)| dx}\over{\int^{2\pi}_{0} \Tilde{\Psi}^2_0(x) dx}} \\
              & \leq {{\max_{x\in[0,2\pi]}|\Tilde{\Psi}_0(x)|}\over{||\Tilde{\Psi}_{0}||_{L^2}}} {{\sum^{n}_{j=1} ||\Tilde{\Psi}_{n-j}||_{L^2} ||\Tilde{V}_j||_{L^2}}\over{||\Tilde{\Psi}_{0}||_{L^2}}}.
\end{aligned}
\end{equation}
Here, the first inequality is based on the triangle inequality. The second inequality stems from the fact that $\max_{x\in[0,2\pi]}|\Tilde{\Psi}_0(x)|$ is the maximum value of the zeroth order of the wave function $\Tilde{\Psi}_0(x)$ over the spatial domain $[0,2\pi]$ and H\"{o}lder's inequality \cite{Rudin1986}. Physically, the quantity $\max_{x\in[0,2\pi]}|\Tilde{\Psi}_0(x)|/||\Tilde{\Psi}_{0}||_{L^2}$ represents the square root of the overdensity. Eq. (\ref{equ: the bounded of the n-th order eigenergy}) indicates that the upper bound for $\Tilde{E}_n$ is related to $||\Tilde{V}_j||_{L^2}$ for $j \leq n$ and $||\Tilde{\Psi}_{j}||_{L^2}$ for $j < n$.

The upper bound for the $L^2$-norm of the background gravitational potential can be derived using the Poisson equation. Recall that each order of the background gravitational potential has a Fourier series given by Eq. (\ref{equ: Fourier series of each order of background wave function and gravitational potential}). Therefore, in combination with Eq. (\ref{equ: background configuration for each order}), the Fourier coefficient $\Tilde{v}_{n,l}$ is given by $-4\int^{2\pi}_{0}\cos{\Big ( (2l) n^{(0)}_{eq}x \Big )}\sum^{n}_{j=1}\Tilde{\Psi}_{n-j}(x)\Tilde{\Psi}_{j-1}(x)dx/(2 \pi l^2)$. Thus, $||\Tilde{V}_n||_{L^2}$ can be bounded by:

\begin{equation}
\label{equ: the bounded of the n-th order gravitational potential}
\begin{aligned}
||\Tilde{V}_n||_{L^2} = \sqrt{{{\sum^{n}_{l=1}|\Tilde{v}_{n,l}|^2}\over{2}}} & \leq \sqrt{\sum^{n}_{l=1}{{8}\over{l^4}}}\sum^{n}_{j=1}{{1}\over{2\pi}}\int^{2\pi}_{0}|\Tilde{\Psi}_{n-j}(x)||\Tilde{\Psi}_{j-1}(x)|dx \\
& \leq \sqrt{\sum^{\infty}_{l=1}{{8}\over{l^4}}}\sum^{n}_{j=1} ||\Tilde{\Psi}_{n-j}||_{L^2} ||\Tilde{\Psi}_{j-1}||_{L^2} = {{2 \pi^2}\over{3\sqrt{5}}} \sum^{n}_{j=1} ||\Tilde{\Psi}_{n-j}||_{L^2} ||\Tilde{\Psi}_{j-1}||_{L^2}.
\end{aligned}
\end{equation}
Here, the first inequality follows from the triangular inequality and the fact that $|\cos{(x)}| \leq 1$. The second inequality is based on H\"{o}lder equality. Consequently, the upper bound for $||\Tilde{V}_n||_{L^2}$ depends on $||\Tilde{\Psi}_{j}||_{L^2}$ for $j < n$.

Finally, a similar estimation can be applied to the Schr\"{o}dinger equation to provide an upper bound for the $L^2$-norm of each order of the background wave function. Combining Eqs. (\ref{equ: background configuration for each order}) and (\ref{equ: Fourier series of each order of background wave function and gravitational potential}), the Fourier coefficient $\Tilde{\psi}_{n,l}$ satisfies $-2l(l+1)\Tilde{\psi}_{n,l} = \int^{2\pi}_{0}\cos{\Big ( (2l+1)n^{(0)}_{eq}x\Big )}\sum^{n}_{j=1} \Tilde{\Psi}_{n=j}(x)\Big [ \Tilde{V}_{j}(x)-\Tilde{E}_j\Big ]dx/(2\pi)$. Therefore, $||\Tilde{\Psi}_n||_{L^2}$ can be bounded by:

\begin{equation}
\label{equ: the bounded of the n-th order wave function}
\begin{aligned}
||\Tilde{\Psi}_n||_{L^2} = \sqrt{{{\sum^{n}_{l=1}|\Tilde{\psi}_{n,l}|^2}\over{2}}} & \leq \sqrt{\sum^{n}_{l=1}{{1}\over{8l^2(l+1)^2}}}\sum^{n}_{j=1}{{1}\over{2\pi}}\int^{2\pi}_{0}|\Tilde{\Psi}_{n-j}(x)|\Big ( |\Tilde{V}_{j}(x)| + |\Tilde{E}_j| \Big ) dx \\
& \leq \sqrt{\sum^{\infty}_{l=1}{{1}\over{8l^2(l+1)^2}}}\sum^{n}_{j=1} ||\Tilde{\Psi}_{n-j}||_{L^2} \Big ( ||\Tilde{V}_{j}||_{L^2} + |\Tilde{E}_j| \Big ) = \sqrt{{{\pi^2-9}\over{24}}} \sum^{n}_{j=1} ||\Tilde{\Psi}_{n-j}||_{L^2} \Big ( ||\Tilde{V}_{j}||_{L^2} + |\Tilde{E}_j| \Big ).
\end{aligned}
\end{equation}
Here, the first inequality follows from the triangular inequality and the fact that $|\cos{(x)}| \leq 1$. The second inequality is based on H\"{o}lder equality. Substitute Eqs. (\ref{equ: the bounded of the n-th order eigenergy}) and (\ref{equ: the bounded of the n-th order gravitational potential}) into Eq. (\ref{equ: the bounded of the n-th order wave function}), we obtain:

\begin{equation}
\label{equ: the recursive inequality for the n-th order wave function}
\begin{aligned}
||\Tilde{\Psi}_n||_{L^2} \leq  \Bigg ( \sqrt{{{\pi^2-9}\over{24}}} \Bigg ) \Bigg ( {{2 \pi^2}\over{3\sqrt{5}}}  \Bigg )\Bigg ( &  {{\max_{x\in[0,2\pi]}|\Tilde{\Psi}_0(x)|}\over{||\Tilde{\Psi}_0||_{L^2}}} {{\sum^{n}_{i=1} \sum^{i}_{j=1} \sum^{j}_{l=1} ||\Tilde{\Psi}_{n-i}||_{L^2} ||\Tilde{\Psi}_{i-j}||_{L^2} ||\Tilde{\Psi}_{j-l}||_{L^2} ||\Tilde{\Psi}_{l-1}||_{L^2}}\over{||\Tilde{\Psi}_0||_{L^2}}} + \\
& \sum^{n}_{i=1} \sum^{i}_{j=1} ||\Tilde{\Psi}_{n-i}||_{L^2} ||\Tilde{\Psi}_{i-j}||_{L^2} ||\Tilde{\Psi}_{j-1}||_{L^2} \Bigg ).
\end{aligned}
\end{equation}

To demonstrate the non-zero radius of convergence, it is sufficient to show that $\lim_{n \rightarrow \infty} \sqrt[n]{||\Tilde{\Psi}_n||_{L^2}}$ is finite.  If this limit is finite, then there exist two finite positive constants $C$ and $r$ such that $||\Tilde{\Psi}_n||_{L^2} \leq Cr^n$ for all $n \in \mathbb{N} \cup \{0\}$. Consequently, $||\Tilde{V}_n||_{L^2} \leq C_1 n r^{n-1}$ for some constant $C_1$, as given by Eq. (\ref{equ: the bounded of the n-th order gravitational potential}). Similarly, $|\Tilde{E}_n| \leq C_2 n(n+1) r^n$ for some constant $C_2$ in accordance with Eqs. (\ref{equ: the bounded of the n-th order eigenergy}) and (\ref{equ: the bounded of the n-th order gravitational potential}), establishing the non-zero radius of convergence for the background eigenenergy. Finally, the non-zero radius of convergence for the background wave function and gravitational potential is given by $\lim_{n \rightarrow \infty} \sqrt[n]{\mathop{\max}_{x \in [0,2\pi]}|\Tilde{\Psi}_n(x)|}=\lim_{n \rightarrow \infty} \sqrt[n]{||\Tilde{\Psi}_n||_{L^2}}$ and $\lim_{n \rightarrow \infty} \sqrt[n]{\mathop{\max}_{x \in [0,2\pi]}|\Tilde{V}_n(x)|}=\lim_{n \rightarrow \infty} \sqrt[n]{||\Tilde{V}_n||_{L^2}}$. This follows because $||\Tilde{V}_n||_{L^2} \leq \mathop{\max}_{x \in [0,2\pi]}|\Tilde{V}_n(x)| \leq \sum^{n}_{l=1}|\Tilde{v}_{n,l}| \leq \sqrt{2n} ||\Tilde{V}_n||_{L^2}$ and $||\Tilde{\Psi}_n||_{L^2} \leq \mathop{\max}_{x \in [0,2\pi]}|\Tilde{\Psi}_n(x)| \leq \sum^{n}_{l=1}|\Tilde{\psi}_{n,l}| \leq \sqrt{2n} ||\Tilde{\Psi}_n||_{L^2}$.

The proof of the finiteness of $\lim_{n \rightarrow \infty} \sqrt[n]{||\Tilde{\Psi}_n||_{L^2}}$ is based on identifying a corresponding geometric sequence that is always larger than the sequence $\{||\Tilde{\Psi}_n||_{L^2}\}$. This approach is formalized in the following proposition:
\\\\

\begin{proposition}
\label{Pro: the upper bound of gemoetrical sequence for background wave function}
Fixed $s > 1$, $||\Tilde{\Psi}_n||_{L^2} \leq {{||\Tilde{\Psi}_0||_{L^2}}\over{(n+1)^s}} r^n(s)$ for all $n \in \mathbb{N} \cup \{ 0\}$, where $r(s) \equiv \Big ( \sqrt{{{\pi^2-9}\over{24}}} \Big ) \Big ( {{2 \pi^2}\over{3\sqrt{5}}} \Big )\Bigg [ {{\max_{x\in [0,2\pi]}|\Tilde{\Psi}_0(x)|}\over{||\Tilde{\Psi}_0||_{L^2}}} \times \Big ( 2^s \zeta(s) \Big )^3 + \Big ( 2^s \zeta(s) \Big )^2\Bigg ]||\Tilde{\Psi}_0||^2_{L^2}$ with Riemann-zeta function $\zeta(s) \equiv \sum^{\infty}_{n=1}{{1}\over{n^s}}$.
\end{proposition}

\renewcommand\qedsymbol{QED}
\begin{proof}
This proposition can be proved by mathematical induction. For $j=1$, Eq. (\ref{equ: the recursive inequality for the n-th order wave function}) shows that $||\Tilde{\Psi}_1||_{L^2} \leq \Big ( \sqrt{{{\pi^2-9}\over{24}}} \Big ) \Big ( {{2 \pi^2}\over{3\sqrt{5}}} \Big ) \Big ( 2 ||\Tilde{\Psi}_0||^4_{L^2}  + ||\Tilde{\Psi}_0||^3_{L^2} \Big ) \leq {{||\Tilde{\Psi}_0||_{L^2}}\over{(1+1)^s}}r(s)$ since $\zeta(s) \geq 1$. Assume this proposition is valid for all $j \leq n - 1$ for any $n \in \mathbb{N} \backslash \{1\}$, then when $j=n$, Eq. (\ref{equ: the recursive inequality for the n-th order wave function}) becomes:

\begin{equation}
\label{equ: the estimation of the upper bound for the n-th order of background wave function}  
\begin{aligned}
||\Tilde{\Psi}_n||_{L^2} \leq & ||\Tilde{\Psi}_0||_{L^2} \Bigg \{ \Bigg ( \sqrt{{{\pi^2-9}\over{24}}} \Bigg ) \Bigg ( {{2 \pi^2}\over{3\sqrt{5}}} \Bigg )||\Tilde{\Psi}_0||^2_{L^2} \Bigg \}^n \Bigg \{ \Bigg [ {{\max_{x\in[0,2\pi]}|\Tilde{\Psi}_0(x)|}\over{||\Tilde{\Psi}_0||_{L^2}}} \Big ( 2^s \zeta(s) \Big )^3 + \Big ( 2^s \zeta(s) \Big )^2\Bigg ]  \Bigg \}^{n-1} \times \\
&  \Bigg \{ {{\max_{x\in[0,2\pi]}|\Tilde{\Psi}_0(x)|}\over{||\Tilde{\Psi}_0||_{L^2}}} \sum^{n}_{i=1} \sum^{i}_{j=1} \sum^{j}_{l=1} \Bigg [ {{1}\over{(n-i+1)(i-j+1)(j-l+1)m}} \Bigg ]^s + \sum^{n}_{i=1} \sum^{i}_{j=1} \Bigg [ {{1}\over{(n-i+1)(i-j+1)j}} \Bigg ]^s  \Bigg \}.
\end{aligned}
\end{equation}
Here, the double summations in Eq. (\ref{equ: the estimation of the upper bound for the n-th order of background wave function}) can be estimated as follows:

\begin{equation}
\label{equ: the estimation of the upper bound for double summation}  
\begin{aligned}
\sum^{n}_{i=1} \sum^{i}_{j=1} \Bigg [ {{1}\over{(n-i+1)(i-j+1)j}} \Bigg ]^s & = \sum^{n}_{i=1} \Bigg [ {{1}\over{(n-i+1)(i+1)}} \Bigg ]^s \sum^{i}_{j=1} \Bigg ( {{1}\over{i-j+1}} + {{1}\over{j}} \Bigg )^s \\
& \leq 2^s \sum^{n}_{i=1} \Bigg [ {{1}\over{(n-i+1)(i+1)}} \Bigg ]^s \sum^{i}_{j=1} \Bigg [ {{1}\over{2}}{{1}\over{(i-j+1)^s}} + {{1}\over{2}}{{1}\over{j^s}} \Bigg ] \\
& \leq 2^s \zeta(s) \sum^{n}_{i=1} \Bigg [ {{1}\over{(n-i+1)i}} \Bigg ]^s \leq {{\Big [ 2^s \zeta(s) \Big ]^2}\over{(n+1)^s}}.
\end{aligned}
\end{equation}
Here, the first inequality follows from the fact that $[(x+y)/2]^s \leq (x^s + y^s)/2$ for all $x$, $y > 0$ and any $s>1$. The second inequality is obtained using the definition of the Riemann-zeta function and noting that $i+1 >i$. The final inequality is achieved by repeating the same argument used to obtain the second inequality. Similarly, the triple summations in Eq. (\ref{equ: the estimation of the upper bound for the n-th order of background wave function}) is also bounded by $[ 2^s \zeta(s) ]^3/(n+1)^s$. Therefore, it is straightforward to conclude that this proposition is valid by mathematical induction.
\end{proof}

Proposition \ref{Pro: the upper bound of gemoetrical sequence for background wave function} establishes that $\lim_{n \rightarrow \infty} \sqrt[n]{||\Tilde{\Psi}_n||_{L^2}} \leq r(s) < \infty$ for all $s>1$.  It should be noted that this proposition confirms the non-zero radius of convergence, rather than providing an accurate estimation. To see this, consider that the optimal estimation from the above inequality is the minimum value of $r(s)$ for $s>1$, which is approximately $124$. Consequently, the corresponding radius of convergence is around $0.01$. However, Fig. (\ref{fig: convergence_of_radius}) suggests that the radius of convergence is about $4.5$, which is roughly two orders of magnitude larger than the aforementioned optimal estimation.

The proof in this part establishes the validity of the non-zero convergence of radius. However, Fig. (\ref{fig: convergence_of_radius}) also suggests that the radii of convergence for the background eigenenergy, gravitational potential, and wave function are identical. Demonstrating the exact identity of these radii of convergence is left for future work.  
\\\\
\textbf{\textit{(3): Number of Nodes for All Nonlinear Background Wave Functions}}
\\

The background wave function is given by Eq. (\ref{equ: Fourier series of the background wave and gravitational potential}). Due to the periodic property, the spatial domain can be confined to $[0,2\pi/n^{(0)}_{eq}]$ without losing generality. It is evident that the background wave function already has nodes at $x=\pi/(2n^{(0)}_{eq})$ and $x=3\pi/(2n^{(0)}_{eq})$ for any parameter  $\Tilde{\rho}^{(0)}_{eq}$ smaller than the radius of convergence $\Tilde{\rho}^{(c)}_{eq}$. The following paragraphs will provide a rigorous proof that the background wave function only has these two nodes in $[0,2\pi/n^{(0)}_{eq}]$. Before proceeding with the proof, it is important to treat the background wave function $\Tilde{\Psi}_{eq} = \Tilde{\Psi}_{eq}(\Tilde{\rho}^{(0)}_{eq},x)$, where the range of $\Tilde{\rho}^{(0)}_{eq}$ is $[0, \Tilde{\rho}^{(c)}_{eq})$. In the $(\Tilde{\rho}^{(0)}_{eq},x)$-space, the background wave function vanishes along two lines $l_1$ and $l_2$, where $l_1 = \{ (\Tilde{\rho}^{(0)}_{eq}, x= \pi/(2n^{(0)}_{eq} )\}$ and $l_2 = \{ (\Tilde{\rho}^{(0)}_{eq},x = 3\pi/(2n^{(0)}_{eq} )\}$.

Suppose there is another node, say $(\Tilde{\rho}_0 , x_0)$, which is not located on either $l_1$ or $l_2$. It is straightforward to find that $\partial \Tilde{\Psi}_{eq}(\Tilde{\rho}_0 , x_0)/\partial x  \neq 0$ because the background wave function satisfies the Schr\"{o}dinger equation, which is a second-order differential equation with respect to $x$. By applying the implicit function theorem \cite{Rudin1976}, we can find an open interval in $[0,\Tilde{\rho}^{(c)}_{eq})$ containing $\Tilde{\rho}_0$ on which there exists a unique differentiable function $g$ such that $g(\Tilde{\rho}_0) = x_0$ and $\Tilde{\Psi}_{eq}( \Tilde{\rho}^{(0)}_{eq}, g(\Tilde{\rho}^{(0)}_{eq}) )=0$ for all $\Tilde{\rho}^{(0)}_{eq}$ in this open interval.

It is important to note that this open interval can be extended to $[0,\Tilde{\rho}^{(c)}_{eq})$ because the endpoints of this open interval are also nodes for the background wave function. Therefore, the curve $C= \{ (\Tilde{\rho}^{(0)}_{eq},g(\Tilde{\rho}^{(0)}_{eq}) ): \Tilde{\rho}^{(0)}_{eq} \in [0,\Tilde{\rho}^{(c)}_{eq}) \}$ in the $(\Tilde{\rho}^{(0)}_{eq},x)$-space is well-defined. Moreover, the curve $C$ cannot intersect with either $l_1$ or $l_2$. otherwise, it would lead to a contradiction that violates the uniqueness property of the differentiable function $g$.

Hence, $g(0) \neq \pi/(2n^{(0)}_{eq})$ and $3\pi/(2n^{(0)}_{eq})$ and $\Tilde{\Psi}_{eq}(0, g(0)) = 0$. However, the construction mentioned in Sec. (\ref{sec: Nonlinear periodic equilibrium construction}) indicates that $\Tilde{\Psi}_{eq}(0, x) = \cos{(n^{(0)}_{eq}x)}$, which only vanishes at $x=\pi/(2n^{(0)}_{eq})$ and $3\pi/(2n^{(0)}_{eq})$. This contradicts the above argument, proving that the background wave function has only two nodes in $[0,2\pi/n^{(0)}_{eq}]$. 

\renewcommand{\theequation}{B\arabic{equation}}
\setcounter{equation}{0}
\section{Generalized Bloch's Theorem for the Operator $\hat{\mathbf{\Theta}}\hat{\mathbf{H}}$}
\label{appendix: Generalized Bloch's Theorem for the Operator ThetaH}
As demonstrated in Sec. (\ref{subsec: Taylor expansion construction}), the background wave function $\Psi_{eq}(x)$ and gravitational potential $V_{eq}(x)$  satisfy the relations $V_{eq}(x + \pi/n^{(0)}_{eq}) = V_{eq}(x)$ and $\Psi_{eq}(x + \pi/n^{(0)}_{eq}) = -\Psi_{eq}(x)$, as shown in Eq. (\ref{equ: Fourier series of the background wave and gravitational potential}). Although the periodicity of this quantum system is governed by the period of the background wave function, which is longer than that of the gravitational potential, the lattice cell size is still determined by the period of the gravitational potential. Consequently, the eigenfunction of the operator $\hat{\mathbf{\Theta}}\hat{\mathbf{H}}$ resembles the Bloch function, represented as $e^{\textbf{\textit{i}}qx}p(q,x)$, where $q$ is the quasi-momentum, and $p(q,x)$ is a periodic function satisfying $p(q,x+\pi/n^{(0)}_{eq}) = p(q,x)$. Unlike the conventional Bloch theorem, however, the corresponding perturbed gravitational potential is modulated by an anti-periodic function.

To arrive at the above statement, the eigenvalue problem for the operator $\hat{\mathbf{\Theta}}\hat{\mathbf{H}}$, given in Eq. (\ref{equ: perturbed Schrodiger-Poisson equation for single component by real and imagine part}), with $\nabla^2 \equiv d^2/dx^2 - |\vec{k}_{\perp}|^2$, where $\vec{k}_{\perp}$ represents the wave vector perpendicular to the $x$-axis, must be reformulated as a system of first-order linear differential equations. To facilitate this transformation, the following new variables are introduced: $\eta_0(x) \equiv \eta(x)$, $\eta_1(x) \equiv d\eta(x)/dx$, $R_0(x) \equiv R(x)$, $R_1(x) \equiv dR(x)/dx$, $\delta V_0(x) \equiv \delta V(x)$, and $\delta V_1(x) \equiv d\delta V(x)/dx$. Therefore, Eq. (\ref{equ: perturbed Schrodiger-Poisson equation for single component by real and imagine part}) can be rewritten as follows:

\begin{equation}
\label{equ: the system of first order differential equation form for perturbed Schrodiger-Poisson equation for single component}
{{d}\over{dx}}\vec{y}(x) \equiv {{d}\over{dx}} \begin{bmatrix} R_1(x) \\ R_0(x) \\ \eta_1(x) \\ \eta_0(x) \\\delta V_1(x) \\ \delta V_0(x) \end{bmatrix} = 
\begin{bmatrix}
0 & 2(V_{eq}(x) - E_{eq}) + |\vec{k}_{\perp}|^2 & 0 & -2\omega^2                                  & 0 & 2\Psi_{eq}(x)       \\
1 & 0                                           & 0 & 0                                           & 0 & 0                   \\
0 & -2                                          & 0 & 2(V_{eq}(x) - E_{eq}) + |\vec{k}_{\perp}|^2 & 0 & 0                   \\
0 & 0                                           & 1 & 0                                           & 0 & 0                   \\
0 & 2\Psi_{eq}(x)                               & 0 & 0                                           & 0 & |\vec{k}_{\perp}|^2 \\
0 & 0                                           & 0 & 0                                           & 1 & 0
\end{bmatrix}
\begin{bmatrix} R_1(x) \\ R_0(x) \\ \eta_1(x) \\ \eta_0(x) \\\delta V_1(x) \\ \delta V_0(x) \end{bmatrix} 
\equiv \mathbf{A}(x)\vec{y}(x).
\end{equation}

Before proceeding to the proof, we introduce a $6 \times 6$ constant diagonal matrix $\mathbf{M} \equiv \begin{bmatrix}
1 & 0 & 0 & 0 & 0  & 0  \\
0 & 1 & 0 & 0 & 0  & 0  \\
0 & 0 & 1 & 0 & 0  & 0  \\
0 & 0 & 0 & 1 & 0  & 0  \\
0 & 0 & 0 & 0 & -1 & 0  \\
0 & 0 & 0 & 0 & 0  & -1
\end{bmatrix}$ to simplify the notation. The Bloch functional-like form for the perturbed quantities is given by the following proposition:
\\\\

\begin{proposition}
\label{Pro: modified Floquet theory for linear Schrodinger Poisson equation}
There exists a non-zero constant $\sigma$ and a solution vector $\vec{y}(x)$ of Eq. (\ref{equ: the system of first order differential equation form for perturbed Schrodiger-Poisson equation for single component}) such that $\mathbf{M}\vec{y}(x+ \pi/n^{(0)}_{eq}) = \sigma \vec{y}(x)$.
\end{proposition}

\renewcommand\qedsymbol{QED}
\begin{proof}

Let $\mathbf{Y}(x)$ be a $6 \times 6$ solution matrix, where each column vector is a solution of Eq. (\ref{equ: the system of first order differential equation form for perturbed Schrodiger-Poisson equation for single component}) and all column vectors are linearly independent of each other. A simple calculation yields:
\begin{equation}
\label{equ: calculation for the Floquet theory with ThetaH}
{{d}\over{dx}} \mathbf{M}\mathbf{Y}(x+\pi/n^{(0)}_{eq}) = \mathbf{M}\mathbf{A}(x+\pi/n^{(0)}_{eq})\mathbf{M}\mathbf{M}\mathbf{Y}(x+\pi/n^{(0)}_{eq}),
\end{equation}
where $\mathbf{M}^2 =\mathbf{I}$(the identity matrix) is applied. 

Next, based on the fact that $V_{eq}(x + \pi/n^{(0)}_{eq}) = V_{eq}(x)$ and $\Psi_{eq}(x + \pi/n^{(0)}_{eq}) = -\Psi_{eq}(x)$,  it is straightforward to find that $\mathbf{M}\mathbf{A}(x+\pi/n^{(0)}_{eq})\mathbf{M} = \mathbf{A}(x)$. Consequently, $\mathbf{M}\mathbf{Y}(x+\pi/n^{(0)}_{eq})$ is also a solution matrix. Thus, any column vector of $\mathbf{M}\mathbf{Y}(x+\pi/n^{(0)}_{eq})$ should be the linear combination of the column vectors of $\mathbf{Y}(x)$. The coefficients of this linear combination form a constant $6 \times 6$ matrix $\mathbf{B}$ with non-zero determinant, satisfying $\mathbf{M}\mathbf{Y}(x+\pi/n^{(0)}_{eq})=\mathbf{Y}(x)\mathbf{B}$.

Let $\sigma$ be the non-zero eigenvalue of the constant matrix $\mathbf{B}$ with the corresponding eigenvector $\vec{b}$. Then, $\vec{y}(x) \equiv \mathbf{Y}(x)\vec{b}$ is the solution of Eq. (\ref{equ: the system of first order differential equation form for perturbed Schrodiger-Poisson equation for single component}). Moreover, $\vec{y}(x)$ satisfies the condition $\mathbf{M}\vec{y}(x+\pi/n^{(0)}_{eq}) = \mathbf{M}\mathbf{Y}(x+\pi/n^{(0)}_{eq})\vec{b}=\mathbf{Y}(x)\mathbf{B}\vec{b}=\sigma \vec{y}(x)$. This completes the proof. 
\end{proof}

The constant matrix $\mathbf{B}$ is known as the monodromy matrix. Moreover, if $|\sigma|=1$, the quasi-momentum $q$ can be introduced such that $\sigma = e^{\textbf{\textit{i}}\pi q/n^{(0)}_{eq}}$. Consequently, the solution vector can be expressed as $\vec{y}(x) = e^{\textbf{\textit{i}}q x} \vec{p}_{\vec{y}}(x)$, where the vector function $\vec{p}_{\vec{y}}(x)$ satisfies $\mathbf{M}\vec{p}_{\vec{y}}(x+\pi/n^{(0)}_{eq}) = \vec{p}_{\vec{y}}(x)$. This form is similar to the Bloch function, except that the corresponding perturbed gravitational potential associates with an anti-periodic function. 

Furthermore, if the background configuration exhibits even symmetry, the eigenvalue of the monodromy matrix possesses the following property:
\\\\

\begin{proposition}
\label{Pro: the even symmetry for the monodromy matrix}
Let $\sigma$ be a non-zero eigenvalue of the monodromy matrix $\mathbf{B}$. If the background configuration satisfies $V_{eq}(-x) = V_{eq}(x)$ and $\Psi_{eq}(-x) = \Psi_{eq}(x)$, then $1/\sigma$ is also an eigenvalue of the monodromy matrix $\mathbf{B}$.
\end{proposition}

\renewcommand\qedsymbol{QED}
\begin{proof}

Let a $6 \times 6$ constant matrix $\mathbf{J} \equiv \begin{bmatrix}
-1 & 0 &  0 & 0 & 0  & 0  \\
0  & 1 &  0 & 0 & 0  & 0  \\
0  & 0 & -1 & 0 & 0  & 0  \\
0  & 0 &  0 & 1 & 0  & 0  \\
0  & 0 &  0 & 0 & -1 & 0  \\
0  & 0 &  0 & 0 & 0  & 1
\end{bmatrix}$ and let the solution matrix $\mathbf{Y}(x)$ satisfy $\mathbf{M}\mathbf{Y}(x+\pi/n^{(0)}_{eq}) = \mathbf{Y}(x)\mathbf{B}$, where $\mathbf{B}$ is the monodromy matrix. A simple calculation yields that:

\begin{equation}
\label{equ: calculation the Floquet theory with even symmetry}    
{{d}\over{dx}} \mathbf{J}\mathbf{\Phi}(-x)=-\mathbf{J}\mathbf{A}(-x)\mathbf{\Phi}(-x) = -\mathbf{J}\mathbf{A}(x)\mathbf{J}\mathbf{J}\mathbf{\Phi}(-x) = \mathbf{A}(x)\mathbf{J}\mathbf{\Phi}(-x),
\end{equation}
where $\mathbf{J}\mathbf{J} = \mathbf{I}$ is used. Therefore, $\mathbf{J}\mathbf{Y}(-x)$ is also a solution matrix. Consequently, there exists another constant matrix $\mathbf{C}$ such that $\mathbf{J}\mathbf{Y}(-x)= \mathbf{Y}(x)\mathbf{C}$.

When $x=0$, $\mathbf{J}\mathbf{Y}(0)= \mathbf{Y}(0)\mathbf{C}$. This implies that $\mathbf{C} = \mathbf{Y}(0)^{-1} \mathbf{J}\mathbf{Y}(0)$ and hence, $C^{-1} = C$. Evaluating at $x=\pi/n^{(0)}_{eq}$ gives $\mathbf{J}\mathbf{Y}(-\pi/n^{(0)}_{eq})= \mathbf{Y}(\pi/n^{(0)}_{eq})\mathbf{C}$. Given that $\mathbf{M}\mathbf{Y}(\pi/n^{(0)}_{eq}) = \mathbf{Y}(0)\mathbf{B}$ and $\mathbf{M}\mathbf{Y}(0) = \mathbf{Y}(-\pi/n^{(0)}_{eq})\mathbf{B}$, it follows that $\mathbf{B}^{-1} = \mathbf{C}\mathbf{B}\mathbf{C}$.

Finally, given a non-zero eigenvalue $\sigma$ of the monodromy matrix $\mathbf{B}$, it satisfies $det(\mathbf{B} - \sigma \mathbf{I})=0$. A simple calculation shows that $det(\mathbf{B} - \mathbf{I}/\sigma) \propto det(\sigma \mathbf{B} - \mathbf{I}) =  det(\mathbf{B})det(\sigma \mathbf{I} - \mathbf{B}^{-1}) = det(\mathbf{B})det(\sigma \mathbf{I} - \mathbf{C}\mathbf{B}\mathbf{C}) =det(\mathbf{B})det(\mathbf{C})det(\mathbf{C})det(\sigma \mathbf{I}-\mathbf{B})=0$. This result implies that $1/\sigma$ is also the eigenvalue of $\mathbf{B}$, thereby completing the proof.
\end{proof}

Again, if $|\sigma|=1$, the above proposition implies that two linearly independent solution vectors exist, corresponding to the quasi-momenta $q$ and $-q$. Consequently, any eigenvalue $\omega^2$ is degenerate unless $q=0$ or $n^{(0)}_{eq}$.

\renewcommand{\theequation}{C\arabic{equation}}
\setcounter{equation}{0}
\section{Properties for the Matrix Representation of the Operator $\hat{\mathbf{\Theta}}\hat{\mathbf{H}}$ Using the Standing Wave Basis}
\label{appendix: The General Property of the Matrix Representation}

Before proceeding to the results, recall that the standing wave basis $u_j$ in either the subspace $\mathbb{T}^{(cos)}_{q}[0,2\pi]$ or $\mathbb{T}^{(sin)}_{q}[0,2\pi]$ is given by Eq. (\ref{equ: the definition of the standing wave basis}) for $q=0,1,...,n^{(0)}_{eq}$. A simple calculation yields the matrix representation for the Hamiltonian operator $\hat{\mathbf{H}}$ and the operator $\hat{\mathbf{\Theta}}$ in Eq. (\ref{equ: definition of H and Theta with normalization factor}) as follows:

\begin{equation}
\label{equ: the matrix element for H and Theta}
\left\{\begin{aligned}
& \langle u_i | \hat{\mathbf{H}} | u_j \rangle = \Big ( \Tilde{k}^2_j+ \Tilde{k}^2_{\perp} - \Tilde{E}_{eq} \Big )\langle u_i | u_j \rangle \delta_{ij} +  \langle u_i | \Tilde{V}_{eq} | u_j \rangle; \\
& \langle u_i | \hat{\mathbf{\Theta}} | u_j \rangle  = {{1}\over{4}}\Big ( \langle u_i | \hat{\mathbf{H}} | u_j \rangle + \langle u_i | \hat{\mathbf{G}} | u_j \rangle  \Big )
= {{1}\over{4}}\langle u_i | \hat{\mathbf{H}} | u_j \rangle - 4\Tilde{\rho}^{(0)}_{eq}\sum_{l} \Big ( 1 - \delta_{\Tilde{k}^2_{\perp}0}\delta_{\Tilde{\kappa}^2_{l}0} \Big ) {{\langle u_i |  
    \Tilde{\Psi}_{eq} | w_l \rangle \langle w_l | \Tilde{\Psi}_{eq} | u_j \rangle }\over{ \Big ( \Tilde{\kappa}^2_l+\Tilde{k}^2_{\perp} \Big )\langle w_l |w_l \rangle}}.
\end{aligned}
\right.
\end{equation}
Here, $\delta$ denotes the Kronecker delta, while $w_l$ represents the standing wave basis for the perturbed density and is defined as follows:

\begin{equation}
\label{equ: the definition of the standing wave basis in the dual sub-space}
w_l(x) \equiv 
\left\{\begin{aligned}
& \sqrt{2} \cos{(  \kappa_l x ) }\text{, for } \mathbb{T}^{(cos)}_{p}[0,2\pi], \\
& \sqrt{2} \sin{(  \kappa_l x  ) }\text{, for } \mathbb{T}^{(sin)}_{p}[0,2\pi], \\
\end{aligned}
\right. \text{where } \kappa_l \equiv 2l n^{(0)}_{eq} + p \text{, }p \equiv n^{(0)}_{eq} - q. 
\end{equation}
Additionally, $\Tilde{k}_j \equiv  k_j /n^{(0)}_{eq} = 2j+q/n^{(0)}_{eq}$ and $\Tilde{\kappa}_l \equiv  \kappa_l /n^{(0)}_{eq} = 2l+p/n^{(0)}_{eq} = (2l+1)-q/n^{(0)}_{eq}$ represent the momenta for the standing wave bases $u_j$ and $w_l$, respectively. To satisfy the mass conservation constraint and eliminate any contributions to the mass when $\Tilde{k}^2_{\perp} = \Tilde{\kappa}^2_l=0$, the term $\delta_{\Tilde{k}^2_{\perp}0}\delta_{\Tilde{\kappa}^2_{l}0}$ is incorporated into $\langle u_i | \hat{\mathbf{\Theta}} | u_j \rangle$. Consequently, the expression $(1 - \delta_{\Tilde{k}^2_{\perp}0}\delta_{\Tilde{\kappa}^2_{l}0})/( \Tilde{\kappa}^2_{l}+ \Tilde{k}^2_{\perp})$  is explicitly set to zero under the condition $\Tilde{k}^2_{\perp} = \Tilde{\kappa}^2_l = 0$. 

Using Eq. (\ref{equ: the matrix element for H and Theta}), it is straightforward to find that the matrix element $\mathbf{F}_{ij}$ becomes:

\begin{equation}
\label{equ: the matrix element for Theta times H operator}
\begin{aligned}
\mathbf{F}_{ij}  & \equiv  \langle u_i | \hat{\mathbf{\Theta}}\hat{\mathbf{H}} | u_j \rangle = \sum_{m}  {{\langle u_i | \hat{\mathbf{\Theta}} | u_m \rangle \langle u_m | \hat{\mathbf{H}} | u_j \rangle}\over{\langle u_m| u_m \rangle}} =  \sum_{m}{{1}\over{4}} \Bigg ( {{\langle u_i | \hat{\mathbf{H}} |u_m \rangle \langle u_m| \hat{\mathbf{H}} | u_j \rangle}\over{\langle u_m | u_m \rangle}} + {{\langle u_i | \hat{\mathbf{G}} | u_m \rangle \langle u_m | \hat{\mathbf{H}} | u_j \rangle}\over{\langle u_m | u_m \rangle}} \Bigg ) \\
&
\begin{aligned}
= & {{1}\over{4}}\Big ( \Tilde{k}^2_i+ \Tilde{k}^2_{\perp} - \Tilde{E}_{eq} \Big ) \Big ( \Tilde{k}^2_j+ \Tilde{k}^2_{\perp} - \Tilde{E}_{eq} \Big ) \langle u_i | u_j \rangle \delta_{ij} +  {{1}\over{4}}\Big ( \Tilde{k}^2_i + \Tilde{k}^2_j+ 2\Tilde{k}^2_{\perp} - 2\Tilde{E}_{eq} \Big ) \langle u_i | \Tilde{V}_{eq} | u_j \rangle + \\
& {{1}\over{4}} \sum_{m} {{\langle u_i | \Tilde{V}_{eq} |u_m \rangle \langle u_m |\Tilde{V}_{eq} | u_j \rangle}\over{\langle u_m | u_m \rangle }}  - 4\Tilde{\rho}^{(0)}_{eq}  \sum_{l} \Big ( 1- \delta_{\Tilde{k}^2_{\perp}0}\delta_{\Tilde{\kappa}^2_{l}0}\Big ) {{\langle u_i | \Tilde{\Psi}_{eq} | w_l \rangle \langle w_l | \Tilde{\Psi}_{eq} | u_j \rangle }\over{ \Big (\Tilde{\kappa}^2_l+\Tilde{k}^2_{\perp} \Big ) \langle w_l | w_l \rangle }} \Big ( \Tilde{k}^2_j+ \Tilde{k}^2_{\perp} - \Tilde{E}_{eq} \Big ) -\\
& 4\Tilde{\rho}^{(0)}_{eq}\sum_{m,l} \Big ( 1- \delta_{\Tilde{k}^2_{\perp}0}\delta_{\Tilde{\kappa}^2_{l}0} \Big ) {{\langle u_i | \Tilde{\Psi}_{eq} | w_l \rangle \langle w_l | \Tilde{\Psi}_{eq} | u_m \rangle \langle u_m | \Tilde{V}_{eq} | u_j \rangle  }\over{ \Big (\Tilde{\kappa}^2_l+\Tilde{k}^2_{\perp} \Big ) \langle w_l | w_l \rangle \langle u_m | u_m \rangle }}.    
\end{aligned} 
\end{aligned}
\end{equation}
It is evident that the matrix element $\mathbf{F}_{ij} = \mathbf{F}_{ij} \Big ( q/n^{(0)}_{eq}, \Tilde{k}^2_{\perp}, \Tilde{\rho}^{(0)}_{eq} \Big )$. To understand the behavior of this matrix representation, it is essential to estimate the contribution from the self-gravity. 

Use the fact that $2\cos{(x)}\cos{(y)} = \cos{(x-y)} + \cos{(x+y)}$ and $2\sin{(x)}\sin{(y)} = \cos{(x-y)} - \cos{(x+y)}$, $\langle u_i | \Tilde{V}_{eq} | u_j \rangle$ becomes:

\begin{equation}
\label{equ: the matrix element of V_eq}
\langle u_i | \Tilde{V}_{eq} | u_j \rangle = 
\left\{\begin{aligned}
& (\delta_{0i}\Tilde{v}_{j}+\delta_{0j}\Tilde{v}_{i}) + {{(1-\delta_{0i})(1-\delta_{0j})}\over{2}}\Big ( \Tilde{v}_{|i-j|} + \Tilde{v}_{i+j} \Big) , \text{ if }u_i\text{, }u_j \in \mathbb{T}^{(cos)}_{q=0}[0,2\pi], \text{ }\forall i,j \in \mathbb{N}\cup \{0\}; \\
& {{1}\over{2}}\Tilde{v}_{|i-j|}, \text{ if }u_i\text{, }u_j \in \mathbb{T}^{(cos)}_q[0,2\pi] \text{ with }q=1,..., n^{(0)}_{eq}-1, \text{ }\forall i,j \in \mathbb{Z}; \\
& {{1}\over{2}}\Big ( \Tilde{v}_{|i-j|} + \Tilde{v}_{i+j+1} \Big ), \text{ if }u_i\text{, }u_j \in \mathbb{T}^{(cos)}_{q=n^{(0)}_{eq}}[0,2\pi], \text{ }\forall i,j \in \mathbb{N}\cup \{0\};\\
& {{1}\over{2}}\Big ( \Tilde{v}_{|i-j|} - \Tilde{v}_{i+j} \Big ), \text{ if }u_i\text{, }u_j \in \mathbb{T}^{(sin)}_{q=0}[0,2\pi], \text{ }\forall i,j \in \mathbb{N}; \\
& {{1}\over{2}}\Tilde{v}_{|i-j|}, \text{ if }u_i\text{, }u_j \in \mathbb{T}^{(sin)}_q[0,2\pi] \text{ with }q=1,..., n^{(0)}_{eq}-1, \text{ }\forall i,j \in \mathbb{Z}; \\
& {{1}\over{2}}\Big ( \Tilde{v}_{|i-j|} - \Tilde{v}_{i+j+1} \Big ), \text{ if }u_i\text{, }u_j \in \mathbb{T}^{(sin)}_{q=n^{(0)}_{eq}}[0,2\pi], \text{ }\forall i,j \in \mathbb{N}\cup \{0\}, 
\end{aligned}
\right.
\end{equation}
where $\Tilde{v}_m$ is the Fourier coefficient of the background gravitational potential $\Tilde{V}_{eq}$ given by Eq. (\ref{equ: Fourier series of the background wave and gravitational potential}) and $\Tilde{v}_0=0$ is introduced to allow Eq. (\ref{equ: the matrix element of V_eq}) to have a compact form. 

Similarly, it is straightforward to conclude that $\langle w_l | \Tilde{\Psi}_{eq} | u_j \rangle$ is:

\begin{equation}
\label{equ: the matrix element of the background wave function}
\begin{split}
& \langle w_l | \Tilde{\Psi}_{eq} | u_j \rangle = \langle u_j | \Tilde{\Psi}_{eq} | w_l \rangle =\\
& \left\{\begin{aligned}
&\delta_{0l}\Tilde{\psi}_{j} + {{(1-\delta_{0l})}\over{2}} \Big ( \Tilde{\psi}_{ | j-l+{{1}\over{2}}  | - {{1}\over{2}} } + \Tilde{\psi}_{j+l}\Big ) \text{ if }u_j \in \mathbb{T}^{(cos)}_{q=n^{(0)}_{eq}}[0,2\pi], \text{ }w_l \in \mathbb{T}^{(cos)}_{p=0}[0,2\pi]; \text{ }j,l \in \mathbb{N} \cup \{ 0\}; \\
& {{1}\over{2}}\Tilde{\psi}_{|j+l + {{1}\over{2}}| - {{1}\over{2}} }, \text{ if }u_j \in \mathbb{T}^{(cos)}_{q}[0,2\pi], \text{ }w_l \in \mathbb{T}^{(cos)}_{p=n^{(0)}_{eq}-q}[0,2\pi], \text{ with }q=1,...,n^{(0)}_{eq}-1, \text{ }\forall l,j \in \mathbb{Z}; \\
& \delta_{0j}\Tilde{\psi}_{l} + {{(1-\delta_{0j})}\over{2}} \Big ( \Tilde{\psi}_{ | l-j+{{1}\over{2}}  | - {{1}\over{2}} } + \Tilde{\psi}_{j+l}\Big ) \text{ if }u_j \in \mathbb{T}^{(cos)}_{q=0}[0,2\pi], \text{ }w_l \in \mathbb{T}^{(cos)}_{p=n^{(0)}_{eq}}[0,2\pi]; \text{ }j,l \in \mathbb{N}\cup \{ 0\}; \\
& {{(1-\delta_{0l})}\over{2}} \Big ( \Tilde{\psi}_{ | j-l+{{1}\over{2}}  | - {{1}\over{2}} } - \Tilde{\psi}_{j+l}\Big ) \text{ if }u_j \in \mathbb{T}^{(sin)}_{q=n^{(0)}_{eq}}[0,2\pi], \text{ }w_l \in \mathbb{T}^{(sin)}_{p=0}[0,2\pi]; \text{ }j,l \in \mathbb{N}\cup \{ 0\}; \\
& -{{1}\over{2}}\Tilde{\psi}_{|j+l + {{1}\over{2}}| - {{1}\over{2}} }, \text{ if }u_j \in \mathbb{T}^{(sin)}_{q}[0,2\pi], \text{ }w_l \in \mathbb{T}^{(sin)}_{p=n^{(0)}_{eq}-q}[0,2\pi], \text{ with }q=1,...,n^{(0)}_{eq}-1, \text{ }\forall l,j \in \mathbb{Z}; \\
& {{(1-\delta_{0j})}\over{2}} \Big ( \Tilde{\psi}_{ | l-j+{{1}\over{2}}  | - {{1}\over{2}} } - \Tilde{\psi}_{j+l}\Big ) \text{ if }u_j \in \mathbb{T}^{(sin)}_{q=0}[0,2\pi], \text{ }w_l \in \mathbb{T}^{(sin)}_{p=n^{(0)}_{eq}}[0,2\pi]; \text{ }j,l \in \mathbb{N} \cup \{ 0\}. \\
\end{aligned}
\right.   
\end{split}
\end{equation}
Here, $\Tilde{\psi}_m$ is the Fourier coefficient of the background wave function, given by Eq. (\ref{equ: Fourier series of the background wave and gravitational potential}).

The following paragraphs will list the properties of the matrix representation of the operator $\hat{\mathbf{\Theta}}\hat{\mathbf{H}}$. First, it is easy to find that both subspaces $\mathbb{T}^{(cos)}_{q}[0,2\pi]$ and $\mathbb{T}^{(sin)}_{q}[0,2\pi]$ should have the same eigenvalue for $q=1,...,n^{(0)}_{eq}-1$.
\\\\

\begin{proposition}
\label{Pro: The same eigenvalues for sine and cos sub-space far from the band gap}
For fixed $q=1,...,n^{(0)}_{eq}-1$, both subspaces $\mathbb{T}^{(cos)}_{q}[0,2\pi]$ and $\mathbb{T}^{(sin)}_{q}[0,2\pi]$ possess the same eigenvalue $\Tilde{\omega}^2$.
\end{proposition}

\renewcommand\qedsymbol{QED}
\begin{proof}
Fixed $q=1,...,n^{(0)}_{eq}-1$,  it is straightforward to show that $\langle u_i | \Tilde{V}_{eq} | u_j \rangle$ is identical in both subspaces $\mathbb{T}^{(cos)}_q[0,2\pi]$ and $\mathbb{T}^{(sin)}_q[0,2\pi]$, as established by Eq. (\ref{equ: the matrix element of V_eq}). Consequently, the matrix representations of the operator $\hat{\mathbf{H}}$ are also identical, as indicated by Eq. (\ref{equ: the matrix element for H and Theta}). A similar conclusion applies to the operator $\hat{\mathbf{G}}$(and equivalently to $\hat{\mathbf{\Theta}}$), where $\Tilde{\Psi}_{eq}$ appears twice, and $|\langle w_l | \Tilde{\Psi}_{eq} | u_j \rangle|$ is identical in both subspaces $\mathbb{T}^{(cos)}_q[0,2\pi]$ and $\mathbb{T}^{(sin)}_q[0,2\pi]$, based on Eq. (\ref{equ: the matrix element of the background wave function}). Thus, the conclusion follows directly, as the matrix representation for the operator $\hat{\mathbf{\Theta}}\hat{\mathbf{H}}$ is also identical in both subspaces.
\end{proof}

Second, we demonstrate that the eigenvalue is continuous at $q/n^{(0)}_{eq}=0$ and $1$, despite the fact that the matrix representation in the standing wave basis is not smooth in these cases (see Eqs. (\ref{equ: the matrix element of V_eq}) and (\ref{equ: the matrix element of the background wave function})). To establish this, we introduce the plane wave basis, denoted as $v^{(q)}_j(x) = e^{\textbf{\textit{i}}(2jn^{(0)}_{eq}+q)x}$. It is straightforward to show that the set $\{v^{(q)}_j(x), {v^{(q)}_j}^{*}(x)\}^{\infty}_{j=-\infty}$ forms a complete orthonormal basis for the subspace $\mathbb{T}^{(cos)}_{q}[0,2\pi] \bigoplus \mathbb{T}^{(sin)}_{q}[0,2\pi]$ with $q=1,...,n^{(0)}_{eq}-1$, where $\bigoplus$ denotes the direct sum. In contrast, $\{v^{(q)}_j(x)\}^{\infty}_{j=-\infty}$ is a complete orthonormal basis for the subspaces $\mathbb{T}^{(cos)}_{q}[0,2\pi] \bigoplus \mathbb{T}^{(sin)}_{q}[0,2\pi]$ for $q=0$ and $n^{(0)}_{eq}$. Since the eigenvalue $\Tilde{\omega}^2$ is independent of the chosen basis, its continuity follows from the continuity of the matrix representation in the plane wave basis.

The matrix element $\mathbf{F}_{ij}$ is considered as a function of $q/n^{(0)}_{eq}$. The equivalence of the matrix element for both subspaces $\mathbb{T}^{(cos)}_{q}[0,2\pi]$ and $\mathbb{T}^{(sin)}_{q}[0,2\pi]$ with $q=1,...,n^{(0)}_{eq}-1$ implies $\mathbf{F}_{ij}(q/n^{(0)}_{eq}) = \langle v^{(q)}_i | \hat{\mathbf{\Theta}}\hat{\mathbf{H}} |  v^{(q)}_j \rangle$ for $q/n^{(0)}_{eq} \in (0,1)$, meaning that $\mathbf{F}_{ij}$ also represents the matrix element in the plane wave basis for $q=1,...,n^{(0)}_{eq}-1$. Additionally, $\langle {v^{(q)}_i}^{*} | \hat{\mathbf{\Theta}}\hat{\mathbf{H}} |  v^{(q)}_j \rangle = \langle v^{(q)}_i | \hat{\mathbf{\Theta}}\hat{\mathbf{H}} |  {v^{(q)}_j}^{*} \rangle = 0$. Therefore, the continuity of the eigenvalue is supported by the following proposition:
\\\\

\begin{proposition}
\label{Pro: Continuity of the matrix element for the plane wave basis}
The matrix element for the plane wave basis is smooth, in other words:

\noindent1. $\lim_{q/n^{(0)}_{eq} \rightarrow 0}\mathbf{F}_{ij}(q/n^{(0)}_{eq}) =  \langle v^{(0)}_i | \hat{\mathbf{\Theta}}\hat{\mathbf{H}} |  v^{(0)}_j \rangle$. 

\noindent2. $\lim_{q/n^{(0)}_{eq} \rightarrow 1}\mathbf{F}_{ij}(q/n^{(0)}_{eq}) =  \langle v^{(n^{(0)}_{eq})}_i | \hat{\mathbf{\Theta}}\hat{\mathbf{H}} |  v^{(n^{(0)}_{eq})}_j \rangle$ if $\Tilde{k}_{\perp} \neq 0$.
\end{proposition}

\renewcommand\qedsymbol{QED}
\begin{proof}
Simple calculations yield the following results: $2\langle v^{(q-n^{(0)}_{eq})}_l | \Tilde{\Psi}_{eq} |  v^{(q)}_j \rangle = \Tilde{\psi}_{|j-l + {{1}\over{2}}| - {{1}\over{2}}}$ and $2\langle v^{(q)}_i | \Tilde{V}_{eq} |  v^{(q)}_j \rangle = \Tilde{v}_{|i-j|}$ for $q=0,1,...,n^{(0)}_{eq}$. These expressions imply that the quasi-momentum dependence in the matrix element of the operator $\hat{\mathbf{\Theta}}\hat{\mathbf{H}}$ in the plane wave basis is exclusively determined by the Laplacian operator. Consequently, the conclusion follows directly, as the matrix element for the Laplacian operator is continuous with respect to the parameter $q/n^{(0)}_{eq}$.
\end{proof}

Third, the diagonal element $\mathbf{F}_{jj}$ possesses the following property:
\\\\

\begin{proposition}
\label{Pro: the montonic unbounded property of diagonal elements}
Given $\Tilde{\rho}^{(0)}_{eq} \in \Big ( 0, \Tilde{\rho}^{(c)}_{eq} \Big )$ within the radius of convergence $\Tilde{\rho}^{(c)}_{eq}$, $\Tilde{k}^2_{\perp}>0$, and $q = 0,1,...,n^{(0)}_{eq}$, there exists a positive integer $N= N \Big ( \Tilde{\rho}^{(0)}_{eq}, \Tilde{k}^2_{\perp}, q/n^{(0)}_{eq} \Big )$ such that:

\noindent1. $\mathbf{F}_{jj} > 0$ $\forall |j|>N$.

\noindent2. The set $\{ \mathbf{F}_{jj}\}_{|j|>N}$ has no upper bound.
\end{proposition}

\renewcommand\qedsymbol{QED}
\begin{proof}
Equation (\ref{equ: the matrix element for Theta times H operator}) implies that:

\begin{equation}
\label{equ: diagonal element form}
\begin{aligned}
\mathbf{F}_{jj} = & {{1}\over{4}}\langle u_j | u_j \rangle \Big ( \Tilde{k}^2_j + \Tilde{k}^2_{\perp} - \Tilde{E}_{eq} \Big )^2 + \\
                  & \Bigg \{  {{1}\over{2}} \langle u_j |\Tilde{V}_{eq}| u_j \rangle + 4\Tilde{\rho}^{(0)}_{eq}\langle u_j | \Tilde{\Psi}_{eq}  ( 2E^{(0)}_{eq} \nabla^{-2}  ) \Tilde{\Psi}_{eq} | u_j \rangle   \Bigg \} \Big ( \Tilde{k}^2_j + \Tilde{k}^2_{\perp} - \Tilde{E}_{eq} \Big )  + \\
                  & \Bigg \{ {{1}\over{4}} \langle u_j |\Tilde{V}^2_{eq}| u_j \rangle + 4\Tilde{\rho}^{(0)}_{eq}\sum_{m}  
                    {{\langle u_j | \Tilde{\Psi}_{eq}  ( 2E^{(0)}_{eq} \nabla^{-2}  ) \Tilde{\Psi}_{eq} | u_m \rangle \langle u_m | \Tilde{V}_{eq} | u_j \rangle}\over{\langle u_m | u_m \rangle }} \Bigg \}.
\end{aligned}
\end{equation}
This equation indicates that the diagonal element can be approximated as a second-degree polynomial with the argument $\Tilde{k}^2_j + \Tilde{k}^2_{\perp} - \Tilde{E}_{eq}$, except that the coefficients still depend on the index $j$. Once these coefficients have an upper bound that is independent of the index $j$, the conclusion follows immediately because the normalized wave number $\Tilde{k}_j$ is proportional to the absolute value of the index $j$.

The validity of the above claim is supported by the following considerations: 

Firstly, the normalization factor satisfies $1 \leq \langle u_j | u_j \rangle = 2\delta_{\Tilde{k}^2_{j}0}+ \Big ( 1-\delta_{\Tilde{k}^2_{j}0} \Big ) \leq 2$. 

Secondly, the terms involving the background gravitational potential can be estimated as follows: $\langle u_j |\Tilde{V}_{eq}|u_j \rangle \leq 2\sqrt{\langle \Tilde{V}_{eq} | \Tilde{V}_{eq} \rangle}$ and $\langle u_j |\Tilde{V}^2_{eq}| u_j \rangle \leq 2\langle \Tilde{V}_{eq} | \Tilde{V}_{eq} \rangle$. These estimates are derived using H\"{o}lder's inequality and the condition $|u_j(x)|\neq \sqrt{2}$ for all $x \in [0,2\pi]$.

Finally, according to Appendix \ref{appendix: Uniform Boundedness of the Perturbed Self-Gravity Energy Tensor in the Operator Theta}, the terms related to the perturbed self-gravity can be estimated as follows: $|\langle u_j | \Tilde{\Psi}_{eq}  ( 2E^{(0)}_{eq} \nabla^{-2}  ) \Tilde{\Psi}_{eq} | u_j \rangle | \leq 2 h(q/n^{(0)}_{eq}, \Tilde{k}^2_{\perp} )\langle \Tilde{\Psi}_{eq}|\Tilde{\Psi}_{eq} \rangle \langle u_j |u_j \rangle \leq 4 h(q/n^{(0)}_{eq}, \Tilde{k}^2_{\perp} )\langle \Tilde{\Psi}_{eq}|\Tilde{\Psi}_{eq}\rangle$ and $|\sum_{m} \langle u_j | \Tilde{\Psi}_{eq}  ( 2E^{(0)}_{eq} \nabla^{-2}  ) \Tilde{\Psi}_{eq} | u_m \rangle \langle u_m | \Tilde{V}_{eq} | u_j \rangle/\langle u_m | u_m \rangle | = |\langle u_j | \Tilde{\Psi}_{eq}  ( 2E^{(0)}_{eq} \nabla^{-2}  ) \Tilde{\Psi}_{eq}|\Tilde{V}_{eq}u_j \rangle| \leq 2 h(q/n^{(0)}_{eq}, \Tilde{k}^2_{\perp} )\langle \Tilde{\Psi}_{eq}|\Tilde{\Psi}_{eq} \rangle \sqrt{\langle u_j |u_j \rangle \langle u_j | \Tilde{V}^2_{eq} | u_j \rangle} \leq 4 h(q/n^{(0)}_{eq}, \Tilde{k}^2_{\perp} )\langle \Tilde{\Psi}_{eq}|\Tilde{\Psi}_{eq}\rangle \sqrt{\langle \Tilde{V}_{eq} | \Tilde{V}_{eq} \rangle}$, where $h(q/n^{(0)}_{eq}, \Tilde{k}^2_{\perp} )$ is defined by Eq. (\ref{equ: definition of function h}). 

Combining the above conditions, the proposition is fully established, as the derived upper bounds depend solely on $h(q/n^{(0)}_{eq}, \Tilde{k}^2_{\perp} )$, and $\langle \Tilde{\Psi}_{eq} | \Tilde{\Psi}_{eq} \rangle$ and $\langle \Tilde{V}_{eq} | \Tilde{V}_{eq} \rangle$, all of which are independent of the index $j$.   
\end{proof}

Finally, the off-diagonal element $\mathbf{F}_{ij}$ has the following approximate upper bound:     
\\\\

\begin{proposition}
\label{Pro: the boundness of off-diaagonal element}
Given $\Tilde{\rho}^{(0)}_{eq} \in \Big ( 0, \Tilde{\rho}^{(c)}_{eq} \Big )$ within the radius of convergence $\Tilde{\rho}^{(c)}_{eq}$, $\Tilde{k}^2_{\perp}>0$, and $q = 0,1,...,n^{(0)}_{eq}$, and for $i \neq j$:

\noindent $|\mathbf{F}_{ij}| \lesssim \Big \{ {{1}\over{4}}|\Tilde{k}^2_{i}+\Tilde{k}^2_{\perp} - \Tilde{E}_{eq}| +\Big [ {{1}\over{4}} + 4\Tilde{\rho}^{(c)}_{eq} h \Big ( {{q}\over{n^{(0)}_{eq}}}, \Tilde{k}^2_{\perp} \Big )  \Big ]  \Big [ |\Tilde{k}^2_{j}+\Tilde{k}^2_{\perp} - \Tilde{E}_{eq}| + |i-j| + {{( \Tilde{\rho}^{(c)}_{eq} )^2 +  (\Tilde{\rho}^{(0)}_{eq} )^2}\over{ ( \Tilde{\rho}^{(c)}_{eq} )^2 -  ( \Tilde{\rho}^{(0)}_{eq} )^2}} \Big ]  \Big \}\Big ( {{\Tilde{\rho}^{(0)}_{eq}}\over{\Tilde{\rho}^{(c)}_{eq}}} \Big )^{|i-j|}$ 
where the function $h$ is defined by Eq. (\ref{equ: definition of function h}).
\end{proposition}

\renewcommand\qedsymbol{QED}
\begin{proof}
Equation (\ref{equ: the matrix element for Theta times H operator}) implies the following formula:

\begin{equation}
\label{equ: the off-diagonal element formula}
\begin{aligned}
\mathbf{F}_{ij} = & {{1}\over{4}}\Big ( \Tilde{k}^2_i + \Tilde{k}^2_j+ 2\Tilde{k}^2_{\perp} - 2\Tilde{E}_{eq} \Big ) \langle u_i | \Tilde{V}_{eq} | u_j \rangle +  4\Tilde{\rho}^{(0)}_{eq} \langle u_i |\Tilde{\Psi}_{eq}(2E^{(0)}_{eq}\nabla^{-2})\Tilde{\Psi}_{eq} | u_j \rangle \Big ( \Tilde{k}^2_j+ \Tilde{k}^2_{\perp} - \Tilde{E}_{eq} \Big ) + \\
& {{1}\over{4}} \langle u_i | \Tilde{V}^2_{eq} | u_j \rangle  + 4\Tilde{\rho}^{(0)}_{eq}\sum_{m} {{\langle u_i |\Tilde{\Psi}_{eq}(2E^{(0)}_{eq}\nabla^{-2})\Tilde{\Psi}_{eq} | u_m \rangle \langle u_m | \Tilde{V}_{eq} | u_j \rangle}\over{\langle u_m | u_m \rangle }}.   
\end{aligned}
\end{equation}
It is important to note that the off-diagonal element arises solely from self-gravity, including both the background and the perturbed gravitational potential. Unlike Proposition \ref{Pro: the montonic unbounded property of diagonal elements}, contributions from both the background and perturbed self-gravity are of the same order. Therefore, these contributions must be estimated with greater precision to avoid incorrect conclusions.

To begin with, Combining with Eq. (\ref{equ: the matrix element of V_eq}) and Sec. (\ref{subsec: Taylor expansion construction}), it follows that $|\langle u_i | \Tilde{V}_{eq} | u_j \rangle| \sim \Big ( \Tilde{\rho}^{(0)}_{eq}/\Tilde{\rho}^{(c)}_{eq} \Big )^{|i-j|}$. Applying this approximation to $\langle u_i | \Tilde{V}^2_{eq} | u_j \rangle$, we obtain:

\begin{equation}
\label{equ: the upper bound of the matrix element of the square of background gravitational potential}
\begin{aligned}
|\langle u_i | \Tilde{V}^2_{eq} | u_j \rangle| & = \Big |\sum_{m} {{\langle u_i | \Tilde{V}_{eq} |u_m \rangle \langle u_m| \Tilde{V}_{eq} | u_j \rangle}\over{\langle u_m | u_m \rangle}} \Big | \leq  \sum_{m} |\langle u_i | \Tilde{V}_{eq} |u_m \rangle| |\langle u_m| \Tilde{V}_{eq} | u_j \rangle| \sim \sum_{m} \Bigg ( {{\Tilde{\rho}^{(0)}_{eq}}\over{\Tilde{\rho}^{(c)}_{eq}}}\Bigg )^{|i-m| + |m-j| } \\
& = \Bigg [ \sum_{m \in (\min{\{i,j\}}, \max{\{i,j\}}) } \Bigg ( {{\Tilde{\rho}^{(0)}_{eq}}\over{\Tilde{\rho}^{(c)}_{eq}}}\Bigg )^{|i-m| + |m-j| } + \sum_{m \notin (\min{\{i,j\}}, \max{\{i,j\}}) } \Bigg ( {{\Tilde{\rho}^{(0)}_{eq}}\over{\Tilde{\rho}^{(c)}_{eq}}}\Bigg )^{|i-m| + |m-j| } \Bigg ] \\
& = \Bigg [ \Bigg ( |i-j| - 1 \Bigg ) \Bigg ( {{\Tilde{\rho}^{(0)}_{eq}}\over{\Tilde{\rho}^{(c)}_{eq}}}\Bigg )^{|i-j|} + 2 \Bigg ( {{\Tilde{\rho}^{(0)}_{eq}}\over{\Tilde{\rho}^{(c)}_{eq}}}\Bigg )^{|i-j|} \sum^{\infty}_{m=0} \Bigg ( {{\Tilde{\rho}^{(0)}_{eq}}\over{\Tilde{\rho}^{(c)}_{eq}}}\Bigg )^{2m}  \Bigg ]  = \Bigg [ |i-j| + {{\Big ( \Tilde{\rho}^{(c)}_{eq} \Big )^2 + \Big (\Tilde{\rho}^{(0)}_{eq} \Big )^2}\over{ \Big ( \Tilde{\rho}^{(c)}_{eq} \Big )^2 - \Big ( \Tilde{\rho}^{(0)}_{eq} \Big )^2}} \Bigg ]\Bigg ( {{\Tilde{\rho}^{(0)}_{eq}}\over{\Tilde{\rho}^{(c)}_{eq}}}\Bigg )^{|i-j|}.
\end{aligned}
\end{equation}
Here, the inequality follows from $|\langle u_m | u_m \rangle| \geq 1$. Additionally, we use the fact that $|i-m| + |m-j|= |i-j|$ if $m \in (\min{\{i,j\}}, \max{\{i,j\}})$, and $|i - m| + |m - j|= |i - j| +2|m|$ if $m \notin (\min{\{i,j\}}, \max{\{i,j\}})$, to derive the penultimate equality. This completes the estimation of the contribution from the background gravitational potential.

Similarly, from Sec. (\ref{subsec: Taylor expansion construction}) and Eq. (\ref{equ: the matrix element of the background wave function}), we can find $|\langle w_l | \Tilde{\Psi}_{eq} | u_j \rangle| = |\langle u_j | \Tilde{\Psi}_{eq} | w_l \rangle|\sim 
\Big ( \Tilde{\rho}^{(0)}_{eq}/\Tilde{\rho}^{(c)}_{eq} \Big )^{|j-l+1/2|-1/2}$ for $q=0$,
$\Big ( \Tilde{\rho}^{(0)}_{eq}/\Tilde{\rho}^{(c)}_{eq} \Big )^{|j+l+1/2|-1/2}$ for $q=1,...,n^{(0)}_{eq}-1$, and
$\Big ( \Tilde{\rho}^{(0)}_{eq}/\Tilde{\rho}^{(c)}_{eq} \Big )^{|l-j+1/2|-1/2}$ for$q=n^{(0)}_{eq}$.
We take the case $q=1,...,n^{(0)}_{eq}-1$ as an example to estimate the perturbed self-gravity energy tensor. The other cases follow a similar approach for evaluation.

A simple calculation yields that:

\begin{equation}
\label{equ: the upper bound of perturbed self-gravity energy tensor}   
\begin{aligned}
|\langle u_i |\Tilde{\Psi}_{eq}(2E^{(0)}_{eq}\nabla^{-2})\Tilde{\Psi}_{eq} | u_j \rangle| &  \leq \sum_{l} \Big ( 1 - \delta_{\Tilde{k}^2_{\perp}0}\delta_{\Tilde{\kappa}^2_{l}0} \Big ) {{|\langle u_i | \Tilde{\Psi}_{eq} | w_l \rangle | | \langle w_l | \Tilde{\Psi}_{eq} | u_j \rangle | }\over{ \Big ( \Tilde{\kappa}^2_l+\Tilde{k}^2_{\perp}\Big )\langle w_l | w_l \rangle }} \\
& \sim \sum_{l} {{1 - \delta_{\Tilde{k}^2_{\perp}0}\delta_{\Tilde{\kappa}^2_l0}}\over{ \Big ( \Tilde{\kappa}^2_l+\Tilde{k}^2_{\perp}\Big )\langle w_l | w_l \rangle }} \Bigg ( {{\Tilde{\rho}^{(0)}_{eq}}\over{\Tilde{\rho}^{(c)}_{eq}}}\Bigg )^{ \Big |i + l + {{1}\over{2}} \Big |+ \Big |j+l+{{1}\over{2}} \Big| - 1} \leq h \Bigg ( {{q}\over{n^{(0)}_{eq}}}, \Tilde{k}^2_{\perp} \Bigg ) \Bigg ( {{\Tilde{\rho}^{(0)}_{eq}}\over{\Tilde{\rho}^{(c)}_{eq}}}\Bigg )^{|i-j| - 1}.
\end{aligned} 
\end{equation}
Here, the function $h$ is defined by Eq. (\ref{equ: definition of function h}). The last inequality follows from the conditions $\Tilde{\rho}^{(0)}_{eq} < \Tilde{\rho}^{(c)}_{eq}$ and $|i + l +1/2|+|j+l+1/2| \geq |i + l + 1/2 - (j+l+1/2)| = |i-j|$.

Finally, the last term in Eq. (\ref{equ: the off-diagonal element formula}) can be estimated as follows:

\begin{equation}
\label{equ: : the upper bound of perturbed self-gravity energy tensor times background gravitational potential}
\begin{aligned}
 \Big |\sum_{m} {{\langle u_i |\Tilde{\Psi}_{eq}(2E^{(0)}_{eq}\nabla^{-2})\Tilde{\Psi}_{eq} | u_m \rangle \langle u_m | \Tilde{V}_{eq} | u_j \rangle}\over{\langle u_m | u_m \rangle}} \Big |  & \leq  \sum_{m} |\langle u_i |\Tilde{\Psi}_{eq}(2E^{(0)}_{eq}\nabla^{-2})\Tilde{\Psi}_{eq} | u_m \rangle| | \langle u_m | \Tilde{V}_{eq} | u_j \rangle| \\
& \lesssim h \Bigg ( {{q}\over{n^{(0)}_{eq}}}, \Tilde{k}^2_{\perp} \Bigg ) \sum_{m} \Bigg ( {{\Tilde{\rho}^{(0)}_{eq}}\over{\Tilde{\rho}^{(c)}_{eq}}}\Bigg )^{|i-m| + |m-j| - 1} \\
 & = h \Bigg ( {{q}\over{n^{(0)}_{eq}}}, \Tilde{k}^2_{\perp} \Bigg )  \Bigg [ |i-j| + {{\Big ( \Tilde{\rho}^{(c)}_{eq} \Big )^2 + \Big (\Tilde{\rho}^{(0)}_{eq} \Big )^2}\over{ \Big ( \Tilde{\rho}^{(c)}_{eq} \Big )^2 - \Big ( \Tilde{\rho}^{(0)}_{eq} \Big )^2}} \Bigg ]\Bigg ( {{\Tilde{\rho}^{(0)}_{eq}}\over{\Tilde{\rho}^{(c)}_{eq}}}\Bigg )^{|i-j|-1}.
\end{aligned}
\end{equation}
Here, the first inequality follows from the fact $|\langle u_m | u_m \rangle| \geq 1$. It is noteworthy that the final equality holds by the same reasoning as presented in Eq. (\ref{equ: the upper bound of perturbed self-gravity energy tensor}).

Consequently, substituting the above estimations into Eq. (\ref{equ: the off-diagonal element formula}) leads to the final conclusion. 
\end{proof}

\renewcommand{\theequation}{D\arabic{equation}}
\setcounter{equation}{0}
\section{Upper Bound for Perturbed Self-Gravity Energy Tensor $\langle g |\hat{\mathbf{G}}|f \rangle$}
\label{appendix: Uniform Boundedness of the Perturbed Self-Gravity Energy Tensor in the Operator Theta}

This appendix presents a proof establishing the existence of an upper bound for the perturbed self-gravity energy tensor. We have known that the operator $\hat{\mathbf{G}}$ preserves the decoupling of eigen-functions within  subspaces $\mathbb{T}^{(cos)}_{q}[0,2\pi]$ and $\mathbb{T}^{(sin)}_{q}[0,2\pi]$ separately for all $q = 0,1,...,n^{(0)}_{eq}$ (see Sec. (\ref{subsubsec: Standing Wave Basis Expansion})), we next want to show that the perturbed self-gravity energy has an upper bound.

Given any two periodic functions $f$ and $g \in \mathbb{T}^{(cos)}_{q}[0,2\pi]$ or $\mathbb{T}^{(sin)}_{q}[0,2\pi]$, the normalized perturbed self-gravity energy tensor is given by:

\begin{equation}
\label{equ: perturbed self-gravity energy tensor}
\begin{aligned}
{{\langle g |\hat{\mathbf{G}} | f \rangle }\over{16\Tilde{\rho}^{(0)}_{eq}}} & = \langle g | \Tilde{\Psi}_{eq} (2E^{(0)}_{eq}\nabla^{-2}) \Tilde{\Psi}_{eq} | f \rangle \\
& = \sum_{l'}\sum_{l} {{\langle g | \Tilde{\Psi}_{eq} | w_{l'} \rangle \langle w_{l'} | 2E^{(0)}_{eq} \nabla^{-2} | w_l \rangle \langle w_l | \Tilde{\Psi}_{eq} | f \rangle}\over{\langle w_{l'} | w_{l'} \rangle \langle w_l | w_l \rangle }} = - \sum_{l} \Big ( 1- \delta_{\Tilde{k}^2_{\perp}0}\delta_{\Tilde{\kappa}^2_l0}\Big )  {{\langle g | \Tilde{\Psi}_{eq} | w_l \rangle \langle w_l | \Tilde{\Psi}_{eq} | f \rangle }\over{ \Big ( \Tilde{\kappa}^2_{l} +\Tilde{k}^2_{\perp} \Big )\langle w_l | w_l \rangle } }.    
\end{aligned}
\end{equation}
Here, the first equality is based on the completeness of the basis set $\{w_l\}$ given by Eq. (\ref{equ: the definition of the standing wave basis in the dual sub-space}), and the last equality holds because $\langle w_{l'} | \nabla^2 | w_l \rangle /(2E^{(0)}_{eq})=-\Big ( \Tilde{\kappa}^2_{l} +\Tilde{k}^2_{\perp} \Big )\langle w_{l'}| w_l \rangle \delta_{l'l}$ with $\Tilde{\kappa}_l \equiv \kappa_l/n^{(0)}_{eq} = 2l+p/n^{(0)}_{eq} = (2l+1) - q/n^{(0)}_{eq}$, which represents the momentum of the standing wave basis $w_l$. The same with Appendix \ref{appendix: The General Property of the Matrix Representation}, the term $(1 - \delta_{\Tilde{k}^2_{\perp}0}\delta_{\Tilde{\kappa}^2_{l}0})/( \Tilde{\kappa}^2_{l}+ \Tilde{k}^2_{\perp})$ is explicitly set to zero when $\Tilde{k}^2_{\perp} = \Tilde{\kappa}^2_l = 0$. This ensures the elimination of mass contributions under this condition and satisfies the mass conservation constraint.

Using H\"{o}lder equality, we have $|\langle g | \Tilde{\Psi}_{eq} | w_l \rangle| \leq \sqrt{2\langle \Tilde{\Psi}_{eq} | \Tilde{\Psi}_{eq} \rangle \langle g | g \rangle }$ and $|\langle w_l | \Tilde{\Psi}_{eq} | f \rangle| \leq \sqrt{2\langle \Tilde{\Psi}_{eq} | \Tilde{\Psi}_{eq} \rangle \langle f | f \rangle }$. Consequently, the above perturbed self-gravity energy tensor can be estimated as follows:

\begin{equation}
\label{equ: the final uniform boundness of the perturbed self-gravity energy tensor}
\begin{aligned}
\Bigg |{{\langle g |\hat{\mathbf{G}} | f \rangle }\over{16\Tilde{\rho}^{(0)}_{eq}}}  \Bigg | & \leq \sum_{l} \Big ( 1- \delta_{\Tilde{k}^2_{\perp}0}\delta_{\Tilde{\kappa}^2_{l}0}\Big ) {{| \langle g | \Tilde{\Psi}_{eq} | w_l \rangle | | \langle w_l | \Tilde{\Psi}_{eq} | f \rangle | }\over{ \Big ( \Tilde{\kappa}^2_l+\Tilde{k}^2_{\perp} \Big )\langle w_l |w_l \rangle }} \\
& \leq 2  \sum_{l} {{1- \delta_{\Tilde{k}^2_{\perp}0}\delta_{\Tilde{\kappa}^2_{l}0}}\over{ \Big ( \Tilde{\kappa}^2_l+\Tilde{k}^2_{\perp} \Big ) \langle w_l |w_l \rangle }} \langle \Tilde{\Psi}_{eq} | \Tilde{\Psi}_{eq} \rangle \sqrt{\langle f | f \rangle \langle g | g \rangle} \equiv 2 h\Bigg ( {{q}\over{n^{(0)}_{eq}}}, \Tilde{k}^2_{\perp} \Bigg ) \langle \Tilde{\Psi}_{eq} | \Tilde{\Psi}_{eq} \rangle \sqrt{\langle f | f \rangle \langle g | g \rangle}.
\end{aligned}
\end{equation}
Here, the first inequality follows from the triangle inequality, and the function $h$ is defined as follows:

\begin{equation}
\label{equ: definition of function h}
h \Bigg ( {{q}\over{n^{(0)}_{eq}}}, \Tilde{k}^2_{\perp} \Bigg ) \equiv \sum_{l} {{1- \delta_{\Tilde{k}^2_{\perp}0}\delta_{\Tilde{\kappa}^2_{l}0}}\over{ \Big (\Tilde{\kappa}^2_l+\Tilde{k}^2_{\perp} \Big )\langle w_l | w_l \rangle }}.
\end{equation}
Therefore, the function $h$ is well-defined and positive because $h( q/n^{(0)}_{eq}, \Tilde{k}^2_{\perp} ) \leq  (1-\delta_{\Tilde{k}^2_{\perp}0})/\Tilde{k}^2_{\perp} + \sum^{\infty}_{m=1} 1/m^2 < \infty$ with $(1-\delta_{\Tilde{k}^2_{\perp}0})/\Tilde{k}^2_{\perp}$ to be zero when $\Tilde{k}^2_{\perp}=0$. Moreover, we utilize the property $\langle w_l | w_l \rangle = 2\delta_{\Tilde{\kappa}^2_{l}0}+ \Big ( 1-\delta_{\Tilde{\kappa}^2_{l}0} \Big ) \geq 1$ to estimate the above inequality. Consequently, the perturbed self-gravity energy tensor is found to be bounded, as shown in Eq. (\ref{equ: the final uniform boundness of the perturbed self-gravity energy tensor}).

\renewcommand{\theequation}{E\arabic{equation}}
\setcounter{equation}{0}
\section{Dispersion Relation in Two-Component Wave Dark Matter}
\label{appendix: General Property for the dispersion relation of two-component wave dark matter}

This appendix analyzes the dispersion relation for the two-component wave dark matter, a simplified model for demonstrating propagating instability in quantum systems, which also provides a useful connection to the propagating instability in single-component wave dark matter. In particular, the general dispersion relation for the multi-component wave dark matter is also derived. Supposed there are $N$ components in the wave dark matter. Each component is labeled by $j$, $j=1,2,...,N$, with mass $m_j$ and described by its own wave function $\Psi_j$, which obeys Schr\"{o}dinger-Poisson equation: 

\begin{equation}
\label{equ: Schrodiger-Poisson equation}
\left\{\begin{aligned}
& \textbf{\textit{i}}{{\partial \Psi_j(\vec{x},t)}\over{\partial t}} = \Big ( -{{\nabla^2}\over{2m_j}} + m_jV(\vec{x},t) \Big )\Psi_j(\vec{x},t), \\
& {\nabla^2}V(\vec{x},t) = \sum^{N}_{j=1}|\Psi_j(\vec{x},t)|^2 - \Lambda,
\end{aligned}
\right.
\end{equation}
where $V$ is the gravitational potential and the square of the $j$-th wave function $|\Psi_j(\vec{x},t)|^2$ represents the mass density for the $j$-th component. The quantity $\Lambda$ represents the cosmological constant. To achieve a consistent equilibrium background, the cosmological constant should be the mean mass density, as explained in Sec. (\ref{sec: General Properties of the linearized Schrodinger-Poisson equation}). This process is also analogous to the global neutralization of quantum plasma with multi-stream, as described by \cite{HMF2000}.

Inspired by the two-fluid gravitational system with homogeneous flows \cite{FP_second1984}, each component is assigned a plane wave with constant mass density $\rho^{(0)}_j$ and constant momentum $\vec{p}^{(0)}_j$, i.e., $\Psi^{(0)}_j(\vec{x},t)=\sqrt{\rho^{(0)}_j}\exp {\Big [\textbf{\textit{i}}\Big ( \vec{p}^{(0)}_j \cdot \vec{x}-E^{(0)}_jt \Big ) \Big ]}$ with $E^{(0)}_j=|\vec{p}^{(0)}_j|^2/2m_j$, as the background. This configuration produces a uniform background density, resulting in $\Lambda = \sum^{N}_{j=1}\rho^{(0)}_j$. Consequently, this setup leads to a force-equilibrium configuration, resulting in a constant gravitational potential, which can be shifted to zero.

To study the stability, the small perturbed wave function for the $j$-th component, denoted as $\delta \Psi_j$, is introduced. These perturbed wave functions satisfy the linearized Schr\"{o}dinger equation, given by:

\begin{equation}
\label{equ: linear Schrodiger equation}
\textbf{\textit{i}}{{\partial \delta \Psi_j(\vec{x},t)}\over{\partial t}} = -{{\nabla^2}\over{2m_j}}\delta\Psi_j(\vec{x},t) + m_j\delta V(\vec{x},t)\Psi^{(0)}_j(\vec{x},t), 
\end{equation}
where $\delta V$ is the perturbed gravitational potential, which satisfies the linearized Poisson equation:

\begin{equation}
\label{equ: linear Poisson equation}
{\nabla^2}\delta V(\vec{x},t) = \sum^{N}_{j=1}\Big ( \Psi^{(0)}_j(\vec{x},t) \delta\Psi_j^{*}(\vec{x},t) + {\Psi^{(0)}_j}^{*}(\vec{x},t) \delta\Psi_j(\vec{x},t) \Big ).
\end{equation}
It should be noted that Eq. (\ref{equ: linear Poisson equation}) implies that such perturbations do not introduce additional mass into the system and adhere to the mass conservation constraint. 

The perturbed wave function for each component can be expressed as $\delta \Psi_j= ( \textit{R}_j + \textbf{\textit{i}}\textit{I}_j ) \Psi^{(0)}_j$, where $\textit{R}_j$ and $\textit{I}_j$ are real functions. Specifically, $\sqrt{\rho^{(0)}_j} ( \textit{R}_j + \textbf{\textit{i}}\textit{I}_j )$ represents the perturbed wave function for the $j$-th component in the rest frame, where the background of the $j$-th component is static. Thus, $\textit{R}_j$ and $\textit{I}_j$ are the real and imaginary parts, respectively, of the perturbed wave function, normalized with respect to $\sqrt{\rho^{(0)}_j}$.

The above expression is connected to fluid variables, facilitating the derivation of conservation laws. To identify the corresponding fluid variables, Madelung transformation is applied to the wave function for each component, i.e., $\Psi_j(\vec{x},t)=\sqrt{\rho_j(\vec{x},t)}\exp{(\textbf{\textit{i}}S_j(\vec{x},t))}$, where $\rho_j(\vec{x},t)$ is the mass density and $S_j(\vec{x},t)$ is the action. The velocity is $\nabla S_j/m_j$. Next, let $\rho_j(\vec{x},t)=\rho^{(0)}_j+\delta\rho_j(\vec{x},t)$ and $S_j(\vec{x},t)=\Big ( \vec{p}^{(0)}_j \cdot \vec{x}-E^{(0)}_jt \Big )+\delta S_j(\vec{x},t)$ and expand $\Psi_j$ to the first order, it finds that $\delta\rho_j=2\rho^{(0)}_j\textit{R}_j$ and $\delta S_j= \textit{I}_j$. Consequently, Eqs. (\ref{equ: linear Schrodiger equation}) and (\ref{equ: linear Poisson equation}) become:

\begin{equation}
\label{equ: linear Madelung equtaion}
\left\{\begin{aligned}
& {{\partial \delta \rho_j(\vec{x},t)}\over{\partial t}} +  {{\vec{p}^{(0)}_j}\over{m_j}} \cdot \nabla \delta \rho_j(\vec{x},t) + \rho^{(0)}_j \nabla^2 \Bigg ( {{\delta S_j(\vec{x},t)}\over{m_j}}\Bigg ) =0, \\
& {{\partial \delta S_j(\vec{x},t)}\over{\partial t}} + {{\vec{p}^{(0)}_j}\over{m_j}} \cdot \nabla \delta S_j(\vec{x},t)-{{\nabla^2 \delta \rho_j(\vec{x},t)}\over{4m_j\rho^{(0)}_j}}  + m_j \delta V(\vec{x},t)=0, \\
& {\nabla^2}\delta V(\vec{x},t) = \sum^{N}_{j=1}\delta \rho_j(\vec{x},t).
\end{aligned}
\right.
\end{equation}
The first equation in Eq. (\ref{equ: linear Madelung equtaion}) is the linearized continuity equation, where $\nabla^2(\delta S_j/m_j)$ indicates the compression term. The second equation is Hamilton-Jacobi equation, in which $-\nabla^2 \delta \rho_j/4m_j\rho^{(0)}_j=-\nabla^2 \textit{R}_j/2m_j$ denotes Bohm quantum potential. Lastly, the third equation in Eq. (\ref{equ: linear Madelung equtaion}) is the Poisson equation, with the total perturbed density as the source. It is noteworthy that $(\vec{p}^{(0)}_j/m_j)\cdot \nabla$ appearing in the continuity and Hamilton-Jacobi equations represents the convection term. 

Furthermore, applying the Laplacian operator($\nabla^2$) on the Hamilton-Jacobi equation and using the continuity and Poisson equation to express $\nabla^2 \delta S_j$ and $\delta V$ as functions of the perturbed density, we obtain:

\begin{equation}
\label{equ: differential equation of the perturbed density}
-{{1}\over{\rho^{(0)}_j}} \Bigg [ \Big ( {{\partial}\over{\partial t}} + {{\vec{p}^{(0)}_j }\over{m_j}} \cdot \nabla \Big )^2 + {{\nabla^2\nabla^2}\over{4m^2_j}} \Bigg ] \delta \rho_j(\vec{x},t) + \sum^{N}_{l=1} \delta \rho_l(\vec{x},t) = 0.
\end{equation}
Here, the operator within the parentheses represents the material derivative with respect to the $j$-th component background velocity ($\vec{p}^{(0)}_j/m_j$). The Bohm quantum potential introduces the operator $\nabla^2\nabla^2/4m^2_l$.

To determine the dispersion relation, a Fourier transformation is applied to Eq. (\ref{equ: differential equation of the perturbed density}), yielding:

\begin{equation}
\label{equ: algebraic equation for the dispersion relation}
\sum^{N}_{l=1} \left\{ 1 + {{1}\over{\rho^{(0)}_l}}\Bigg [ \Bigg ( \omega + {{\vec{p}^{(0)}_l \cdot \vec{k}}\over{m_l}}\Bigg )^2 - {{|\vec{k}|^4}\over{4m^2_l}} \Bigg ] \delta_{jl} 
\right\} \delta \rho_l(\vec{k},\omega) = 0.
\end{equation}
Here, $\omega$ is the angular frequency, $\vec{k}$ is the wave vector, and $\delta_{jl}$ is Kronecker delta, with $j=1,2,...,N$. It should be noted that the same notation is used for physical quantities in both real and Fourier space. This mixed usage does not introduce any ambiguity in the context of identifying propagating instability. This convention is applied throughout this appendix unless otherwise specified. 

Equation (\ref{equ: algebraic equation for the dispersion relation}) forms a system of linear equations that has a non-trivial solution if and only if the determinant of the coefficient matrix is zero. This condition provides the dispersion relation. A straightforward calculation yields the dispersion relation as:

\begin{equation}
\label{equ: dispersion relation for N-component}
\sum^{N}_{j=1}{{\rho^{(0)}_j}\over{\Bigg ( \omega + {{\vec{p}^{(0)}_j \cdot \vec{k}}\over{m_j}}\Bigg )^2 - {{|\vec{k}|^4}\over{4m^2_j}}}}=-1.
\end{equation}
It is noteworthy that the classical multi-fluid gravitational system with homogeneous flows shares the same dispersion relation as Eq. (\ref{equ: dispersion relation for N-component}), except that the term $|\vec{k}|^4/4m^2_j$ is replaced by $|\vec{k}|^2$ multiplied by the square sound speed, assuming all components have the same sound speed \cite{FP_second1984}.

To examine the existence of the propagating instability, we consider a two-component wave dark matter system ($N=2$) with the identical particle mass ($m_1=m_2=m$) and the same mass density ($\rho^{(0)}_1=\rho^{(0)}_2=\rho$). Without losing of generality, e choose a specific mass scale such that $m=1$. Additionally, an appropriate center of mass frame is selected, so that the background plane waves propagate along the $z$-axis, i.e., $\vec{p}^{(0)}_1=-\vec{p}^{(0)}_2=p_{bg}\hat{z}$, where $p_{bg}>0$. In this scenario, the dispersion relation becomes:

\begin{equation}
\label{equ: dispersion relation for two-component}
\Tilde{\omega}^4-2\Big ( \Tilde{p}^2 + {{\Tilde{k}^4}\over{4}} - \Tilde{\rho}\Big )\Tilde{\omega}^2+ \Big ( \Tilde{p}^2 - {{\Tilde{k}^4}\over{4}} \Big ) \Big ( \Tilde{p}^2 - {{\Tilde{k}^4}\over{4}} +2\Tilde{\rho} \Big )=0,
\end{equation}
where $\Tilde{\omega}=\omega/p^2_{bg}$, $\Tilde{\rho}=\rho/p^4_{bg}$, $\Tilde{p} = p/p_{bg}$ with $p \equiv \vec{k} \cdot \hat{z}$, and $\Tilde{k} = \lvert\vec{k}/p_{bg} \rvert$.

It is important to note that the parameter $\Tilde{\rho}$ represents the square of the ratio between the quantum and Jeans wavelengths and serves the same role as $\Tilde{\rho}^{(0)}_{eq}$ mentioned in the main text (see Secs. (\ref{sec: Nonlinear periodic equilibrium construction}), (\ref{sec: General Version of Bloch's theorem for the Stability analysis to the Periodic equilibrium Background}), and (\ref{sec: Comparison Between Theoretical Prediction and Simulating result})).   

Equation (\ref{equ: dispersion relation for two-component}) has the following two roots:

\begin{equation}
\label{equ: solution of dispersion relation for two-component}
\Tilde{\omega}_{\pm}^2 = \Big ( \Tilde{p}^2 + {{\Tilde{k}^4}\over{4}} - \Tilde{\rho} \Big )\pm \sqrt{\Tilde{p}^2 \Tilde{k}^4-4\Tilde{p}^2 \Tilde{\rho}+\Tilde{\rho}^2}.
\end{equation}
The significance of the propagating instability increases with the absolute value of the imaginary part of $\Tilde{\omega}_{\pm}^2$. Consequently, it is of interest to determine the conditions under which the term under the square root in Eq. (\ref{equ: solution of dispersion relation for two-component}) becomes maximally negative. With $\Tilde{\rho}=2\Tilde{p}^2$, this expression achieves its most negative value as $-\Tilde{p}^2\beta$ with $\beta\equiv 4\Tilde{p}^2-\Tilde{k}^4$. The eigenvalue becomes $\omega_\pm^2=-\beta/4 \pm \textbf{\textit{i}}|\Tilde{p}|\sqrt{\beta}$. Note that $\Tilde{k}^2 = \Tilde{p}^2+\Tilde{k}^2_\perp$, where $\Tilde{k}_\perp$ represents the normalized magnitude of the perpendicular wave number. When $\Tilde{k}_\perp=0$, $\beta=\Tilde{p}^2(4-\Tilde{p}^2)$. Under this condition, the maximum value of $|\Tilde{p}|\sqrt{\beta}$ is achieved at $\Tilde{p}^2=8/3$.

It is noted that the classical two-fluid gravitational system with equal density and pressure also shares the same solution as Eq. (\ref{equ: solution of dispersion relation for two-component}). However, in the classical system, the natural normalization factor is the sound speed rather than the flow velocity. Consequently, the Mach number, defined as the ratio of the flow speed to the sound speed, is introduced in the dispersion relation for the classical two-fluid gravitational system \cite{FP_second1984}. In this quantum system, we use the Bohm quantum potential $|\vec{k}|^2/4$ to represent the squared sound speed. As $\Tilde{k}_\perp=0$, the effective quantum Mach number is found to be $2/|\Tilde{p}|$. The proceeding result for the propagating instability indicates that the maximum absolute value of the imaginary part of $\Tilde{\omega}_{\pm}^2$ occurs when the quantum Mach number satisfies $2/|\Tilde{p}| = \sqrt{1.5} \sim 1.2$. When the Mach number is sufficiently large, the propagating instability will be weaken.

To identify the stability type derived from Eqs. (\ref{equ: dispersion relation for two-component}) and (\ref{equ: solution of dispersion relation for two-component}), the following procedure is introduced. The key idea is to treat Eqs. (\ref{equ: dispersion relation for two-component}) as the characteristic polynomial of $2 \times 2$ matrix with $\Tilde{\omega}^2$ as the argument. In this context, the coefficients for $\Tilde{\omega}^2$ and $\Tilde{\omega}^0$ correspond to the negative trace and determinant of this $2 \times 2$ matrix, respectively. For clarity, the trace, denoted as $T$, and the determinant, denoted as $D$, are defined as follows:

\begin{equation}
\label{equ: the definition of trace and determinant for the two-component case}
\left \{ \begin{aligned}
& T( \Tilde{p}^2, \Tilde{k}^2_\perp, \Tilde{\rho} ) = 2\Big ( \Tilde{p}^2 + {{\Tilde{k}^4}\over{4}} - \Tilde{\rho}\Big ); \\
& D( \Tilde{p}^2, \Tilde{k}^2_\perp, \Tilde{\rho} ) = \Big ( \Tilde{p}^2 - {{\Tilde{k}^4}\over{4}} \Big ) \Big ( \Tilde{p}^2 - {{\Tilde{k}^4}\over{4}} +2\Tilde{\rho} \Big ).
\end{aligned}
\right.
\end{equation}

It is straightforward to find $\Tilde{\omega}^2_{+}+\Tilde{\omega}^2_{-} = T$ and $\Tilde{\omega}^2_{+}\Tilde{\omega}^2_{-} = D$. Additionally, the term inside the square root in Eq. (\ref{equ: solution of dispersion relation for two-component}) is proportional to $T^2-4D$, known as the discriminant. Using these relationships, the procedure for identifying the stability type can be summarized as follows:
\\\\
\noindent\textbf{1.If the discriminant $\bm{T^2-4D<0}$, then $\bm{\Tilde{\omega}^2_{\pm} \in \mathbb{C} \backslash \mathbb{R}}$, indicating the presence of propagating instability.}

\noindent\textbf{2. If $\bm{T^2-4D>0}$ with the negative determinant $\bm{D<0}$, then $\bm{\Tilde{\omega}^2_{-} < 0 <  \Tilde{\omega}^2_{+}}$. Since the eigenvalue $\bm{\Tilde{\omega}^2_{-}}$ is negative, this indicates the presence of Jeans instability.}

\noindent\textbf{3. If $\bm{T^2-4D>0}$ with $\bm{D>0}$ and the negative trace $\bm{T<0}$, then both eigenvalues $\bm{\Tilde{\omega}^2_{\pm}}$ are negative, indicating the presence of Jeans instability.}

\noindent\textbf{4. If $\bm{T^2-4D>0}$ with $\bm{D>0}$ and $\bm{T>0}$, then $\bm{\Tilde{\omega}^2_{\pm} > 0}$. Therefore, this indicates a stable mode.}
\\\\

We will derive the general properties of the dispersion relation for two-component wave dark matter, as given in Eq. (\ref{equ: solution of dispersion relation for two-component}), in the following paragraphs. Recall that the stability can be characterized by a three-dimensional phase diagram, where any point is labeled as $\Big (\Tilde{p}^2, \Tilde{k}^2_\perp, \Tilde{\rho} \Big )$, and eigenvalues $\Tilde{\omega}^2_{\pm} = \Tilde{\omega}^2_{\pm}\Big (\Tilde{p}^2, \Tilde{k}^2_\perp, \Tilde{\rho} \Big )$.

Followed by the above argument, the stability types in $\Big (\Tilde{p}^2, \Tilde{k}^2_\perp, \Tilde{\rho} \Big )$-phase diagram are summarized as follows: 

\begin{equation}
\label{equ: phase diagram for two-component}
\begin{split}
& \text{If } \Big (\Tilde{p} - 1\Big )^2 + \Tilde{k}^2_\perp < 1 \text{ or } \Big (\Tilde{p} + 1\Big )^2 + \Tilde{k}^2_\perp < 1:
\left \{ \begin{aligned}
& \Tilde{\omega}^2_{\pm}\Big ( \Tilde{p}^2, \Tilde{k}^2_\perp, \Tilde{\rho} \Big ) > 0 \text{ if } 0<  \Tilde{\rho} < \Tilde{\rho}_{-}(\Tilde{p}^2, \Tilde{k}^2_\perp), \\
& \Tilde{\omega}^2_{\pm}\Big ( \Tilde{p}^2, \Tilde{k}^2_\perp, \Tilde{\rho} \Big ) \in \mathbb{C} \backslash \mathbb{R} \text{ if } \Tilde{\rho}_-(\Tilde{p}^2, \Tilde{k}^2_\perp) < \Tilde{\rho} < \Tilde{\rho}_+(\Tilde{p}^2, \Tilde{k}^2_\perp), \\
& \Tilde{\omega}^2_{\pm}\Big ( \Tilde{p}^2, \Tilde{k}^2_\perp, \Tilde{\rho} \Big ) < 0 \text{ if } \Tilde{\rho} > \Tilde{\rho}_+(\Tilde{p}^2, \Tilde{k}^2_\perp),
\end{aligned}
\right. \\
& \text{ }\\
& \text{If } \Big (\Tilde{p} - 1\Big )^2 + \Tilde{k}^2_\perp > 1 \text{ and } \Big (\Tilde{p} + 1\Big )^2 + \Tilde{k}^2_\perp > 1:
\left \{\begin{aligned}
& \Tilde{\omega}^2_{\pm}\Big ( \Tilde{p}^2, \Tilde{k}^2_\perp, \Tilde{\rho} \Big ) > 0 \text{ if } 0 < \Tilde{\rho} < \Tilde{\rho}_{D}(\Tilde{p}^2, \Tilde{k}^2_\perp), \\
& \Tilde{\omega}^2_{+}\Big ( \Tilde{p}^2, \Tilde{k}^2_\perp, \Tilde{\rho} \Big ) > 0 \text{ and } \Tilde{\omega}^2_{-}\Big ( \Tilde{p}, \Tilde{k}_\perp, \Tilde{\rho} \Big ) < 0 \text{ if } \Tilde{\rho} > \Tilde{\rho}_{D}(\Tilde{p}^2, \Tilde{k}^2_\perp),  \\
\end{aligned}
\right.\\
& \text{where: }
\left \{\begin{aligned}
& \Tilde{\rho}_\pm \Big (\Tilde{p}^2,\Tilde{k}^2_\perp \Big ) \equiv  2\Tilde{p}^2 \pm\sqrt{4\Tilde{p}^4 -  \Tilde{p}^2\Big (\Tilde{p}^2+\Tilde{k}^2_\perp \Big )^2}, \\
& \Tilde{\rho}_{D}\Big ( \Tilde{p}^2, \Tilde{k}^2_\perp\Big ) \equiv {{1}\over{8}}\Big [ \Big ( \Tilde{p}^2 + \Tilde{k}^2_\perp \Big ) ^2 - 4\Tilde{p}^2\Big ].
\end{aligned}
\right. \\
\end{split} 
\end{equation}
Here, the functions $\Tilde{\rho}_{\pm}(\Tilde{p}^2, \Tilde{k}^2_\perp)$ and $\Tilde{\rho}_{D}(\Tilde{p}^2, \Tilde{k}^2_\perp)$ satisfy $T^2 \Big (\Tilde{p}^2, \Tilde{k}^2_\perp, \Tilde{\rho}_{\pm}(\Tilde{p}^2, \Tilde{k}^2_\perp) \Big ) - 4D\Big ( \Tilde{p}^2, \Tilde{k}^2_\perp, \Tilde{\rho}_{\pm}(\Tilde{p}^2, \Tilde{k}^2_\perp) \Big )=0$ and $D\Big ( \Tilde{p}^2, \Tilde{k}^2_\perp, \Tilde{\rho}_{D}(\Tilde{p}^2, \Tilde{k}^2_\perp) \Big )=0$, respectively. Eq. (\ref{equ: phase diagram for two-component}) shows $\Tilde{\omega}^2_+( \Tilde{p}, \Tilde{k}_\perp, \Tilde{\rho} )$ is always positive definite when $\Big ( \Tilde{p} \pm 1\Big )^2 + \Tilde{k}^2_\perp > 1$. 

We have found that propagating instability occurs in the two-component wave dark matter system. However, this instability only arises when $\Big (\Tilde{p} - 1\Big )^2 + \Tilde{k}^2_\perp < 1$ or $\Big (\Tilde{p} + 1\Big )^2 + \Tilde{k}^2_\perp < 1$. These inequalities are physically connected to classical two-fluid gravitational systems, particularly when $\Tilde{k}^2_{\perp}=0$. Once $\Tilde{k}^2_{\perp}=0$, the inequalities simplify to $\Tilde{p}^2 < 4$, which corresponds to the supersonic regime, as previously discussed. This is consistent with classical two-fluid gravitational systems, where only supersonic flows exhibit propagating instability. \cite{FP_second1984, CMS1998}. 

There is always at least one zero frequency root for $\Tilde{\omega}^2_{\pm}$, i.e., $\Tilde{\omega}^2_{-} = 0$ or $\Tilde{\omega}^2_{+} = 0$, when either $\Big ( \Tilde{p} - 1\Big )^2 + \Tilde{k}^2_\perp = 1$ or $\Big ( \Tilde{p} + 1\Big )^2 + \Tilde{k}^2_\perp = 1$, which is independent with $\Tilde{\rho}$. This state will be discussed in the following paragraph. Without loss of generality, we further assume $\Tilde{p}>0$ in the following paragraph, which means that only the relation $\Big ( \Tilde{p} - 1\Big )^2 + \Tilde{k}^2_\perp = 1$ is considered. The same argument can be applied to the other $\Tilde{p}$.

These perturbed wave functions with zero frequency correspond to another force-equilibrium configuration. The wave functions for these two components are labeled by $\pm$, and the background wave functions are expressed as $\Psi^{(0)}_{\pm}=\sqrt{\rho} \exp{\Big [\textbf{\textit{i}} \Big (\pm p_{bg}\hat{z} \cdot \vec{x} - E_{bg}t \Big )\Big ]}$ with $p_{bg}>0$ and $E_{bg} = p^2_{bg}/2$.  The wave vector for the perturbed density can be expressed as $\vec{k} = p\hat{z} + \vec{k}_\perp$, and the equality implies $(p_{bg}-p)^2 + |\vec{k}_\perp|^2=p^2_{bg}$. Combining this fact with Eq. (\ref{equ: linear Madelung equtaion}),  these zero frequency perturbed wave functions are $2\sqrt{\rho}\delta\Psi_{\pm}=\pm \delta\rho \exp {\Big \{ \textbf{\textit{i}} \Big [ \pm\Big ( q\hat{z} - \vec{k}_\perp \Big ) \cdot \vec{x} - E_{bg}t\Big ]\Big \}}$ with $q \equiv p_{bg}-p$. Therefore, the perturbed wave functions are independent collided plane waves. The leading order effect of each perturbed wave function to the corresponding background wave function is to create a random distortion to the constant phase front of the background wave, which can also be an equilibrium state. 

\twocolumngrid

\label{lastpage}

\end{document}